\documentclass[11pt]{article}
\usepackage[T1]{fontenc}
\usepackage[utf8]{inputenc}
\usepackage{lmodern}
\usepackage[margin=1in]{geometry}
\usepackage{amsmath,amssymb,amsthm,mathtools,bm}
\usepackage{microtype,booktabs,tabularx,array,authblk}
\usepackage{xcolor,tikz,pgfplots}
\definecolor{boundblue}{RGB}{0,92,153}
\definecolor{rateorange}{RGB}{196,77,0}
\definecolor{keygreen}{RGB}{0,125,89}
\definecolor{accentpurple}{RGB}{126,69,143}
\definecolor{mutedgray}{RGB}{88,96,107}
\definecolor{auxblue}{RGB}{66,137,174}
\usetikzlibrary{arrows.meta,positioning,calc,fit}
\pgfplotsset{compat=1.18}
\usepackage{enumitem,aliascnt,needspace,etoolbox,placeins}
\makeatletter
\pretocmd{\l@section}{\Needspace{5\baselineskip}}{}{}
\makeatother
\BeforeBeginEnvironment{theorem}{\Needspace{6\baselineskip}}
\BeforeBeginEnvironment{proposition}{\Needspace{6\baselineskip}}
\BeforeBeginEnvironment{lemma}{\Needspace{6\baselineskip}}
\BeforeBeginEnvironment{corollary}{\Needspace{5\baselineskip}}
\usepackage{hyperref}
\hypersetup{colorlinks=true,linkcolor=blue!45!black,citecolor=blue!45!black,
 urlcolor=blue!45!black,pdfauthor={Hemant Mishra and Mark M. Wilde},pdftitle={Gaussian entanglement and quantum communication: geodesic optimization and multimode bounds}}
\usepackage[nameinlink,noabbrev]{cleveref}
\numberwithin{equation}{section}
\newtheorem{theorem}{Theorem}[section]
\theoremstyle{plain}
\newaliascnt{lemma}{theorem}
\newtheorem{lemma}[lemma]{Lemma}
\aliascntresetthe{lemma}
\theoremstyle{plain}
\newaliascnt{proposition}{theorem}
\newtheorem{proposition}[proposition]{Proposition}
\aliascntresetthe{proposition}
\theoremstyle{plain}
\newaliascnt{corollary}{theorem}
\newtheorem{corollary}[corollary]{Corollary}
\aliascntresetthe{corollary}
\theoremstyle{definition}
\newaliascnt{definition}{theorem}

\aliascntresetthe{definition}
\theoremstyle{definition}
\newaliascnt{example}{theorem}

\aliascntresetthe{example}
\theoremstyle{remark}
\newaliascnt{remark}{theorem}

\aliascntresetthe{remark}
\crefname{theorem}{Theorem}{Theorems}
\Crefname{theorem}{Theorem}{Theorems}
\crefname{lemma}{Lemma}{Lemmas}
\Crefname{lemma}{Lemma}{Lemmas}
\crefname{proposition}{Proposition}{Propositions}
\Crefname{proposition}{Proposition}{Propositions}
\crefname{corollary}{Corollary}{Corollaries}
\Crefname{corollary}{Corollary}{Corollaries}
\crefname{definition}{Definition}{Definitions}
\Crefname{definition}{Definition}{Definitions}
\crefname{example}{Example}{Examples}
\Crefname{example}{Example}{Examples}
\crefname{remark}{Remark}{Remarks}
\Crefname{remark}{Remark}{Remarks}
\crefformat{equation}{#2(#1)#3}
\Crefformat{equation}{Eq.~#2(#1)#3}
\crefrangeformat{equation}{#3(#1)#4--#5(#2)#6}
\Crefrangeformat{equation}{Eqs.~#3(#1)#4--#5(#2)#6}
\crefmultiformat{equation}{#2(#1)#3}{ and~#2(#1)#3}{, #2(#1)#3}{, and~#2(#1)#3}
\Crefmultiformat{equation}{Eqs.~#2(#1)#3}{ and~#2(#1)#3}{, #2(#1)#3}{, and~#2(#1)#3}

\crefname{section}{Section}{Sections}
\Crefname{section}{Section}{Sections}
\crefname{figure}{Figure}{Figures}
\Crefname{figure}{Figure}{Figures}
\crefname{table}{Table}{Tables}
\Crefname{table}{Table}{Tables}
\crefname{appendix}{Appendix}{Appendices}
\Crefname{appendix}{Appendix}{Appendices}
\DeclareMathOperator{\Tr}{Tr}
\DeclareMathOperator{\diag}{diag}
\DeclareMathOperator{\Rea}{Re}
\DeclareMathOperator{\supp}{supp}
\DeclareMathOperator{\arccoth}{arccoth}
\DeclareMathOperator{\arctanh}{arctanh}
\DeclareMathOperator{\Proj}{Proj}
\DeclareMathOperator{\Log}{Log}
\DeclareMathOperator*{\argmin}{arg\,min}
\newcommand{\R}{\mathbb R}

\newcommand{\id}{\mathrm{id}}
\newcommand{\cP}{\mathcal C_{\mathrm{PPT}}}
\newcommand{\cS}{\mathcal C_{\mathrm{sep}}}
\newcommand{\cF}{\mathcal C}
\newcommand{\cK}{\mathcal K}
\newcommand{\norm}[1]{\left\|#1\right\|}
\newcommand{\inner}[2]{\langle #1,#2\rangle}
\newcommand{\ket}[1]{|#1\rangle}
\newcommand{\bra}[1]{\langle#1|}
\newcommand{\gent}{g_{\mathrm e}}

\newcommand{\I}{I}
\title{\bfseries Gaussian entanglement and 
quantum communication:\\geodesic optimization and multimode bounds}
\author[1]{Hemant K. Mishra\thanks{hemantmishra1124@iitism.ac.in}}
\author[2]{Mark M. Wilde\thanks{wilde@cornell.edu}}
\affil[1]{Department of Mathematics and Computing, Indian Institute of Technology (ISM) Dhanbad, Jharkhand 826004, India}
\affil[2]{School of Electrical and Computer Engineering, Cornell University, Ithaca, New York 14850, USA}
\date{October 13, 2026}
\begin{document}
\maketitle
\begin{abstract}
Loss and noise limit the use of light to distribute entanglement, transmit quantum information, and establish secret keys. Gaussian states have a compact description in terms of means and covariance matrices, but their entanglement still requires global optimization. We develop a framework for computing Gaussian-restricted versions of the relative entropy of entanglement and Rains entanglement. An integral representation reveals a convex geometry under matrix geometric means and reduces Gaussian Rains entanglement to relative entropy minimized over Gaussian states with positive partial transpose. We obtain unique optimal references and additivity for independent Gaussian input states. Our convex-majorization algorithm has an explicit geometric convergence rate, uniform in the faithful regularization parameter and valid for pure input states. Controlled numerical integration and approximate convex subproblem solutions give a bit-complexity bound that is polynomial in accuracy bits for each fixed rational input covariance matrix, with explicit dependence on covariance size. These calculations bound extractable entanglement and, through the relative entropy of entanglement, extractable secret key. Teleportation simulation, energy allocation, and a convex optimization of passive-basis loss cores give bounds on quantum and private communication under an input-energy constraint, allowing arbitrary two-way classically assisted protocols. Two- and three-mode examples compare the bounds with achievable rates, including matching pure-loss benchmarks. Finally, we separate the Gaussian-restricted quantities from their unrestricted counterparts and give a photon-number truncation method with controlled error for computing the latter.
\end{abstract}
\clearpage
\begingroup
\setlength{\parskip}{0pt}
\tableofcontents
\endgroup
\clearpage

\section{Introduction}\label{sec:intro}
\subsection{Background and motivation}
Light is a natural carrier of quantum information, but loss and noise reduce
its usefulness for distributing entanglement and establishing secret keys.
To assess an optical communication scheme, one therefore needs both an
achievable protocol and an upper bound on what every protocol could achieve.
Entanglement measures provide such upper bounds. The difficulty is that even
a few optical modes have an infinite-dimensional state space, so evaluating
these measures directly can be demanding.

Gaussian states are an important setting in which to address this difficulty.
They describe, for example, thermal light, squeezed light, and their evolution
through a broad class of noisy optical devices. A Gaussian state is completely
specified by its mean quadratures and covariance matrix, regardless of the
number of photons it can contain. Serafini's textbook gives a systematic introduction \cite{Serafini2023}; another overview
is \cite{AdessoRagyLee2014}. This compact description suggests a route to
entanglement calculations, but there is a distinction between evaluating
relative entropy for two specified Gaussian states and finding its smallest
value over all separable Gaussian reference states. The first task is a
matrix calculation. The second requires a global optimization, and a
numerical search that stops at a local minimum need not solve it.

The \emph{relative entropy of entanglement (REE)} is the minimum quantum
relative entropy with respect to a separable, unentangled reference state
\cite{VedralEtAl1997,VedralPlenio1998}. It bounds the rate of
secret-key distillation \cite{HorodeckiEtAl2009}. \emph{Rains entanglement} uses a larger reference set
specified by partial transposition and bounds the rate of entanglement
distillation \cite{Rains1999,Rains2001}. The Rains entanglement does not bound distillable secret key: positive partial
transpose (PPT) states have zero Rains entanglement yet can have distillable secret key \cite{HorodeckiEtAl2005}.
For Gaussian input states, restricting the reference states to be Gaussian
gives upper bounds on these unrestricted quantities. Equality with the
unrestricted optimizations is a separate question, not a consequence of the
Gaussian description of the input state.

Several established results make the restricted problems concrete. Chen
studied Gaussian REE and reductions for symmetric two-mode states
\cite{Chen2005}. In his study, he used exponential representations of Gaussian states and formulas
for their relative entropy based on these exponential representations, which also appear in work by Krueger and Holevo
\cite{Krueger2006,Holevo2010,Holevo2012}. Werner and Wolf characterized
separability and positive partial transpose through covariance matrices
\cite{WernerWolf2001}. These formulas specify both the objective and its
constraints, but the objective is not generally convex along straight lines
in the reference covariance matrix. A method that resolves this optimization
obstacle would make the Gaussian restriction substantially more useful for
studying noisy states and communication channels.

\subsection{Summary of contributions}
Our first contribution is a positive integral representation for Gaussian relative entropy that exposes
both an order property and its particular convex geometry.
To state it, let $\rho_V$ and $\sigma_W$ be zero-mean bosonic Gaussian states with
covariance matrices $V$ and $W$, respectively, on $N$ modes
shared between parties $A$ and $B$. The
matrix $\Omega\coloneqq\bigoplus_{j=1}^N
\left(\begin{smallmatrix}0&1\\-1&0\end{smallmatrix}\right)$ records the
quadrature commutation relations. We use the vacuum-covariance-$I$
normalization, explained in \cref{sec:prelim}. With relative entropy $D$ and
entropy $S$ measured in bits, set
\begin{equation}\label{eq:intro-f}
 f_V(W)\coloneqq(\ln2)D(\rho_V\Vert\sigma_W),\qquad
 s(V)\coloneqq(\ln2)S(\rho_V).
\end{equation}
We derive the identity
\begin{equation}\label{eq:intro-resolvent}
 f_V(W)=\frac12\ln\det W-N\ln2-s(V)
 +\frac12\int_0^1\Tr[(V+i\Omega)(W+it\Omega)^{-1}]\,dt.
\end{equation}
The uncertainty relation makes the integrand nonnegative. This combined
representation, proved in \cref{thm:resolvent}, is the starting point for the
results below. Its derivation uses a Hamiltonian integral previously given
by Miller \cite[Eq.~(19) and Appendix~A]{Miller2025}; that earlier integral
and our combined relative-entropy representation are distinguished in
\cref{sec:gibbs-integral}.

First, the integral lets us control what happens when noise is added to a
reference state while the input state is kept fixed. We construct a reference
with positive partial transpose (PPT) whose relative entropy is no larger
than the original Rains objective. Consequently, Gaussian Rains entanglement
$R_G$ equals Gaussian PPT relative entropy $E_{\mathrm{PPT}}^G$:
\begin{equation}\label{eq:intro-rains}
 R_G(A;B)_\rho=E_{\mathrm{PPT}}^G(A;B)_\rho.
\end{equation}
These quantities are defined in \cref{eq:RG,eq:EPG}, and the equality is
proved in \cref{thm:rains}. Thus two apparently different optimizations reduce
to the same problem. When either party holds one mode, PPT is equivalent to
Gaussian separability, so this problem also computes Gaussian REE.

Second, the reference-state optimizations become strictly convex when
covariance matrices are interpolated using the matrix geometric mean instead
of a straight line. Geodesic closure of quantum covariance matrices is
known \cite[Corollary~8]{BhatiaJain2015}, and geometric means have previously
been used for Gaussian entanglement and log-determinant inequalities
\cite[Section~II and Lemma~11]{LamiEtAl2017}. Our contribution concerns the
quantum relative entropy itself: the positive representation in
\cref{eq:intro-resolvent} yields its strict convexity along these paths.
\Cref{sec:geometry} introduces this geometry through a one-mode example
before proving the general result. We obtain unique optimal reference
covariance matrices, including for pure input states, and show that
independent Gaussian input states have additive Gaussian-restricted
entanglement. These results lead to globally convergent algorithms and
justified symmetry reductions (\cref{sec:additivity,sec:algorithms}).
They complement the general infinite-dimensional attainment results of
\cite{LamiShirokov2023} by giving explicit covariance bounds and uniqueness.
The optimization results are quantitative, not only asymptotic.
Our main algorithm replaces the objective at each step by an ordinary
convex upper function and minimizes that function. We prove that each
exact iteration reduces the remaining objective error by a fixed factor;
approximate subproblem solutions are allowed with an explicitly controlled
error (\cref{prop:MM}). The contraction factor remains valid as the
faithful regularization margin approaches zero, including for pure input
states. General surrogate minimization and geodesic optimization are
established frameworks \cite{Mairal2013,ZhangSra2016}; the explicit
Gaussian curvature and contraction estimates are proved here.

We also count the work required by the subproblems. Geometrically graded
quadrature preserves their convexity while sharply improving the dependence
on numerical accuracy. Together with bounded semidefinite optimization,
this gives a finite-precision cost bound that is polynomial in the number
of requested accuracy bits for each fixed rational input covariance
matrix (\cref{thm:runtime}). The bound retains explicit dependence on
covariance size, which can be unfavorable; it is not a uniform
polynomial-time claim in the encoded input size. \Cref{sec:algorithms}
explains these distinctions without assuming optimization expertise,
and \cref{app:runtime} gives the detailed estimates.

Third, we connect the Gaussian optimizations to operational limits.
Let $D_{\leftrightarrow}(\rho)$ and $K_D(\rho)$ denote the optimal rates of
entanglement and secret-key distillation from independent copies of a
bipartite state, using local operations and classical communication (LOCC), with messages
allowed in both directions.
Let $R$ and $E_R$ denote the unrestricted entanglement quantities and let
$E_R^G$ denote Gaussian REE. For every finite-mode bipartite Gaussian state,
 the following bounds are known:
\begin{equation}\label{eq:intro-operational}
 D_{\leftrightarrow}(\rho)\le R(\rho)\le R_G(\rho),\qquad
 K_D(\rho)\le E_R(\rho)\le E_R^G(\rho).
\end{equation}
\Cref{sec:operational-states} defines the tasks precisely and compares these
upper bounds with achievable rates for two- and three-mode states. Symmetric
squeezed thermal states admit a one-variable computation, while less symmetric
examples use the multimode optimization. A full-rank Gaussian example also
shows that restricting the reference to Gaussian states can strictly increase
both entanglement quantities. This distinguishes our problem from minimizing
$D(\sigma\Vert\rho)$ with a separable $\sigma$, for which Gaussian optimality
has been established in a different setting \cite[Lemma~3.2]{RippchenEtAl2026}.
The order of the two states in relative entropy matters.

Finally, we use the state calculations to bound energy-constrained quantum
and private communication through general multimode Gaussian channels.
Teleportation-based simulations replace a channel by local operations,
classical communication, and a shared resource state; bounding the resource's
entanglement then bounds communication. Recent work uses sectors of fixed
photon number to obtain energy-dependent results for single-mode channels
\cite{PirandolaSector2026,PirandolaNoisy2026,FalcoEtAl2026}.
We apply the scalar gain theorems to different mode types and give
matrix inequalities for expressing a general Gaussian channel as simpler
channels followed by Gaussian processing. A convex matrix formulation
optimizes the passive mode basis jointly with the losses, with a closed
solution for the real phase-insensitive channels studied here. An analytic
scalar predecessor also tightens bounds for multiple-output preparations. A complementary construction uses
exact Gaussian teleportation resources, whose reference entanglement can be
computed by the earlier results. The resulting bounds, defined in
\cref{eq:Ucore,eq:UG-def}, apply to arbitrary adaptive
protocols assisted by two-way classical communication; the protocols need
not be Gaussian. Pure-loss benchmarks have matching achievable rates, while
correlated-noise examples illustrate the distinct contributions of the two
upper bounds. A separate photon-number cutoff method gives controlled
approximations to the \emph{unrestricted} entanglement quantities
(\cref{sec:unrestricted}), rather than assuming that Gaussian references
are always optimal.

\subsection{Paper organization}
Our paper develops the main calculation from its mathematical foundations to its
operational applications. \Cref{sec:prelim} introduces bosonic Gaussian states,
relative entropy, and the reference sets, including the distinctions between
Gaussian-restricted and unrestricted quantities. \Cref{sec:resolvent}
derives the positive integral representation and uses it to simplify the
Rains optimization. \Cref{sec:geometry} then explains why the correct matrix
interpolation gives strict convexity and unique optima.

With these tools established, \cref{sec:additivity} studies independent
states and symmetries of their optimal references, and \cref{sec:algorithms}
turns the geometry into numerical procedures with explicit convergence rates,
error controls, and runtime bounds. \Cref{sec:operational-states} brings together the state-distillation
bounds, exactly reducible examples, and comparisons with achievable
protocols. \Cref{sec:channels} develops the channel simulations and
energy-constrained communication bounds. \Cref{sec:unrestricted} explains
how to approximate the unrestricted entanglement measures when Gaussian
references are insufficient, before \cref{sec:conclusion} summarizes the
results and remaining questions. The appendices give the quantitative optimization proofs and finite-precision
analysis, supplementary error estimates, finite-dimensional implementation
details, and the information needed to reproduce the numerical comparisons.

\section{Gaussian states, relative entropy, and reference sets}\label{sec:prelim}
The optimizations in this paper involve quantum states but are performed on
finite covariance matrices. We first explain that correspondence and then define the reference states allowed in each
entanglement quantity. These preliminaries also identify which constraints
can be imposed by linear matrix inequalities.

\subsection{Quadratures and quantum covariance matrices}
We consider a bipartite system $AB$, where $A$ and $B$ contain $m\geq1$ and
$n\geq1$ bosonic modes, respectively, and set $N\coloneqq m+n$.
The quadrature vector and symplectic form are defined as
\begin{equation}\label{eq:omega}
 \hat r\coloneqq(\hat q_1,\hat p_1,\ldots,\hat q_N,\hat p_N)^{\top},
 \qquad \Omega\coloneqq\Omega_A\oplus\Omega_B
 =\bigoplus_{j=1}^N\begin{pmatrix}0&1\\-1&0\end{pmatrix}.
\end{equation}
Here $[\hat r_j,\hat r_k]=i\Omega_{jk}$, with $\hbar=1$, and
$[X,Y]\coloneqq XY-YX$. For a state $\rho$ with finite second moments,
its mean vector and covariance matrix are defined by
\begin{equation}\label{eq:cov}
 d_j\coloneqq\Tr[\hat r_j\rho],\qquad
 V_{jk}\coloneqq\Tr\!\left[
 \{\hat r_j-d_jI,\hat r_k-d_kI\}\rho\right],
 \qquad \{X,Y\}\coloneqq XY+YX.
\end{equation}
The anticommutator makes $V$ real and symmetric. We omit the factor $1/2$
sometimes included in this definition in another convention, so the vacuum covariance matrix is
$I_{2N}$. In our convention used here, a real symmetric matrix $V$ is a
\emph{quantum covariance matrix} exactly when $V+i\Omega\succeq0$.
Here $X\succeq Y$ means that the Hermitian matrix $X-Y$ is positive
semidefinite, and $X\succ Y$ means it is positive definite. Necessity is the
uncertainty relation; sufficiency means that a Gaussian state with those
moments exists \cite{Serafini2023,Holevo2012}. A quantum covariance matrix
is always positive definite as a real matrix, even for a pure state.

Williamson's theorem \cite{Williamson1936} gives the decomposition
\begin{equation}\label{eq:williamson}
 V=S\left(\bigoplus_{j=1}^N\nu_j(V)I_2\right)S^{\top},
 \qquad S\Omega S^{\top}=\Omega,\qquad \nu_j(V)\geq1.
\end{equation}
A real matrix satisfying $S\Omega S^{\top}=\Omega$ is called \emph{symplectic}:
it preserves the quadrature commutation relations. The positive numbers
$\nu_j(V)$ are the \emph{symplectic eigenvalues}, different from the ordinary
eigenvalues of $V$. At the state level, a Gaussian unitary transforms $\rho$
into a product of thermal states with mean photon numbers
$(\nu_j(V)-1)/2$. This gives both the physical meaning of the decomposition
and a way to evaluate functions of a Gaussian state using single-mode
thermal formulas. A Gaussian state is \emph{faithful} when its density
operator has trivial kernel. This is equivalent to every $\nu_j(V)>1$,
or $V+i\Omega\succ0$. Nonfaithful Gaussian states have one or more vacuum
factors in their Williamson representation; they are not necessarily pure.

We write $\rho_V$ and $\sigma_W$ for zero-mean Gaussian states with quantum
covariance matrices $V$ and $W$. When the bipartition matters, these are
states of $AB$. A displacement changes the mean without changing the
covariance matrix. In \cref{sec:entanglement-definitions} we justify using
the same mean for the input and reference states, and then removing that
mean by local displacements.

\subsection{Entropy and support conditions}
We use base-two logarithms in information quantities. The quantum relative
entropy was introduced by Umegaki \cite{Umegaki1962}. For a normalized state
$\rho$ and a nonzero positive trace-class operator $\tau$, it is denoted by
$D(\rho\Vert\tau)$. When the support condition holds and the trace is
well-defined, its expression is
\begin{equation}\label{eq:D}
 D(\rho\Vert\tau)\coloneqq
 \Tr[(\log_2\rho-\log_2\tau)\rho],
 \qquad \supp\rho\subseteq\supp\tau.
\end{equation}
The support of a positive operator is the orthogonal complement of its
kernel. If the support condition fails, we set $D(\rho\Vert\tau)\coloneqq+\infty$:
the reference assigns zero probability to an event that can occur under
$\rho$. We allow $\tau$ to be subnormalized because Rains references need
not have unit trace.

In infinite dimensions even a support-compatible relative entropy can be
infinite. It must not be evaluated by subtracting two infinite numbers.
For all finite-entropy input states used below, one may instead write
\begin{equation}\label{eq:D-cross-entropy}
 D(\rho\Vert\tau)=-S(\rho)-\Tr[\rho\log_2\tau],
 \qquad S(\rho)\coloneqq-\Tr[\rho\log_2\rho].
\end{equation}
The expectation of the positive part of $-\log_2\tau$ is defined as the
increasing limit of bounded spectral truncations. Its negative part is
bounded, since a trace-class $\tau$ is bounded. Thus the right side has an
unambiguous value in the extended real numbers. More generally, relative
entropy is understood in its lower-semicontinuous extension
\cite{LamiShirokov2023}: limits of states cannot have a relative entropy
larger than the lower limit of the approximating relative entropies.
We will prove the particular covariance limit needed here directly in
\cref{thm:resolvent}, rather than rely on a formal cancellation at a pure
reference state.

For thermal entropy, define
\begin{equation}\label{eq:g}
 g(x)\coloneqq(x+1)\log_2(x+1)-x\log_2x,\qquad
 \gent(x)\coloneqq(\ln2)g(x),\qquad x\geq0,
\end{equation}
where $0\log_2 0\coloneqq0$, so $g(0)=0$. Williamson's decomposition gives
\begin{equation}\label{eq:s}
 s(V)\coloneqq(\ln2)S(\rho_V)
 =\sum_{j=1}^N\gent\!\left(\frac{\nu_j(V)-1}{2}\right).
\end{equation}
All covariance matrices in this paper have finite entries, so these
entropies are finite. We continue to use
$f_V(W)\coloneqq(\ln2)D(\rho_V\Vert\sigma_W)$ from \cref{eq:intro-f}.
This rescaling lets the matrix calculations use natural logarithms without
changing the units of the reported entanglement quantities.

The entropy formula \eqref{eq:s} follows directly from Williamson's
representation: Gaussian unitaries preserve entropy, a mode with
symplectic eigenvalue $\nu_j$ becomes a thermal state of mean occupation
$(\nu_j-1)/2$, and entropy adds on the resulting product. The thermal
geometric distribution has entropy $g((\nu_j-1)/2)$
\cite{Holevo2012,Serafini2023}.

For operators $X$ and $Y$, $\norm{X}$ denotes the operator norm, $\norm{X}_1\coloneqq\Tr\sqrt{X^\dagger X}$
 the trace norm,
$\norm{X}_{\mathrm F}\coloneqq(\Tr[X^\dagger X])^{1/2}$ the Frobenius norm,
and $\inner{X}{Y}\coloneqq\Tr[XY]$ the real inner product on symmetric
matrices.   Identity matrices have the dimension required by their
surrounding expression unless a subscript specifies it.

\subsection{PPT and separable reference covariance matrices}
Partial transposition is a test for entanglement, not a physical channel.
Fix local Fock bases and define the linear operation $\Gamma_B$ by its
matrix elements:
\begin{equation}\label{eq:partial-transpose-definition}
 \langle i,j|X^{\Gamma_B}|k,\ell\rangle
 \coloneqq\langle i,\ell|X|k,j\rangle.
\end{equation}
Ordinary matrix or Hilbert-space transposition is written $X^{\top}$.
In the Fock basis $\hat a^{\top}=\hat a^\dagger$, so
$\hat q^{\top}=\hat q$ and $\hat p^{\top}=-\hat p$.
Thus partial transposition reflects the momentum coordinates of $B$
\cite{Simon2000,VidalWerner2002}. Define
\begin{equation}\label{eq:transpose}
 T\coloneqq I_{2m}\oplus\bigoplus_{j=1}^n\diag(1,-1),\qquad
 \Omega_\Gamma\coloneqq T\Omega T.
\end{equation}
The zero-mean operator $\sigma_W^{\Gamma_B}$ is Gaussian in the precise
sense that its characteristic function remains a Gaussian exponential:
\begin{equation}\label{eq:pt-characteristic}
 \chi_{\sigma_W^{\Gamma_B}}(z)
 \coloneqq\Tr[e^{iz^{\top}\hat r}\sigma_W^{\Gamma_B}]
 =\exp[-z^{\top}TWTz/4].
\end{equation}
It is Hermitian, trace class, and has trace one, but it need not be positive.
A Williamson unitary transforms it into a tensor product of trace-one
Gaussian operators with the possibly signed eigenvalues given in
\eqref{eq:signed-thermal}. In particular, its covariance matrix $TWT$
being positive definite as a real matrix does not imply positivity of
the operator. It is positive exactly when $TWT+i\Omega\succeq0$.
Since $T=T^{-1}=T^{\top}$, the covariance characterization of PPT Gaussian
states \cite{Simon2000,WernerWolf2001} is
\begin{equation}\label{eq:ppt-set}
 \cP\coloneqq\left\{W\in\R^{2N\times2N}:W=W^{\top},\quad
 W+i\Omega\succeq0,\quad W+i\Omega_\Gamma\succeq0\right\}.
\end{equation}
The first uncertainty inequality ensures that $W$ describes a Gaussian
state. The second ensures that this Gaussian state's partial transpose is
also positive. Together they characterize covariance matrices of PPT
Gaussian states. For a non-Gaussian state the analogous moment inequalities
are necessary, but not in general sufficient, for positivity of its partial
transpose.

The corresponding set of covariance matrices of separable Gaussian states
has the characterization \cite[Propositions~1 and~2]{WernerWolf2001}
\begin{equation}\label{eq:sep-set}
 \cS\coloneqq\left\{W=W^{\top}:\begin{array}{l}
  \text{there exist real symmetric }G_A,G_B\text{ such that}\\
  W\succeq G_A\oplus G_B,\quad G_A+i\Omega_A\succeq0,\\
  G_B+i\Omega_B\succeq0
 \end{array}\right\}.
\end{equation}
Here $G_A$ and $G_B$ are matrices of dimensions $2m$ and $2n$, respectively.
They are local covariance matrices witnessing separability. Sufficiency has a direct preparation interpretation: start
with a product Gaussian state of covariance matrix $G_A\oplus G_B$ and
apply correlated local random displacements with classical covariance
$[W-(G_A\oplus G_B)]/2$. The resulting covariance matrix is $W$.
The converse in \cite{WernerWolf2001} says that every separable Gaussian
state admits such witnesses. 

We will use two notions of convexity, so we specify the usual one now.
A set $\mathcal X$ of real symmetric matrices is \emph{Euclidean convex}
if $(1-s)X+sY\in\mathcal X$ whenever $X,Y\in\mathcal X$ and $0\le s\le1$.
A real function $h$ is Euclidean convex if
\begin{equation}\label{eq:euclidean-convex}
 h((1-s)X+sY)\le(1-s)h(X)+s h(Y)
 \quad(X,Y\in\mathcal X,\ 0\le s\le1).
\end{equation}
Both sets in \cref{eq:ppt-set,eq:sep-set} are Euclidean convex because their
constraints are linear matrix inequalities. For \cref{eq:sep-set},
interpolating both the reference matrix and its local witnesses proves the
claim. Nevertheless, for fixed $V$ the function $f_V(W)$ is not generally
convex on the convex set of faithful quantum covariance matrices under
ordinary line interpolation. The thermal example in
\eqref{eq:ordinary-second} demonstrates this failure even along product
thermal reference states, which are separable and PPT.
For $m=1$ or $n=1$, PPT and Gaussian separability coincide, so
$\cP=\cS$ \cite[Section~III]{WernerWolf2001}. They differ in general when
both parties hold at least two modes.

\subsection{Entanglement quantities and matching the means}
\label{sec:entanglement-definitions}
We can now specify various entanglement measures followed by their Gaussian
restrictions. Let $\mathrm{SEP}(A;B)$ be the set of separable states, i.e., the trace-norm closed convex hull of
product states on $AB$. The REE is defined as
\begin{equation}\label{eq:ER}
 E_R(A;B)_\rho\coloneqq
 \inf\left\{D(\rho_{AB}\Vert\sigma_{AB}):
                   \sigma_{AB}\in\mathrm{SEP}(A;B)\right\}.
\end{equation}
To define Rains entanglement precisely, let $\mathcal T_+(AB)$ denote the
positive trace-class operators, and set
\begin{equation}\label{eq:PPTprime}
 \mathrm{PPT}'(A;B)\coloneqq
 \left\{\tau_{AB}\in\mathcal T_+(AB):
        \tau_{AB}^{\Gamma_B}\text{ is trace class},\quad
        \norm{\tau_{AB}^{\Gamma_B}}_1\le1\right\}.
\end{equation}
A partial-transpose trace norm is assigned $+\infty$ when the partial
transpose is not trace class. The reference operator in \cref{eq:PPTprime}
need not be normalized; it is subnormalized because $\Tr\tau\le\norm{\tau^{\Gamma_B}}_1\le1$. The Rains entanglement is defined as
\begin{equation}\label{eq:R}
 R(A;B)_\rho\coloneqq
 \inf\left\{D(\rho_{AB}\Vert\tau_{AB}):
                         \tau_{AB}\in\mathrm{PPT}'(A;B)\right\}.
\end{equation}
Rains' alternative normalized-reference expression is
\begin{equation}\label{eq:R-normalized}
 R(A;B)_\rho=
 \inf_{\sigma_{AB}}\left\{D(\rho_{AB}\Vert\sigma_{AB})+
                         \log_2\norm{\sigma_{AB}^{\Gamma_B}}_1\right\},
\end{equation}
where the infimum is over normalized states of $AB$
\cite{Rains1999,Rains2001}. For a fixed $\sigma$, the largest multiple
$c\sigma$ in $\mathrm{PPT}'$ has
$c\coloneqq1/\norm{\sigma^{\Gamma_B}}_1$. The identity
$D(\rho\Vert c\sigma)=D(\rho\Vert\sigma)-\log_2c$ then proves
\cref{eq:R-normalized}. Smaller multiples cannot improve the objective.

For a Gaussian input state, define the three restricted quantities
\begin{align}
 E_R^G(A;B)_{\rho_V}
 &\coloneqq\frac1{\ln2}\inf\left\{f_V(W):W\in\cS\right\},\label{eq:ERG}\\
 E_{\mathrm{PPT}}^G(A;B)_{\rho_V}
 &\coloneqq\frac1{\ln2}\inf\left\{f_V(W):W\in\cP\right\},\label{eq:EPG}\\
 R_G(A;B)_{\rho_V}
 &\coloneqq\frac1{\ln2}\inf\left\{f_V(W)+e_N(W):
                          W=W^{\top},\ W+i\Omega\succeq0\right\},\label{eq:RG}
\end{align}
where the logarithmic negativity in natural units is
\begin{equation}\label{eq:negativity}
 e_N(W)\coloneqq\ln\norm{\sigma_W^{\Gamma_B}}_1
 =\sum_{j=1}^N\max\left\{0,-\ln\widetilde\nu_j(W)\right\}.
\end{equation}
The numbers $\widetilde\nu_j(W)>0$ are the symplectic eigenvalues of $TWT$;
values below one signal failure of the PPT condition. To recall the
trace-norm calculation of Vidal and Werner \cite[Section~V.C]{VidalWerner2002},
Williamson diagonalization of the partially transposed Gaussian operator
reduces it to a tensor product of single-mode trace-one Gaussian operators.
A factor of symplectic eigenvalue $a>0$ has eigenvalues
\begin{equation}\label{eq:signed-thermal}
 p_k(a)\coloneqq\frac2{a+1}\left(\frac{a-1}{a+1}\right)^k,
 \qquad k=0,1,\ldots.
\end{equation}
For $a\ge1$ they are nonnegative and sum to one. For $0<a<1$ they alternate
in sign, and summing their absolute geometric series gives
$\sum_k|p_k(a)|=1/a$. Unitary invariance and multiplicativity of the trace
norm give \cref{eq:negativity}. The Williamson unitary in this calculation
is used to evaluate an operator, not to implement partial transposition.

The definitions in \cref{eq:ERG,eq:EPG,eq:RG} suppress the means of the two states. That restriction is
legitimate because an optimal Gaussian reference can always be chosen to
have the same mean as the input state. To see the point, let $H(W)\succ0$
be the quadratic Hamiltonian matrix of a faithful Gaussian reference,
defined by the exponential representation in \cref{eq:gibbs-state}.
Changing the reference mean from $d_\rho$ to $d_\sigma$ changes relative
entropy by
\begin{equation}\label{eq:means}
 \frac1{2\ln2}(d_\rho-d_\sigma)^{\top} H(W)(d_\rho-d_\sigma)\ge0.
\end{equation}
The extra term in \cref{eq:means} follows by expanding the quadratic exponent and taking the input
expectation; its linear centered term has zero mean. Displacements do not
change separability, PPT, or logarithmic negativity. Thus setting
$d_\sigma=d_\rho$ cannot worsen any of the three objectives. The nonfaithful
case follows from the explicit thermal limit in \cref{thm:resolvent} and
its discussion. A common displacement can then be removed locally from
both states, giving the zero-mean notation used throughout.

The optimizations allow nonfaithful reference states, with relative entropy
interpreted as above. Restricting to faithful Gaussian references gives the
same infima by the scaling limit proved below, but need not preserve
attainment. Before proving any equality between reference problems, their
definitions give
\begin{equation}\label{eq:orders}
 R\le E_R\le E_R^G,\qquad
 R\le R_G\le E_{\mathrm{PPT}}^G\le E_R^G,
\end{equation}
with all quantities evaluated on the same Gaussian state and bipartition.
In particular, imposing PPT in the normalized Gaussian Rains problem is
not part of its definition; proving that it loses nothing is a main result
of \cref{sec:rains}.

\section{An integral representation and the Gaussian Rains reduction}
\label{sec:resolvent}
Our aim is to replace the apparent competition between terms in Gaussian
relative entropy by an expression with a definite matrix sign. The reason
is practical: a nonnegative inverse-matrix integral can be compared under
added noise and differentiated along matrix paths. The noise comparison
will prove the Rains reduction in this section; the derivatives will prove
global convexity in \cref{sec:geometry}. We begin with the known Gaussian
relative-entropy formula, derive the Hamiltonian integral with its prior
attribution, and then establish the combined representation used here.

\subsection{From the Gaussian Gibbs formula to a Hamiltonian integral}
\label{sec:gibbs-integral}
A faithful zero-mean Gaussian reference state has a quadratic Gibbs
representation. Let $W$ be its $2N\times2N$ quantum covariance matrix,
with all symplectic eigenvalues greater than one, and write
\begin{equation}\label{eq:gibbs-state}
 \sigma_W=Z(W)^{-1}\exp[-\hat r^{\top} H(W)\hat r/2],\qquad
 Z(W)\coloneqq\Tr\exp[-\hat r^{\top} H(W)\hat r/2].
\end{equation}
The matrix $H(W)$ is real symmetric and positive definite. It specifies
the quadratic operator in the exponent; it is not the density operator.
Taking the logarithm of \cref{eq:gibbs-state}, then its expectation in the
zero-mean Gaussian state $\rho_V$, gives the established formula
\begin{equation}\label{eq:gibbs-f}
 f_V(W)=-s(V)+\ln Z(W)+\frac14\Tr[VH(W)].
\end{equation}
Indeed, symmetry of $H(W)$ makes its contraction with the antisymmetric
commutator vanish, so
$\Tr[(\hat r^{\top}H(W)\hat r)\rho_V]=\Tr[VH(W)]/2$.
Thus \cref{eq:gibbs-f} is simply input negative entropy plus reference
cross entropy. Gaussian exponential and relative-entropy formulas of this
type are treated in \cite{Chen2005,Krueger2006,Holevo2010,Holevo2012,
SeshadreesanLamiWilde2018}.

To evaluate the two reference terms, write
$W=S D_W S^{\top}$, where
$D_W\coloneqq\bigoplus_{j=1}^N\nu_j(W)I_2$ and $S$ is symplectic.
The transformed reference is a product of thermal states. Its $j$th
quadratic Hamiltonian parameter is
$h_j\coloneqq\ln[(\nu_j(W)+1)/(\nu_j(W)-1)]
=2\arccoth\nu_j(W)$, where
$\arccoth x\coloneqq\tfrac12\ln[(x+1)/(x-1)]$ for $x>1$.
Transforming the quadratic form back gives
\begin{equation}\label{eq:H-Williamson}
 H(W)=S^{-\top}\left(\bigoplus_{j=1}^N
                 2\arccoth\nu_j(W)I_2\right)S^{-1}.
\end{equation}
For one thermal mode,
$\Tr\exp[-h_j(\hat a_j^\dagger\hat a_j+1/2)]
=[2\sinh(h_j/2)]^{-1}=\sqrt{\nu_j(W)^2-1}/2$.
Multiplying these factors and using $\det S=1$ proves
\begin{equation}\label{eq:Z}
 \ln Z(W)=\frac12\ln\det(W+i\Omega)-N\ln2,
 \qquad \det(W+i\Omega)=\prod_{j=1}^N(\nu_j(W)^2-1).
\end{equation}
\Cref{eq:gibbs-f,eq:H-Williamson,eq:Z} already evaluate the relative
entropy. The next representation of $H(W)$ is useful because it replaces
a symplectic diagonalization by ordinary matrix inverses.

\begin{samepage}
\begin{lemma}[Hamiltonian inverse integral; Miller]\label{lem:H-integral}
Let $W$ be the quantum covariance matrix of a faithful zero-mean Gaussian
state on $N$ modes, and let $H(W)$ be its Hamiltonian matrix in
\eqref{eq:gibbs-state}. Then
\begin{equation}\label{eq:H-integral}
 H(W)=\int_{-1}^1(W+it\Omega)^{-1}\,dt.
\end{equation}
Formula \eqref{eq:H-integral} appears in \cite[Eq.~(19) and Appendix~A]{Miller2025}, with
covariance matrices normalized to have vacuum value $I/2$.
\end{lemma}
\end{samepage}
\begin{proof}
The symplectic identity $S\Omega S^{\top}=\Omega$ implies
$W+it\Omega=S(D_W+it\Omega)S^{\top}$. Taking the inverse therefore gives a
congruence by $S^{-\top}$ and $S^{-1}$. It remains to integrate each $2\times2$
block. Set $J\coloneqq\left(\begin{smallmatrix}0&1\\-1&0\end{smallmatrix}\right)$,
so $J^2=-I_2$. Direct multiplication shows that, for $\nu>1$,
\begin{equation}\label{eq:thermal-inverse-integral}
 \begin{aligned}
 (\nu I_2+itJ)^{-1}&=\frac{\nu I_2-itJ}{\nu^2-t^2},\\
 \int_{-1}^1(\nu I_2+itJ)^{-1}dt
 &=I_2\int_{-1}^1\frac{\nu}{\nu^2-t^2}\,dt
 =2\arccoth(\nu)I_2.
 \end{aligned}
\end{equation}
The term proportional to $tJ$ is odd and integrates to zero. Integrating
all blocks and undoing the congruence reproduces \cref{eq:H-Williamson}.
\end{proof}

The integral in \cref{lem:H-integral} is a known ingredient, not a novelty
claim of this paper. We next combine it with the logarithmic determinant
in \cref{eq:Z}. The resulting integral has the positive matrix
$V+i\Omega$ as its weight. That combination is what allows the order and
convexity arguments below, and it also avoids separately evaluating
singular terms at nonfaithful reference states.

\subsection{A positive relative-entropy integral, including pure-state limits}
We now keep the input state fixed and reorganize its relative entropy into
one positive integral. A faithful reference is the simplest case, but the
extension to nonfaithful references matters: even the optimal reference for
a product vacuum state is nonfaithful. The theorem states the formula on
the full physical domain and explains how its finite or infinite value
encodes the quantum support condition.

\begin{samepage}
\begin{theorem}[Positive-resolvent representation]\label{thm:resolvent}
Let $\rho_V$ and $\sigma_W$ be zero-mean Gaussian states of $N$ modes
with quantum covariance matrices $V$ and $W$. Let $f_V$ and $s(V)$ be
as in \eqref{eq:intro-f}, \eqref{eq:s}, and let $\Omega$ be the symplectic form
in \eqref{eq:omega}. Set
\begin{equation}\label{eq:CQ}
 C_V\coloneqq V+i\Omega,\qquad Q_t(W)\coloneqq(W+it\Omega)^{-1},
 \qquad 0\le t<1.
\end{equation}
Then
\begin{equation}\label{eq:resolvent}
 f_V(W)=\frac12\ln\det W-N\ln2-s(V)
          +\frac12\int_0^1\Tr[C_VQ_t(W)]\,dt.
\end{equation}
The integrand is real and nonnegative. The integral is an ordinary finite
integral when $W$ is faithful, and otherwise is its nonnegative improper
integral, with value in $[0,+\infty]$. In particular, \eqref{eq:resolvent}
includes the support condition for relative entropy. Moreover,
\begin{equation}\label{eq:scaling-limit}
 \lim_{\kappa\downarrow1} f_V(\kappa W)=f_V(W)
\end{equation}
with equality in the extended real numbers.
\end{theorem}
\end{samepage}

\begin{proof}
The proof has two steps. For faithful $W$, we combine the logarithmic
determinant with the Hamiltonian integral. For a reference with vacuum
factors, we calculate the thermal limits explicitly and then use monotone
convergence.

First, the inverse $Q_t(W)$ exists and is positive definite for every
physical $W$ and $t<1$. To verify this, use $W=S D_W S^{\top}$ as in
\cref{eq:williamson}. Then
\begin{equation}\label{eq:positive-resolvent-blocks}
 W+it\Omega=S\left(\bigoplus_j[\nu_j(W)I_2+itJ]\right)S^{\top}.
\end{equation}
Each Hermitian block in brackets has ordinary eigenvalues
$\nu_j(W)\pm t>0$, since $\nu_j(W)\ge1$. Congruence by the invertible $S$
preserves positivity, though it does not generally preserve these
ordinary eigenvalues. Hence $Q_t(W)\succ0$. Also $C_V=V+i\Omega\succeq0$
by the input uncertainty relation. In particular,
\begin{equation}\label{eq:positive-trace}
 \Tr[C_VQ_t(W)]=\Tr[C_V^{1/2}Q_t(W)C_V^{1/2}]\ge0,
\end{equation}
which proves both reality and nonnegativity without assuming that the two
matrices commute.

Suppose now that $W$ is faithful. We recall the logarithmic determinant
derivative to make the reorganization explicit. For an invertible smooth
matrix $A(t)$,
\begin{equation}\label{eq:Jacobi-proof}
 \begin{aligned}
 \det A(t+\varepsilon)
 &=\det A(t)\det\!\left[I+\varepsilon A(t)^{-1}A'(t)
                         +o(\varepsilon)\right]\\
 &=\det A(t)\left[1+\varepsilon\Tr(A(t)^{-1}A'(t))
                         +o(\varepsilon)\right].
 \end{aligned}
\end{equation}
Here the linear term in the determinant expansion is the trace.
When $A(t)\succ0$, division and differentiation of the ordinary real
logarithm give $\frac{d}{dt}\ln\det A(t)=\Tr[A(t)^{-1}A'(t)]$.
Apply this identity to $A(t)=W+it\Omega$, whose derivative is $i\Omega$,
and integrate from $0$ to $1$:
\begin{equation}\label{eq:logdet-integral}
 \ln\det(W+i\Omega)-\ln\det W
 =\int_0^1\Tr[i\Omega Q_t(W)]\,dt.
\end{equation}
Because $W$ and $\Omega$ are real, $Q_{-t}(W)=\overline{Q_t(W)}$.
Splitting the Hamiltonian integral at zero and changing variables on its
negative half consequently yields
\begin{equation}\label{eq:H-real}
 H(W)=\int_0^1[Q_t(W)+\overline{Q_t(W)}]dt
      =2\Rea\int_0^1Q_t(W)\,dt.
\end{equation}
The trace $\Tr[VQ_t(W)]$ is real: $V$ and $Q_t(W)$ are Hermitian, so its
complex conjugate is $\Tr[Q_t(W)V]=\Tr[VQ_t(W)]$.
Substituting \cref{eq:logdet-integral,eq:H-real} into the known formula
\cref{eq:gibbs-f} combines its two integrals into
$\frac12\int_0^1\Tr[(V+i\Omega)Q_t(W)]dt$ and proves
\cref{eq:resolvent} for faithful references.

For an arbitrary physical $W$, write its Williamson decomposition as above
and set $V'\coloneqq S^{-1}VS^{-\top}$. Define the mean variance of its $j$th canonical
pair by
\begin{equation}\label{eq:bj}
 b_j\coloneqq\frac{V'_{2j-1,2j-1}+V'_{2j,2j}}2\geq 1.
\end{equation}
To verify the inequality, write $a=V'_{2j-1,2j-1}$,
$c=V'_{2j,2j}$, and $z=V'_{2j-1,2j}$. The physical one-mode marginal
satisfies $ac-z^2\ge1$, so
$b_j=(a+c)/2\ge\sqrt{ac}\ge1$.
In these coordinates the input mean photon number of mode $j$ is
$(b_j-1)/2$, since $\hat a_j^\dagger\hat a_j
=(\hat q_j^2+\hat p_j^2-1)/2$.

The cross entropy
against the faithful reference $\kappa W$, for $\kappa>1$, separates into
thermal-mode contributions. The transformed input need not be a product:
the logarithm of the reference is a sum of single-mode Hamiltonians, so
only these marginal second moments enter. More explicitly, a thermal
reference of covariance $aI_2$, with $a>1$, has the geometric spectrum
$p_k(a)$ in \cref{eq:signed-thermal}. Its negative logarithm is
$\ln[(a+1)/2]+\hat a^\dagger\hat a\ln[(a+1)/(a-1)]$.
Taking this expectation in each transformed input marginal, with
$a=\kappa\nu_j(W)$, gives
\begin{equation}\label{eq:thermal-cross}
 f_V(\kappa W)+s(V)
 =\sum_{j=1}^N\left[
   \ln\frac{\kappa\nu_j(W)+1}{2}
   +\frac{b_j-1}{2}
       \ln\frac{\kappa\nu_j(W)+1}{\kappa\nu_j(W)-1}
              \right].
\end{equation}
If $\nu_j(W)>1$, this contribution has a finite limit. If $\nu_j(W)=1$,
the reference mode is vacuum. The transformed input has support in that
vacuum factor exactly when $b_j=1$: its photon number expectation is
$(b_j-1)/2$, and zero expectation of the nonnegative number operator forces
the marginal to be vacuum. Moreover, a state with a pure marginal has no correlations with
that subsystem: positivity confines it to the tensor product of the pure
marginal's one-dimensional support and the other subsystem. Thus a vacuum
marginal is an actual vacuum tensor factor of the transformed input state. In this case the term tends to zero. If $b_j>1$, the input has support
outside the vacuum factor, and the term diverges to $+\infty$.
The reference support is the full Hilbert space of its thermal modes
tensored with its vacuum factors. The preceding alternatives therefore
exhaust precisely the possibilities for $\supp\rho_V\subseteq\supp\sigma_W$.
They establish \cref{eq:scaling-limit}: each thermal summand converges to
its finite cross entropy, or a violated vacuum condition makes the sum
diverge. There are only $N$ summands, so no infinite sum of mode limits is
being interchanged here.

Finally, as $\kappa\downarrow1$ the positive matrices
$(\kappa W+it\Omega)^{-1}$ increase to $Q_t(W)$ for each $t<1$. Monotone
convergence of the nonnegative traces in the faithful formula, and
continuity of $\ln\det(\kappa W)$, prove \cref{eq:resolvent} at the boundary.
\end{proof}

The positivity in \cref{thm:resolvent} is specifically quantum: the trace-weighting matrix is
$V+i\Omega$, which is positive semidefinite by the uncertainty relation.
One cannot replace it by an arbitrary Hermitian matrix in the arguments
that follow. At the same time, the representation is numerically and
conceptually simpler than a cancellation of divergent terms. For example,
if $V=W=I_2$, its integral is $\int_0^1 2/(1+t)\,dt=2\ln2$, and the finite
terms give $f_{I_2}(I_2)=0$ without evaluating a divergent Hamiltonian.

The same argument also completes the common-mean reduction at a
nonfaithful reference. In its Williamson coordinates, a mean mismatch
$\delta'$ replaces $b_j$ in \cref{eq:thermal-cross} by
$b_j+\norm{\delta'_j}^2$, where $\delta'_j$ is the two-component mean
mismatch of mode $j$. Each thermal contribution is nondecreasing in this
number, including its vacuum limit. Setting the reference mean equal to
the input mean therefore cannot increase relative entropy.

\begin{samepage}
\begin{corollary}[Lower semicontinuity]\label{cor:lsc}
For a fixed quantum covariance matrix $V$, the extended function $W\mapsto f_V(W)$ is lower
semicontinuous on the set of quantum covariance matrices.
\end{corollary}
\end{samepage}
\begin{proof}
Replace the upper endpoint of the integral in \cref{eq:resolvent} by
$a<1$. Each resulting expression is continuous locally on the physical
set of covariance matrices, because $W+it\Omega$ is uniformly invertible on a compact
neighborhood for $t\le a$. The expressions increase to $f_V(W)$ as
$a\uparrow1$. A supremum of continuous functions is lower semicontinuous.
\end{proof}

\subsection{Adding reference noise and controlling the relative entropy}
The Rains penalty will be removed by adding carefully chosen noise to a
Gaussian reference state. Before constructing that noise, we need a bound
on its cost. The next inequality concerns an increase in the reference covariance matrix, with
the input kept fixed. It is not a monotonicity statement saying that adding
noise must reduce relative entropy. Instead, it quantifies the largest
increase that can occur.

\begin{samepage}
\begin{proposition}[Noise-ordering inequality]\label{prop:noise}
Let $V,W,W'$ be physical $N$-mode covariances with $W'\succeq W$. Then
\begin{equation}\label{eq:noise}
 f_V(W')\le f_V(W)+\frac12\ln\frac{\det W'}{\det W}.
\end{equation}
In particular, for $\kappa\geq 1$,
\begin{equation}\label{eq:scaling-noise}
 f_V(\kappa W)\le f_V(W)+N\ln\kappa.
\end{equation}
These inequalities hold for nonfaithful input states and references, with
extended-value interpretation. If $f_V(W)<\infty$ and $W'\ne W$, then
\eqref{eq:noise} is strict.
\end{proposition}
\end{samepage}
\begin{proof}
For every $t<1$, inverse order gives $Q_t(W')\preceq Q_t(W)$. Multiplying by
the positive semidefinite matrix $C_V$ inside the trace preserves the
inequality. Comparing the two nonnegative integrals in
\cref{eq:resolvent} proves \cref{eq:noise}. For strictness, $Q_0(W)-Q_0(W')=W^{-1}-(W')^{-1}$ is a nonzero real
positive semidefinite matrix when $W'\ne W$. Thus
\begin{equation}\label{eq:strict-noise}
 \Tr[C_V(Q_0(W)-Q_0(W'))]
 =\Tr[V(W^{-1}-(W')^{-1})]>0.
\end{equation}
Continuity preserves this positive difference on an interval near $t=0$;
the difference is nonnegative for every other $t<1$. Since the original
integral is finite, subtracting the integrals is legitimate and gives the
strict inequality. Finally, $\det(\kappa W)=\kappa^{2N}\det W$, so scaling
gives \cref{eq:scaling-noise}.
\end{proof}

The ellipsoid $\{x:x^{\top}W^{-1}x\le1\}$ has volume proportional to
$\sqrt{\det W}$. Thus the rightmost term in \cref{eq:noise} is its
\emph{logarithmic volume expansion}.
The decrease of the inverse-covariance term compensates for every other
contribution except at most this logarithmic volume increase. This precise
compensation will match the logarithmic negativity of a Gaussian reference.

\subsection{Replacing a Gaussian Rains reference by a PPT state}
\label{sec:rains}
The Rains objective combines relative entropy with a partial-transpose
trace-norm penalty. We now eliminate the penalty by an explicit reference
replacement. The partial transpose is used to identify which symplectic
eigenvalues need to be raised; no physical implementation of partial
transposition is asserted.

\begin{samepage}
\begin{theorem}[PPT reference replacement]\label{thm:rains}
Let $\rho_{AB}=\rho_V$ be a zero-mean Gaussian state with $m$ modes on
$A$, $n$ modes on $B$, and $N\coloneqq m+n$. For the quantities defined
in \eqref{eq:RG}, \eqref{eq:EPG},
\begin{equation}\label{eq:rains-equality}
 R_G(A;B)_\rho=E_{\mathrm{PPT}}^G(A;B)_\rho.
\end{equation}
More specifically, let $\sigma_W$ be a zero-mean Gaussian reference,
let $T$ be the momentum-reflection matrix in \eqref{eq:transpose}, and
write its ordinary Williamson decomposition as
\begin{equation}\label{eq:reference-Williamson}
 W=S_W\left(\bigoplus_{j=1}^N\nu_j(W)I_2\right)S_W^{\top},
 \qquad \nu_j(W)\ge1.
\end{equation}
The positive matrix $TWT$ has its own Williamson decomposition,
generally with a different symplectic matrix $S$:
\begin{equation}\label{eq:pt-williamson}
 TWT=S\left(\bigoplus_{j=1}^N\widetilde\nu_j I_2\right)S^{\top}.
\end{equation}
The covariance matrix
\begin{equation}\label{eq:Wplus}
 W_+\coloneqq TS\left(\bigoplus_{j=1}^N
                       \max\left\{1,\widetilde\nu_j\right\}I_2\right)S^{\top}T
\end{equation}
is physical and PPT, and satisfies
\begin{equation}\label{eq:replacement}
 f_V(W_+)\le f_V(W)+e_N(W).
\end{equation}
If $W$ is not PPT and $f_V(W)<\infty$, the inequality is strict.
Consequently every optimal Gaussian reference in the normalized Rains
optimization \eqref{eq:RG} is PPT.
\end{theorem}
\end{samepage}
\begin{proof}
The replacement raises, and does not lower, the Williamson diagonal entries
of $TWT$. Consequently $W_+\succeq W$, so $W_++i\Omega\succeq0$.
Its partial transpose has all symplectic eigenvalues at least one, hence
$TW_+T+i\Omega\succeq0$ as well. Thus $W_+\in\cP$.
Since $\det S=1$ and $|\det T|=1$,
\begin{equation}\label{eq:volume-neg}
 \frac12\ln\frac{\det W_+}{\det W}
 =\sum_{j=1}^N\ln\frac{\max\left\{1,\widetilde\nu_j\right\}}{\widetilde\nu_j}
 =e_N(W).
\end{equation}
\Cref{prop:noise} proves \cref{eq:replacement}. Taking infima shows
$E_{\mathrm{PPT}}^G\le R_G$. Conversely, every PPT Gaussian reference is
allowed in \cref{eq:RG} and has $e_N=0$, so $R_G\le E_{\mathrm{PPT}}^G$.
If $W$ is not PPT, at least one $\widetilde\nu_j<1$ and therefore
$W_+\ne W$. The strict part of \cref{prop:noise} then gives a strictly
better PPT reference whenever the original objective is finite.
An optimal finite-valued Gaussian Rains reference must therefore be PPT.
Existence is established in \cref{thm:attainment} below.
\end{proof}

\begin{figure}[t]
\centering
\begin{tikzpicture}[>=Latex,font=\small,x=1cm,y=1cm]
 \node[font=\normalsize] at (2.0,3.05) {Reference: $TWT$};
 \node[font=\normalsize] at (9.8,3.05) {Replacement: $TW_+T$};
 \draw[->,mutedgray] (.2,0)--(.2,2.45);
 \node[rotate=90,anchor=south] at (-.32,1.1) {Symplectic eigenvalue};
 \draw[densely dashed,mutedgray] (.15,1)--(3.8,1);
 \node[left] at (.2,1) {$1$};
 \draw[thick,rateorange] (1.3,0)--(1.3,.48);
 \fill[rateorange] (1.3,.48) circle(2.3pt);
 \draw[thick,boundblue] (2.8,0)--(2.8,1.95);
 \fill[boundblue] (2.8,1.95) circle(2.3pt);
 \draw[->,thick,keygreen] (1.3,.57)--(1.3,.93);
 \draw[keygreen,thick,fill=white] (1.3,1) circle(2.6pt);
 \node[below=4pt] at (1.3,0) {$\widetilde\nu_1<1$};
 \node[below=4pt] at (2.8,0) {$\widetilde\nu_2>1$};
 \draw[->,thick,boundblue] (4.25,1.4)--(7.45,1.4);
 \node[align=center,above=6pt] at (5.85,1.4) {Raise only\\eigenvalues below $1$};
 \node[below=7pt] at (5.85,1.4) {$W_+\succeq W$};
 \draw[->,mutedgray] (8,0)--(8,2.45);
 \draw[densely dashed,mutedgray] (7.95,1)--(11.6,1);
 \node[left] at (8,1) {$1$};
 \draw[thick,keygreen] (9.1,0)--(9.1,1);
 \draw[keygreen,thick,fill=white] (9.1,1) circle(2.6pt);
 \draw[thick,boundblue] (10.6,0)--(10.6,1.95);
 \fill[boundblue] (10.6,1.95) circle(2.3pt);
 \node[below=4pt] at (9.1,0) {$1$};
 \node[below=4pt] at (10.6,0) {$\widetilde\nu_2$};
 \node[align=center] at (5.85,-1.0)
 {$\displaystyle \frac12\ln\frac{\det W_+}{\det W}=e_N(W)$:
 the logarithmic volume cost equals the negativity penalty.};
\end{tikzpicture}
\caption{The reference replacement in \cref{thm:rains}, shown for two
partially transposed symplectic eigenvalues. Only eigenvalues below the
vacuum threshold are raised. The construction acts on covariance matrices
in a proof; it is not a channel implementing partial transposition.
The corresponding increase $W_+-W\succeq0$ is ordinary Gaussian displacement
noise on the reference.}
\label{fig:replacement}
\end{figure}
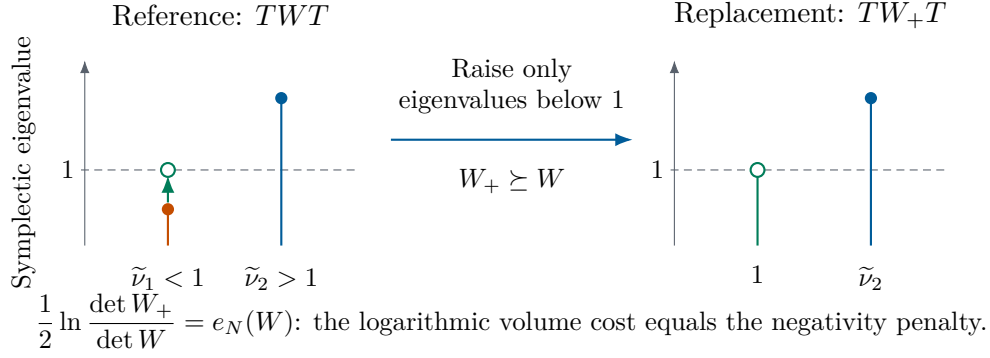

\begin{samepage}
\begin{corollary}[One mode against many modes]\label{cor:one-many}
If $A$ or $B$ consists of one mode, then every Gaussian input state obeys
\begin{equation}\label{eq:one-many}
 R_G(A;B)_\rho=E_R^G(A;B)_\rho.
\end{equation}
\end{corollary}
\end{samepage}
\begin{proof}
For these partitions, $\cP=\cS$ by
\cite[Section~III]{WernerWolf2001}; apply \cref{thm:rains}.
\end{proof}

\Cref{fig:replacement} illustrates the replacement and the matching
volume cost. The reference $W_+$ depends on the chosen $W$. It is not generally the
minimizer of $f_V$ over $\cP$, and setting $W=V$ in \cref{eq:Wplus} does
not solve the remaining minimization. The content of the theorem is that
an entangled Gaussian Rains reference never improves on the best PPT
Gaussian reference. Neither \cref{thm:rains} nor \cref{cor:one-many}
asserts equality with unrestricted Rains entanglement.

\section{Geodesic convexity and the optimal covariance matrix}\label{sec:geometry}
We now address the main optimization obstacle. The preceding section
reduced Gaussian Rains entanglement to the same relative-entropy objective
used for Gaussian REE, but with PPT rather than separable reference states.
For either problem, a local numerical minimum is useful only when we know
that a smaller value cannot be hidden elsewhere. Convexity usually gives
that guarantee. Here ordinary straight-line convexity fails, even for one
thermal mode, so we must choose a different way of connecting reference
covariance matrices.

The section has three steps. We introduce the matrix geometric mean and
show that its interpolation paths stay inside the allowed sets. We then
prove strict convexity of the relative entropy along those paths. Finally,
we prove that the minimum is attained and is unique. No background in
Riemannian geometry is required: the paths and their derivatives are given
explicitly. The interpolation is a mathematical tool for the reference
optimization, not a proposed physical evolution of an optical state.

\subsection{Why ordinary covariance interpolation is insufficient}
Ordinary convexity means convexity along straight-line matrix
interpolations, as defined in \cref{eq:euclidean-convex}.
A single thermal mode already shows why that notion is insufficient for
the relative-entropy objective. Take $V=vI_2$
and $W=wI_2$, with $v\geq 1$ and $w>1$. Then
\begin{equation}\label{eq:one-mode}
 f_{vI_2}(wI_2)=\frac12\ln(w^2-1)-\ln2
                 +v\arccoth w-\gent((v-1)/2).
\end{equation}
Its ordinary second derivative is
\begin{equation}\label{eq:ordinary-second}
 \frac{d^2}{dw^2}f_{vI_2}(wI_2)
 =\frac{-w^2+2vw-1}{(w^2-1)^2},
\end{equation}
which is negative for sufficiently large $w$. In contrast, if $x=\ln w$,
\begin{equation}\label{eq:log-second}
 \frac{d^2}{dx^2}f_{vI_2}(e^xI_2)
 =\frac{w[v(w^2+1)-2w]}{(w^2-1)^2}>0.
\end{equation}
Thus the relevant one-dimensional interpolation is geometric, not
arithmetic. For noncommuting matrices there is no common eigenbasis in
which to take entrywise logarithms; the matrix geometric mean supplies the
basis-independent replacement.

\subsection{The matrix geometric mean and physical constraints}
A positive scalar can be interpolated multiplicatively: between $w_0$
and $w_1$ the path is $w_0^{1-s}w_1^s$. The matrix geometric mean extends
this idea to matrices that do not commute. For positive definite real
symmetric matrices $W_0,W_1$, define \cite[Chapters~4 and~6]{Bhatia2007}
\cite{LawsonLim2021}
\begin{equation}\label{eq:geodesic}
 W_0\#_sW_1
 \coloneqq W_0^{1/2}
       (W_0^{-1/2}W_1W_0^{-1/2})^sW_0^{1/2},
 \qquad 0\le s\le1.
\end{equation}
For commuting matrices, \cref{eq:geodesic} is the scalar geometric
interpolation in their common eigenbasis. For general matrices it is a
basis-independent construction, called the \emph{weighted matrix geometric
mean}. Its path is the geodesic of the affine-invariant, or trace, metric
on positive definite matrices \cite[Chapter~6]{Bhatia2007}. We use the word
``geodesic'' only for this specific path.

A set $\mathcal X$ is \emph{geodesically convex} if $W_0\#_sW_1\in\mathcal X$
for all $W_0,W_1\in\mathcal X$ and $s\in[0,1]$. A function
$h\colon\mathcal X\to\R\cup\{+\infty\}$ is \emph{geodesically convex} if
\begin{equation}\label{eq:gconvex-definition}
 h(W_0\#_sW_1)\le(1-s)h(W_0)+s h(W_1)
 \quad(W_0,W_1\in\mathcal X,\ 0\le s\le1).
\end{equation}
It is strictly geodesically convex on its finite-valued domain when this
inequality is strict for distinct endpoints with finite values and
$0<s<1$. These are the ordinary scalar convexity inequalities applied to
the specified curved paths instead of straight lines. They do not assert
Euclidean convexity of $h$.

Useful properties of the matrix geometric mean include
\begin{align}
 (MW_0M^{\top})\#_s(MW_1M^{\top})&=M(W_0\#_sW_1)M^{\top},\label{eq:mean-congruence}\\
 (W_0\#_sW_1)^{-1}&=W_0^{-1}\#_sW_1^{-1},\label{eq:mean-inverse}\\
 W_0\succeq X_0,\ W_1\succeq X_1
 &\ \Longrightarrow\ W_0\#_sW_1\succeq X_0\#_sX_1,\label{eq:mean-order}\\
 \ln\det(W_0\#_sW_1)&=(1-s)\ln\det W_0+s\ln\det W_1.
 \label{eq:mean-det}
\end{align}
Here $M$ is invertible, and all matrices compared by $\#_s$ are positive
definite. The mean also commutes with direct sums. These properties and
their Gaussian applications are reviewed in \cite[Section~II]{LamiEtAl2017};
see \cite{Vishnoi2018} for the optimization viewpoint.

Before differentiating the objective, we must know that the
interpolation does not leave the physical domain. The inverse form of the
uncertainty relation below is an established fact in Gaussian quantum
information: see \cite[Lemma~1]{GiedkeEtAl2001}, and its use with geometric
means in \cite[Lemma~11]{LamiEtAl2017}. Geodesic closure of quantum
covariance matrices is also established in
\cite[Corollary~8]{BhatiaJain2015}. We give a proof because the same matrix
argument simultaneously handles PPT constraints, local separability
witnesses, and the regularization used by our algorithms.

\begin{samepage}
\begin{lemma}[Uncertainty as an inverse-order inequality]\label{lem:uncertainty}
Let $W\in\R^{d\times d}$ be symmetric and positive definite, and let
$J\in\R^{d\times d}$ be real, nonsingular, and antisymmetric. Then
\begin{equation}\label{eq:inverse-uncertainty}
 W+iJ\succeq0\quad\Longleftrightarrow\quad W\succeq JW^{-1}J^{\top}.
\end{equation}
Consequently the set $\{W\succ0:W+iJ\succeq0\}$ is geodesically convex.
\end{lemma}
\end{samepage}
\begin{proof}
Set $K\coloneqq W^{-1/2}JW^{-1/2}$, so $K$ is real and
antisymmetric. The matrix $iK$ is Hermitian. Its eigenvalues occur in
opposite pairs: complex conjugation changes $iK$ to $-iK$, so an eigenvalue
$\lambda$ is accompanied by $-\lambda$. Congruence by $W^{-1/2}$ gives
$W+iJ\succeq0$ exactly when $I+iK\succeq0$. The paired eigenvalues imply
that the latter condition is equivalent to $-I\preceq iK\preceq I$,
and hence, by the spectral theorem, to $(iK)^2\preceq I$. Since $J^{\top}=-J$,
\begin{equation}\label{eq:square-uncertainty}
 (iK)^2=-W^{-1/2}JW^{-1}JW^{-1/2}
       =W^{-1/2}JW^{-1}J^{\top}W^{-1/2}.
\end{equation}
Congruence by $W^{1/2}$ now gives exactly
$W\succeq JW^{-1}J^{\top}$, proving both directions of
\cref{eq:inverse-uncertainty}. This derivation also explains why the
reality and antisymmetry assumptions on $J$ matter.

If both endpoints satisfy it,
then \cref{eq:mean-congruence,eq:mean-inverse,eq:mean-order} give
\begin{equation}\label{eq:physical-mean}
 W_0\#_sW_1\succeq
 (JW_0^{-1}J^{\top})\#_s(JW_1^{-1}J^{\top})
 =J(W_0\#_sW_1)^{-1}J^{\top}.
\end{equation}
The first inequality follows from monotonicity of the geometric mean
in both arguments. The equality then uses congruence invariance followed
by the inverse identity in \cref{eq:mean-inverse}. Applying
\cref{eq:inverse-uncertainty} to $W_0\#_sW_1$ proves that it satisfies the
same uncertainty constraint as both endpoints.
\end{proof}

\begin{samepage}
\begin{proposition}[Geometry of the reference sets]\label{prop:free-geometry}
Fix the $m$-mode versus $n$-mode bipartition and the corresponding
symplectic forms in \eqref{eq:omega}, \eqref{eq:transpose}. The sets $\cP$ and $\cS$
of PPT and separable Gaussian reference covariance matrices, defined in
\eqref{eq:ppt-set}, \eqref{eq:sep-set}, are closed. For $W_0,W_1$ in either set,
both $(1-s)W_0+sW_1$ and $W_0\#_sW_1$ belong to that same set for every
$s\in[0,1]$. For $\kappa\ge1$ and $M>0$, the intersection of either set with
\begin{equation}\label{eq:extra-constraints}
 \{W:W+i\kappa\Omega\succeq0,\ W\preceq MI\}
\end{equation}
is also closed and retains both convexity properties (and can be empty).
\end{proposition}
\end{samepage}
\begin{proof}
The PPT constraint consists of two uncertainty inequalities, one
with $J=\Omega$ and one with $J=\Omega_\Gamma$. Both matrices are real,
nonsingular, and antisymmetric. \Cref{lem:uncertainty} preserves each
inequality along the same geometric-mean path, proving closure under
that path for $\cP$.

For $W_0,W_1\in\cS$, choose local physical witnesses
$G_{A,j},G_{B,j}$ satisfying
$W_j\succeq G_{A,j}\oplus G_{B,j}$ for $j=0,1$. Monotonicity and the
direct-sum property of the geometric mean give
\begin{equation}\label{eq:sep-mean}
 W_0\#_sW_1\succeq
 (G_{A,0}\#_sG_{A,1})\oplus(G_{B,0}\#_sG_{B,1}).
\end{equation}
Apply \cref{lem:uncertainty} separately with $J=\Omega_A$ and
$J=\Omega_B$. The two diagonal blocks on the right are  local quantum
covariance matrices and therefore witness separability of the matrix on
the left through \cref{eq:sep-set}. This proves the claim without needing
the optimal local witnesses to be unique. For the extra uncertainty constraint use $J=\kappa\Omega$;
for the upper bound use monotonicity and $(MI)\#_s(MI)=MI$.
Euclidean convexity follows directly from the linear matrix inequalities.

Closedness of $\cP$ is also direct. For $\cS$, it needs a little care because
the local witness matrices are optimized. Let $W_j\to W$ and choose
$G_{A,j}\oplus G_{B,j}\preceq W_j$. The bounded sequence $W_j$ gives a
common upper bound $MI$ on each positive witness. For a local witness
$G\preceq MI$, \cref{lem:uncertainty} gives
$G\succeq\Omega_{\rm loc}G^{-1}\Omega_{\rm loc}^{\top}\succeq M^{-1}I$.
Thus the witnesses remain in a compact matrix interval. A subsequence
of the witness pairs converges, and its limits satisfy all three closed
matrix inequalities in
\cref{eq:sep-set}. They witness $W\in\cS$.
\end{proof}

\Cref{prop:free-geometry} supplies the feasibility half of the optimization
argument: the curved paths used below connect allowed reference states to
allowed reference states. Its additional constraints are included now
because numerical optimization will keep reference eigenvalues away from
singularities and prevent the covariance matrices from becoming unbounded.
The same geometry survives both restrictions.

\subsection{Strict convexity of the relative entropy}
Feasibility of a path is only the first half of the argument. We now
show that the relative entropy cannot bend above its chord along that
path. For a smooth curve $W(s)$, the quantity
$\frac{d^2}{ds^2}f_V(W(s))$ is its scalar second derivative. When $W(s)$ is
the geometric-mean path, this derivative is sometimes called a
\emph{directional geodesic Hessian}. We give its exact matrix expression,
rather than only its sign, because the expression both proves strict
convexity and can be checked numerically or used to estimate curvature.
No coordinate formula for a Riemannian Hessian is needed.

\begin{samepage}
\begin{theorem}[Strict geodesic convexity]\label{thm:gconvex}
Fix a zero-mean Gaussian input state $\rho_V$ with 
$2N\times2N$ quantum covariance matrix $V$, and define
$f_V(W)\coloneqq(\ln2)D(\rho_V\Vert\sigma_W)$ for the zero-mean Gaussian
reference $\sigma_W$. This extended function is geodesically convex on
the set of quantum covariance matrices, in the sense of
\eqref{eq:gconvex-definition}. If $W_0\ne W_1$ and
$f_V(W_0),f_V(W_1)<\infty$, then for every $s\in(0,1)$,
\begin{equation}\label{eq:strict-geodesic}
 f_V(W_0\#_sW_1)<(1-s)f_V(W_0)+sf_V(W_1).
\end{equation}
On faithful reference covariance matrices, the exact directional geodesic Hessian is
\begin{equation}\label{eq:geodesic-hessian}
 \frac{d^2}{ds^2}f_V(W(s))
 =\frac12\int_0^1\Tr\!\left[
 C_VQ_tX(2Q_t-W^{-1})XQ_t\right]dt,
 \qquad X\coloneqq\dot W(s),
\end{equation}
Here $W(s)=W_0\#_sW_1$, 
$C_V\coloneqq V+i\Omega$, and $Q_t\coloneqq(W(s)+it\Omega)^{-1}$.
The formula holds at every point of a faithful reference path, and its
right side is strictly positive when $W_0\ne W_1$.
\end{theorem}
\end{samepage}
\begin{proof}
We first differentiate the path explicitly, then show that the
inverse-matrix second derivative is positive semidefinite. Set
$B\coloneqq W_0^{1/2}$ and
$L\coloneqq\ln(W_0^{-1/2}W_1W_0^{-1/2})$. Then
$W(s)=B e^{sL}B$, $\dot W(s)=B e^{sL}LB$, and
$\ddot W(s)=B e^{sL}L^2B$; the factors $L$ and $e^{sL}$ commute.
Inserting $W(s)^{-1}=B^{-1}e^{-sL}B^{-1}$ between the two derivatives
shows that $\dot W W^{-1}\dot W=B e^{sL}L^2B$. Also
$\ln\det W(s)=\ln\det W_0+s\Tr L$. We have therefore proved
\begin{equation}\label{eq:geodesic-derivatives}
 \ddot W=\dot W W^{-1}\dot W,
 \qquad \frac{d^2}{ds^2}\ln\det W=0.
\end{equation}
For fixed $t<1$, differentiating
$(W+it\Omega)Q_t=I$ gives $\dot Q_t=-Q_t\dot WQ_t$.
Differentiate this identity once more, and use
$\ddot W=\dot W W^{-1}\dot W$ from \cref{eq:geodesic-derivatives}:

\begin{equation}\label{eq:Q-second}
 \frac{d^2}{ds^2}Q_t(W(s))
 =2Q_tXQ_tXQ_t-Q_tXW^{-1}XQ_t
 =Q_tX(2Q_t-W^{-1})XQ_t.
\end{equation}
Although the real symmetric matrix $X$ may be indefinite, the second
derivative in \cref{eq:Q-second} is positive semidefinite. To identify its sign, set
$A_t\coloneqq itW^{-1/2}\Omega W^{-1/2}$. The matrix $A_t$ is Hermitian. Its eigenvalues are $\pm t/\nu_j(W)$: indeed,
$iW^{-1/2}\Omega W^{-1/2}$ is similar to $i\Omega W^{-1}$, whose spectrum
is obtained from Williamson's decomposition. Physicality gives
$\nu_j(W)\ge1$, so all these eigenvalues have magnitude less than one
for $t<1$. Factoring $Q_t=W^{-1/2}(I+A_t)^{-1}W^{-1/2}$ yields
\begin{equation}\label{eq:positive-middle}
 2Q_t-W^{-1}
 =W^{-1/2}(I-A_t)(I+A_t)^{-1}W^{-1/2}\succ0.
\end{equation}
For an eigenvalue $a\in(-1,1)$ of $A_t$, the corresponding eigenvalue
of $(I-A_t)(I+A_t)^{-1}$ is $(1-a)/(1+a)>0$. This proves the strict sign
in \cref{eq:positive-middle}. In \cref{eq:Q-second}, set $B_t\coloneqq XQ_t$;
then the second derivative is $B_t^\dagger(2Q_t-W^{-1})B_t\succeq0$.
The matrix $X$ itself need not be positive. Taking the trace of the second
derivative in \cref{eq:Q-second} against $C_V\succeq0$ proves convexity of
each integrand. The determinant term is affine, so integration gives geodesic
convexity. Differentiation under the integral is justified for faithful
paths by uniform invertibility, and gives \cref{eq:geodesic-hessian}.

To complete the proof, strictness must not be lost when $C_V$ has a
kernel, as it does for pure input states. Strictness also holds at the nonfaithful boundary, where differentiation
under the full improper integral is unnecessary. At $t=0$ and on a
nonconstant geodesic,
\begin{equation}\label{eq:strict-tzero}
 \Tr[C_VW^{-1}XW^{-1}XW^{-1}]
 =\Tr[VW^{-1}XW^{-1}XW^{-1}]>0.
\end{equation}
Here $W^{-1}XW^{-1}XW^{-1}$ is a nonzero real positive semidefinite matrix,
while $V\succ0$. The $t=0$ integrand is therefore strictly convex along the
path. Fix $s\in(0,1)$ and the distinct endpoints. Its strictly positive
chord gap persists by continuity for all $t$ in some interval $[0,a]$.
Every remaining integrand has a nonnegative chord gap. Integrating gives
a strictly positive total gap when the endpoint integrals are finite.
This proves \cref{eq:strict-geodesic}; infinite endpoints cause no difficulty
for extended convexity.
\end{proof}

Only finite matrices enter the Hessian proof. The input state can be pure, so
$C_V$ can have a kernel. The contribution near $t=0$, where its real part
$V$ is strictly positive, prevents this kernel from destroying strictness.
This distinction is useful both for uniqueness and for the scalar examples
below.

\subsection{Existence, uniqueness, and boundary behavior}
A strictly convex objective can still approach its infimum only at
infinity. To rule this out, we need a finite region that contains all
reference matrices with sufficiently small objective.
We obtain a compact sublevel set from an elementary measurement: heterodyne
outcomes provide a classical lower bound on the objective.

\begin{samepage}
\begin{lemma}[Explicit sublevel bounds]\label{lem:compact}
Let $V$ be a quantum covariance matrix of $N$ modes, let $f_V$ be defined
in \eqref{eq:intro-f}, and fix $B>0$. Let $a_+(B)>1$ be the unique solution
on $(1,\infty)$ of
\begin{equation}\label{eq:aplus}
 a_+(B)^{-1}-1+\ln a_+(B)=2B.
\end{equation}
Every  $N$-mode quantum covariance matrix $W$ satisfying $f_V(W)\le B$ obeys
\begin{equation}\label{eq:compact-bound}
 M^{-1}I\preceq W\preceq MI,
 \qquad M\coloneqq a_+(B)\lambda_{\max}(V+I)-1.
\end{equation}
The number $M$ can be enlarged without invalidating the conclusion. If the
input state is faithful and $\nu_0\coloneqq\min_j\nu_j(V)>1$, then also
\begin{equation}\label{eq:faithful-lower}
 \nu_j(W)\geq \coth\!\left(\frac{B+s(V)}{\nu_0-1}\right)>1
 \quad\text{for every }j.
\end{equation}
\end{lemma}
\end{samepage}
\begin{proof}
Heterodyne measurement of centered Gaussian states gives classical Gaussian
distributions whose covariance matrices are proportional to $V+I$ and
$W+I$, with the same proportionality constant
\cite{Serafini2023}. Relative-entropy data processing and the classical
Gaussian formula imply
\begin{equation}\label{eq:heterodyne}
 f_V(W)\geq \frac12\sum_{j=1}^{2N}(a_j^{-1}-1+\ln a_j),
 \quad
 \{a_j\}_{j=1}^{2N}=\operatorname{spec}
       [(V+I)^{-1/2}(W+I)(V+I)^{-1/2}].
\end{equation}
Each summand is nonnegative. The scalar function is increasing for $a>1$,
so each $a_j\le a_+(B)$ and hence $W+I\preceq a_+(B)(V+I)$. This proves the
upper bound. By \cref{lem:uncertainty},
$W\succeq\Omega W^{-1}\Omega^{\top}\succeq M^{-1}I$, proving the lower bound.

For the faithful-input assertion, finite relative entropy already excludes
vacuum factors in $W$: the corresponding $b_j$ in \cref{eq:bj} cannot equal
one. More quantitatively, $V/\nu_0$ is physical, so $b_j\geq \nu_0$ in every
Williamson coordinate system. Setting $\kappa=1$ in
\cref{eq:thermal-cross} gives nonnegative terms whose sum is at most
$B+s(V)$. Therefore
\begin{equation}\label{eq:thermal-bound}
 \frac{\nu_0-1}{2}\ln\frac{\nu_j(W)+1}{\nu_j(W)-1}
 \le B+s(V).
\end{equation}
Solving this scalar inequality proves \cref{eq:faithful-lower}.
\end{proof}

\begin{samepage}
\begin{theorem}[Unique optimal Gaussian covariance matrix]\label{thm:attainment}
For every finite-mode Gaussian input covariance matrix $V$, each of the
optimizations defining $E_R^G$ and $E_{\mathrm{PPT}}^G$ has a unique optimal
centered covariance. The normalized Gaussian Rains optimization in
\eqref{eq:RG} has the same unique optimal covariance matrix as the PPT problem.
These optimal covariance matrices are faithful whenever the input is faithful. If the input is not faithful, an optimal covariance matrix can be
nonfaithful. No finite-valued nonglobal local minimum exists in either
covariance optimization.
\end{theorem}
\end{samepage}
\begin{proof}
A sufficiently noisy product thermal covariance matrix is faithful, separable,
and has finite objective. Choose $B$ strictly larger than its objective.
By \cref{lem:compact}, every relevant minimizing sequence lies in a fixed
compact matrix interval. By \cref{prop:free-geometry} the reference set is closed,
and by \cref{cor:lsc} the objective is lower semicontinuous. A convergent
subsequence therefore attains the infimum. Two distinct minimizers would
contradict \cref{eq:strict-geodesic}, because the intervening geodesic remains
feasible. Every Gaussian Rains minimizer must be PPT by the strict
replacement in \cref{thm:rains}; conversely the unique optimal PPT covariance
attains the Gaussian Rains infimum. This proves the additional uniqueness
claim. The faithful-input assertion follows from \cref{eq:faithful-lower}. For a pure product input, the unique minimizer is
the input itself, so nonfaithful optimal references really occur.

Finally, if a feasible finite-valued $W$ is not globally minimizing, take
a feasible $Y$ with smaller objective. For arbitrarily small $s>0$,
geodesic convexity gives $f_V(W\#_sY)<f_V(W)$. These feasible covariances
converge to $W$, ruling out a nonglobal local minimum.
\end{proof}

As a consequence, $E_R^G$ vanishes exactly on separable Gaussian input states,
and $R_G$ vanishes exactly on PPT Gaussian input states. Indeed, an attained zero
of relative entropy forces the reference state to equal the input.
The PPT-entangled Gaussian examples of \cite{WernerWolf2001} consequently
have $R_G=0<E_R^G$. This is a separate distinction from the gap between
Gaussian-restricted and unrestricted quantities developed in
\cref{sec:symmetric}.

\section{Additivity and symmetry of optimal references}\label{sec:additivity}
The geometry established in \cref{sec:geometry} answers two structural
questions about Gaussian entanglement. Do independent input states require
independent optimal references, or can correlations between the reference
copies lower the answer? Which symmetries of an input state may safely be
imposed on its optimal reference? A geometric midpoint and uniqueness of the optimizer answer these
questions. Both tools act within the Gaussian family, unlike an arithmetic
mixture of Gaussian density operators, which is usually not Gaussian.
The consequences will simplify both the state examples and the resources
used to simulate communication channels.

\subsection{Additivity for independent Gaussian input states}
The tensor-product question asks whether correlations between different
copies can improve a Gaussian reference. A parity symmetrization removes
those correlations without increasing the objective.

\begin{samepage}
\begin{theorem}[Gaussian additivity]\label{thm:additivity}
Let $\rho_{A_1B_1}$ and $\omega_{A_2B_2}$ be finite-mode Gaussian
states with finite covariance matrices. For Gaussian REE and Gaussian
Rains entanglement, defined in \eqref{eq:ERG}, \eqref{eq:RG},
\begin{align}
 E_R^G(A_1A_2;B_1B_2)_{\rho\otimes\omega}
 &=E_R^G(A_1;B_1)_\rho+E_R^G(A_2;B_2)_\omega,
 \label{eq:sep-additive}\\
 R_G(A_1A_2;B_1B_2)_{\rho\otimes\omega}
 &=R_G(A_1;B_1)_\rho+R_G(A_2;B_2)_\omega.
 \label{eq:rains-additive}
\end{align}
In the separable and PPT covariance minimizations, the unique optimal
reference for the product input is the product of the unique single-input
optimal references.
\end{theorem}
\end{samepage}
\begin{proof}
We start with a reference that may correlate the two copies, then
remove its cross-copy blocks by a geometric parity average. Let $V_1,V_2$
be the input covariance matrices, with $N_1,N_2$ modes in their respective
bipartite systems. Group quadratures by copy, so the product state has
covariance matrix $V=V_1\oplus V_2$. Set
$P\coloneqq(-I_{2N_1})\oplus I_{2N_2}$. This is the phase-space
action of parity on the first copy and identity on the second. It is local
with respect to $A_1A_2:B_1B_2$, even though the covariance ordering groups
$A_1B_1$ together. In particular, it preserves both reference sets, and $PVP=V$.

For an arbitrary feasible reference $W$ of finite objective, define
\begin{equation}\label{eq:parity-midpoint}
 X\coloneqq W\#_{1/2}(PWP).
\end{equation}
Congruence invariance and symmetry of the midpoint imply
\begin{equation}\label{eq:parity-invariant}
 PXP=(PWP)\#_{1/2}W=X.
\end{equation}
Conjugation by $P$ changes the sign of the off-diagonal copy blocks
and leaves the diagonal blocks fixed. The identity $PXP=X$ therefore forces
$X_{12}=-X_{12}=0$, so $X=X_1\oplus X_2$. The midpoint is feasible by
\cref{prop:free-geometry}, and invariance of relative entropy under parity
gives
\begin{equation}\label{eq:parity-objective}
 f_V(X)\le\frac12[f_V(W)+f_V(PWP)]=f_V(W).
\end{equation}
Each block $X_j$ is a feasible reference for its own bipartition. For PPT this
follows by taking principal blocks in both uncertainty inequalities. For
separability, take the corresponding principal blocks of the local
witnesses in \cref{eq:sep-set}. Relative entropy is additive on product
states, hence
\begin{equation}\label{eq:product-lower}
 f_V(W)\geq  f_{V_1}(X_1)+f_{V_2}(X_2)
 \geq \inf_{Y_1\in\cF_1}f_{V_1}(Y_1)
       +\inf_{Y_2\in\cF_2}f_{V_2}(Y_2),
\end{equation}
Here $\cF_j$ denotes the separable reference set for the $j$th
bipartition in the REE problem, or its PPT reference set in the PPT
problem; the same choice is made for both copies. Conversely, the product of
feasible references is feasible, giving the opposite inequality of the
infima. Use \cref{thm:rains} for the Rains statement. Existence and
uniqueness from \cref{thm:attainment} identify the optimizer itself.
\end{proof}

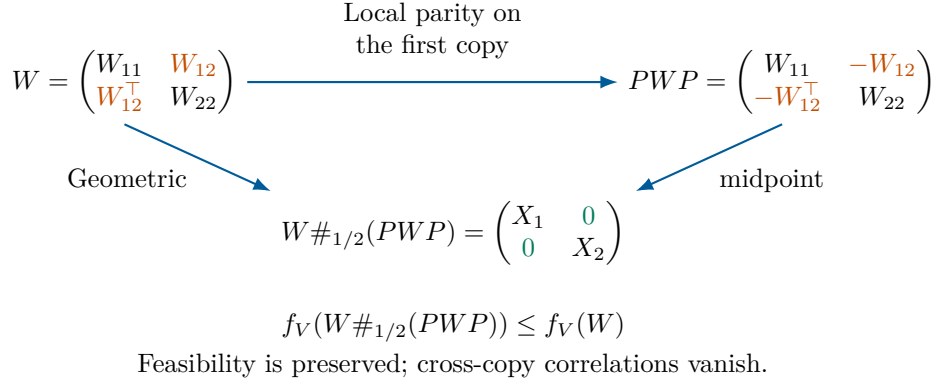
\begin{figure}[t]
\centering
\begin{tikzpicture}[>=Latex,font=\small]
 \node (left) at (0,0)
 {$W=\begin{pmatrix}W_{11}&\textcolor{rateorange}{W_{12}}\\
 \textcolor{rateorange}{W_{12}^{\top}}&W_{22}\end{pmatrix}$};
 \node (right) at (8.7,0)
 {$PWP=\begin{pmatrix}W_{11}&\textcolor{rateorange}{-W_{12}}\\
 \textcolor{rateorange}{-W_{12}^{\top}}&W_{22}\end{pmatrix}$};
 \draw[->,thick,boundblue] (left.east)--
 node[above=5pt,align=center,text=black]{Local parity on\\the first copy}(right.west);
 \node (mid) at (4.35,-2.0)
 {$\displaystyle W\#_{1/2}(PWP)=\begin{pmatrix}X_1&\textcolor{keygreen}{0}\\
 \textcolor{keygreen}{0}&X_2\end{pmatrix}$};
 \draw[->,thick,boundblue] (left.south)--node[below left,text=black]{Geometric}(mid.north west);
 \draw[->,thick,boundblue] (right.south)--node[below right,text=black]{midpoint}(mid.north east);
 \node[below=10pt of mid,align=center]
 {$f_V(W\#_{1/2}(PWP))\le f_V(W)$\\[3pt]
 Feasibility is preserved; cross-copy correlations vanish.};
\end{tikzpicture}
\caption{Geometric parity symmetrization for a product input covariance matrix
$V=V_1\oplus V_2$. Blocks here refer to the two copies, not to the two
parties. Parity on the first copy is a local unitary across the grouped
bipartition $A_1A_2:B_1B_2$. The geometric midpoint is parity invariant and
therefore has no cross-copy blocks. Its diagonal blocks $X_1,X_2$ need not
equal $W_{11},W_{22}$. This construction stays Gaussian and proves additivity without
averaging Gaussian density operators.}
\label{fig:parity}
\end{figure}

\Cref{fig:parity} shows why correlations between copies can be removed.
The proof of \cref{thm:additivity} uses a geometric midpoint of covariance matrices, not a mixture of
states and not an average of Hamiltonians. It includes boundary references
because \cref{thm:gconvex} was proved on the finite-valued extended domain.
No additivity statement for either unrestricted measure follows from it.

\subsection{Inherited symmetries and the free boundary}
A symmetric input state often suggests a smaller reference family,
but a suggestion alone does not justify restricting an optimization.
Uniqueness supplies the missing argument: if a symmetry preserves both
the input state and the allowed references, it must fix the unique optimal
reference. We record the two forms of this observation needed for the
explicit two-mode calculation.

\begin{samepage}
\begin{proposition}[Symmetry and boundary properties]\label{prop:symmetry}
Let $\rho_V$ be a zero-mean Gaussian state, let $f_V$ be defined in
\eqref{eq:intro-f}, and let $\cF$ be either reference set in
\eqref{eq:sep-set}, \eqref{eq:ppt-set}. Let $W_*$ be the unique minimizer supplied
by \cref{thm:attainment}.
If an invertible real matrix $M$ preserves physicality and $\cF$, satisfies
$MVM^{\top}=V$, and obeys $f_V(MWM^{\top})=f_V(W)$ for every feasible $W$, then
\begin{equation}\label{eq:optimizer-symmetry}
 MW_*M^{\top}=W_*.
\end{equation}
In particular, this applies to input-preserving local Gaussian unitaries;
it also applies to a party swap when the setting is symmetric, and to
simultaneous transposition of all modes when that leaves the input fixed.
If $V\notin\cF$, then $W_*$ is not an interior point of $\cF$ in the
real symmetric matrix space. Here ``interior'' means that every symmetric
matrix in some sufficiently small Frobenius-norm ball around $W_*$ is
also in $\cF$.
\end{proposition}
\end{samepage}
\begin{proof}
The transformed covariance is another optimizer, so uniqueness proves
\cref{eq:optimizer-symmetry}. The listed transformations preserve the
corresponding quantum relative entropy and the reference set; full
transposition is a mathematical symmetry, not a quantum channel.
For the last assertion, suppose $W_*$ were interior. The physical geodesic
$W_*\#_sV$ remains free for sufficiently small $s\geq 0$, by continuity and
interiority. Since $f_V(V)=0$ and $f_V(W_*)>0$, geodesic convexity would give
\begin{equation}\label{eq:boundary-descent}
 f_V(W_*\#_sV)\le(1-s)f_V(W_*)<f_V(W_*)
 \qquad(0<s\ll1),
\end{equation}
contradicting optimality.
\end{proof}

The separable-boundary conclusion recovers the property proved by Chen
\cite[Theorem~2]{Chen2005}, here as a consequence of the geometry of the
relative-entropy objective with the input state fixed in its first argument. The symmetry statement does not permit imposing a
reference symmetry that the input lacks. In particular, merely writing an
input in a two-mode standard form does not justify restricting an
arbitrary optimizer to that same standard form.

\subsection{Opposite phase rotations and a multimode separability reduction}
\label{sec:phase-symmetry}
A symmetry useful for communication resources is weaker than exchange symmetry
and does not require either party to have only one mode. It expresses the
phase correlation of a squeezed pair: rotating all of Alice's modes one way
and all of Bob's modes the opposite way leaves the state unchanged.

\begin{samepage}
\begin{proposition}[Opposite-phase-invariant Gaussian states]\label{prop:phase-symmetry}
Let $\rho_V$ be a zero-mean bipartite Gaussian state with quantum
covariance matrix $V$, with $m$ modes on $A$ and $n$ modes on $B$.
Let $R_G,E_R^G$ and $f_V$ have the definitions in
\eqref{eq:RG}, \eqref{eq:ERG}, \eqref{eq:intro-f}.
Suppose that, for every real $\theta$,
\begin{equation}\label{eq:opposite-phase}
 (R_\theta^{\oplus m}\oplus R_{-\theta}^{\oplus n})
 V(R_\theta^{\oplus m}\oplus R_{-\theta}^{\oplus n})^{\top}=V,
 \qquad R_\theta\coloneqq\begin{pmatrix}\cos\theta&\sin\theta\\-\sin\theta&\cos\theta\end{pmatrix}.
\end{equation}
Then the unique optimal Gaussian PPT reference is separable, and
\begin{equation}\label{eq:phase-equality}
 R_G(A;B)_{\rho_V}=E_R^G(A;B)_{\rho_V}
 =\frac1{\ln2}\min\left\{f_V(W):W\succeq I\right\}.
\end{equation}
If $V$ also has no position--momentum cross correlations, order all positions
before all momenta and set $J\coloneqq I_m\oplus(-I_n)$ and let $K_V$ be the position block of $V$. The optimization further reduces to
\begin{equation}\label{eq:paired-reduction}
 V=K_V\oplus JK_VJ,\qquad
 \min\left\{f_V(K\oplus JKJ):K=K^{\top},\ K\succeq I_{m+n}\right\}.
\end{equation}
The value of the latter minimum divided by $\ln2$ is the value in
\eqref{eq:phase-equality}.
\end{proposition}
\end{samepage}
\begin{proof}
By uniqueness and \cref{prop:symmetry}, the optimal PPT covariance $W_*$
inherits \cref{eq:opposite-phase}. Partial transposition changes this symmetry
into an ordinary common phase rotation on all $m+n$ modes. Write the resulting common rotation as $e^{\theta\Omega}$.
Differentiating its invariance at $\theta=0$ gives
$\Omega(TW_*T)-(TW_*T)\Omega=0$, so $TW_*T$ commutes with $\Omega$. A positive real symmetric matrix commuting
with $\Omega$ can be diagonalized by an orthogonal symplectic matrix: identify
it with its complex Hermitian matrix on the mode amplitudes and use unitary
diagonalization. Its ordinary eigenvalues are its symplectic eigenvalues,
each repeated twice. Its uncertainty relation is therefore equivalent to
$TW_*T\succeq I$, or $W_*\succeq I$. Such a covariance is separable: start with
product vacua and add shared Gaussian displacement noise of covariance
$(W_*-I)/2$. Since every $W\succeq I$ is an allowed separable reference,
the three infima coincide. \Cref{thm:rains} supplies the Rains identity.

If the input has no position--momentum cross block, it is also invariant
under simultaneous transposition of all modes. Its unique optimizer inherits
this symmetry, so its position--momentum block vanishes. Differentiating
\cref{eq:opposite-phase} at $\theta=0$ then forces its momentum block to equal
$JKJ$ when its position block is $K$. More explicitly, in the
position-then-momentum order the generator of opposite rotations is
$\left(\begin{smallmatrix}0&J\\-J&0\end{smallmatrix}\right)$.
Its commutator with $K\oplus K_p$ vanishes precisely when
$K_p=JKJ$. Finally, $K\oplus JKJ\succeq I$ is equivalent to $K\succeq I$,
since $J$ is orthogonal. The reduced variable $K$ is a position block,
not the quadratic Hamiltonian matrix $H(W)$.
\end{proof}

The reduction in \cref{prop:phase-symmetry} applies to the full Gaussian optimization, not an
assumption about a convenient reference family. It applies to multimode
channel resource states with correlated, phase-insensitive noise. It does
not say that an arbitrary multimode PPT Gaussian state is separable.
The reduced matrix in \cref{eq:paired-reduction} has size $m+n$, rather than
$2(m+n)$, which makes the examples below particularly transparent.

\section{Covariance algorithms with quantitative runtime guarantees}\label{sec:algorithms}
The preceding results identify the global optimum, but a numerical method
also needs a finite stopping rule. We now show how to compute the Gaussian
reference quantities to a prescribed accuracy. The input state $\rho_V$ is
fixed throughout: the variable is its reference covariance matrix $W$. Our principal algorithm
repeatedly solves an ordinary convex optimization problem that bounds the
original objective from above. Each exact iteration reduces the remaining
objective error by a fixed factor. Approximate subproblem solutions are
also allowed, with an explicit accounting of their errors.

The main text describes the algorithm, its guarantees, and their meaning.
\Cref{app:runtime} proves the quantitative curvature, convergence, and
finite-precision estimates. The general ideas of majorization--minimization
and strongly geodesically convex optimization have established treatments
\cite{Mairal2013,ZhangSra2016}. Our contribution is an explicit analysis of
this Gaussian objective and its covariance constraints, including a
contraction factor that does not deteriorate when the faithful
regularization parameter approaches one.

\subsection{Controlled regularization and an accuracy-independent domain}
A reference with a vacuum factor can have an unbounded Hamiltonian matrix.
We therefore optimize first over reference states whose symplectic
eigenvalues exceed one by a small specified amount. The reference-noise
inequality makes the effect of this restriction quantitative, even for
pure input states.

Fix a quantum covariance matrix $V$ of $N$ modes and choose either
$\cF=\cS$ or $\cF=\cP$, as defined in
\cref{eq:sep-set,eq:ppt-set}. For $1<\kappa\le2$, define the values in bits
\begin{equation}\label{eq:regularized-value}
 \begin{split}
 E_{\cF}&\coloneqq\frac1{\ln2}\inf\left\{f_V(W):W\in\cF\right\},\\
 E_{\cF,\kappa}&\coloneqq\frac1{\ln2}\inf\left\{f_V(W):W\in\cF,
                                      \ W+i\kappa\Omega\succeq0\right\}.
 \end{split}
\end{equation}
Scaling a reference covariance matrix preserves both reference types.
\Cref{prop:noise} therefore gives
\begin{equation}\label{eq:regularization-error}
 0\le E_{\cF,\kappa}-E_{\cF}\le N\log_2\kappa.
\end{equation}
Thus the margin $\kappa-1$ is a controlled approximation, not an assumption
that the original optimum is faithful.

For runtime estimates it is useful to use the \emph{same outer covariance bound}
for every sufficiently small margin. Set
\begin{equation}\label{eq:runtime-initial}
 W_0\coloneqq2(V_A\oplus V_B),\qquad
 \Lambda\ge\norm{V},\quad \Lambda\ge1,\qquad B>f_V(W_0),\quad B\ge1.
\end{equation}
The scalar $\Lambda$ is a known upper bound; using a rational upper bound
rather than an exactly computed eigenvalue is sufficient. The product
reference $W_0$ satisfies every constraint for $1<\kappa\le2$.
Choose $M\ge4$ at least as large as the bound from
\cref{lem:compact}, and define
\begin{equation}\label{eq:K}
 \cK\coloneqq\cF\cap
       \left\{W:W+i\kappa\Omega\succeq0,\ W\preceq MI\right\}.
\end{equation}
This compact set is convex under both ordinary line segments and
matrix-geometric-mean paths. It contains an optimizer of the regularized
problem, and every $W\in\cK$ satisfies $M^{-1}I\preceq W\preceq MI$.
Both $B$ and $M$ can be chosen independently of $\kappa$ and the final
accuracy. \Cref{app:bit-cost} gives explicit choices and explains their
finite-precision construction. For a faithful input state,
\cref{eq:faithful-lower} can alternatively identify a nonzero margin that
removes no optimizer.

We measure optimization errors in natural units until the final division
by $\ln2$. Define
\begin{equation}\label{eq:runtime-gap}
 f_\kappa^\star\coloneqq\min\left\{f_V(W):W\in\cK\right\},\qquad
 e_k\coloneqq f_V(W_k)-f_\kappa^\star,\qquad
 \mu\coloneqq\frac{1}{64M^2\Lambda^2}.
\end{equation}
The initial error is at most $B$, because relative entropy is nonnegative.
The small constant $\mu$ will quantify a worst-case reduction of this
error; it is conservative and does not predict the rate observed in a
numerical implementation.

\subsection{Convex majorization and geometric convergence}
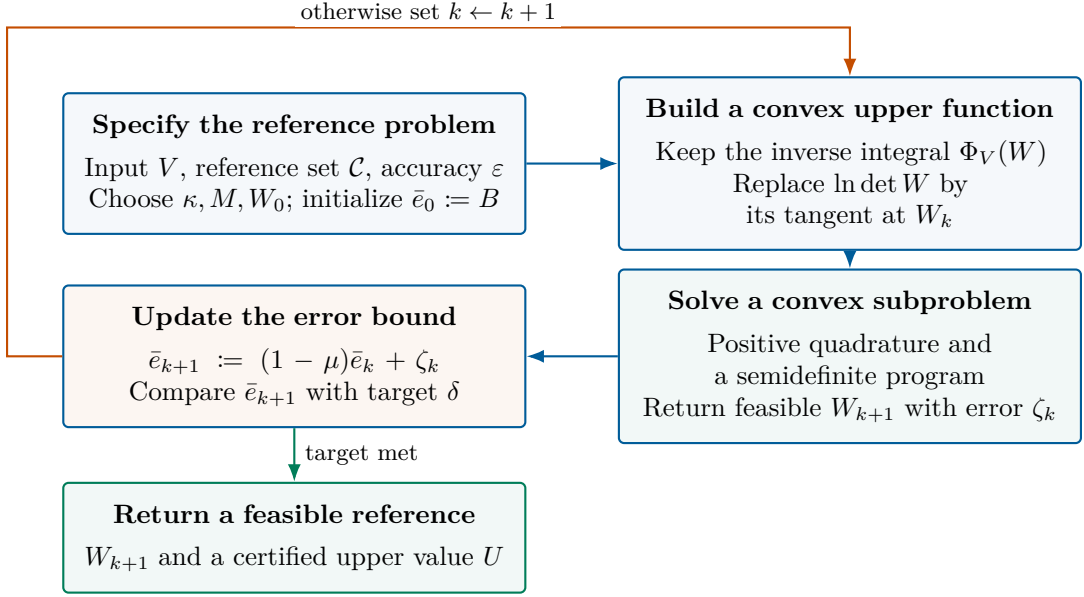
\begin{figure}[t]
\centering
\begin{tikzpicture}[>=Latex,font=\small,
 box/.style={draw=boundblue,rounded corners=2pt,thick,align=center,
              text width=5.55cm,minimum height=1.55cm,inner sep=8pt},
 arr/.style={->,thick,draw=boundblue}]
 \node[box,fill=boundblue!4] (start) at (0,0)
 {\textbf{Specify the reference problem}\\[4pt]
 Input $V$, reference set $\cF$, accuracy $\varepsilon$\\
 Choose $\kappa,M,W_0$; initialize $\bar e_0\coloneqq B$};
 \node[box,fill=boundblue!4] (build) at (7.35,0)
 {\textbf{Build a convex upper function}\\[4pt]
 Keep the inverse integral $\Phi_V(W)$\\
 Replace $\ln\det W$ by its tangent at $W_k$};
 \node[box,fill=keygreen!5] (solve) at (7.35,-2.55)
 {\textbf{Solve a convex subproblem}\\[4pt]
 Positive quadrature and a semidefinite program\\
 Return feasible $W_{k+1}$ with error $\zeta_k$};
 \node[box,fill=rateorange!5] (check) at (0,-2.55)
 {\textbf{Update the error bound}\\[4pt]
 $\bar e_{k+1}\coloneqq(1-\mu)\bar e_k+\zeta_k$\\
 Compare $\bar e_{k+1}$ with target $\delta$};
 \node[box,draw=keygreen,fill=keygreen!5,minimum height=1.05cm]
 (out) at (0,-4.95)
 {\textbf{Return a feasible reference}\\[4pt]
 $W_{k+1}$ and a certified upper value $U$};
 \draw[arr] (start.east)--(build.west);
 \draw[arr] (build.south)--(solve.north);
 \draw[arr] (solve.west)--(check.east);
 \draw[arr,draw=keygreen] (check.south)--node[right,font=\footnotesize]{target met}(out.north);
 \draw[arr,draw=rateorange] (check.west)--(-3.8,-2.55)--(-3.8,1.8)--
 node[above,font=\footnotesize,fill=white,inner sep=2pt]{otherwise set $k\leftarrow k+1$}
 (7.35,1.8)--(build.north);
\end{tikzpicture}
\caption{Classical computation of a Gaussian-reference entanglement bound.
The two right-hand boxes implement one convex-majorization iteration; they
do not represent quantum operations. The maintained scalar $\bar e_k$ is
an upper bound on the unknown objective error $e_k$, provided every inner
error $\zeta_k$ includes quadrature, optimization, and rounding errors and
the returned reference is feasible. The original unregularized value also
requires the separate correction in \cref{eq:regularization-error}.
A certified inner objective gap, not an unverified solver termination flag,
is required for the guarantee.}
\label{fig:algorithm}
\end{figure}

The positive integral separates the objective into an ordinary convex
term and a concave logarithmic determinant. Define
\begin{equation}\label{eq:phi}
 \Phi_V(W)\coloneqq\frac12\int_0^1
             \Tr\!\left[C_V(W+it\Omega)^{-1}\right]dt,
 \qquad C_V\coloneqq V+i\Omega.
\end{equation}
Matrix inversion is operator convex, and $C_V\succeq0$, so $\Phi_V$ is
convex along straight lines in $W$. At the current covariance matrix
$W_k$, replace only $\frac12\ln\det W$ by its tangent upper bound
\begin{equation}\label{eq:logdet-tangent}
 \ln\det W\le\ln\det W_k+\Tr[W_k^{-1}(W-W_k)].
\end{equation}
The next iterate minimizes the resulting convex function:
\begin{equation}\label{eq:MM}
 W_{k+1}\coloneqq\argmin\left\{
 \Phi_V(W)+\frac12\Tr[W_k^{-1}W]:W\in\cK\right\}.
\end{equation}
The minimizer is unique; the positive curvature of the inverse integral
is shown in \cref{eq:phi-hessian}. The full upper function, including its
constant term, is denoted by
\begin{equation}\label{eq:surrogate}
 \begin{split}
 q_k(W)&\coloneqq f_V(W)+\mathcal B(W,W_k),\\
 \mathcal B(W,W_k)&\coloneqq\frac12\left[
       \Tr(W_k^{-1}W)-\ln\det W+\ln\det W_k-2N\right].
 \end{split}
\end{equation}
Here $\mathcal B\ge0$ and $\mathcal B(W_k,W_k)=0$. Consequently, exact
minimization gives $f_V(W_{k+1})\le q_k(W_{k+1})\le q_k(W_k)=f_V(W_k)$.
This construction is called \emph{majorization--minimization}; the convex
upper function is also called a surrogate \cite{Mairal2013}.
\Cref{fig:algorithm} summarizes the computation.

\begin{samepage}
\begin{theorem}[Geometric convergence of convex majorization]\label{prop:MM}
Fix a finite-mode Gaussian input state $\rho_V$, either reference set
$\cF\in\{\cS,\cP\}$, and the domain $\cK$ in \eqref{eq:K}. Let $\mu$ and
$e_k$ be defined in \eqref{eq:runtime-gap}, and initialize at the product
reference $W_0$ in \eqref{eq:runtime-initial}. For exact subproblem
solutions in \eqref{eq:MM},
\begin{equation}\label{eq:MM-rate}
 e_{k+1}\le(1-\mu)e_k,\qquad e_k\le B\exp(-\mu k).
\end{equation}
The iterates converge to the unique optimal covariance matrix in $\cK$.
More generally, suppose $W_{k+1}\in\cK$ has certified subproblem error
\begin{equation}\label{eq:inner-accuracy}
 q_k(W_{k+1})-\min\left\{q_k(W):W\in\cK\right\}\le\zeta_k.
\end{equation}
Then
\begin{equation}\label{eq:inexact-rate}
 e_{k+1}\le(1-\mu)e_k+\zeta_k.
\end{equation}
For $0<\delta<B$, errors $\zeta_k\le\mu\delta/2$ and at least
\begin{equation}\label{eq:iteration-count}
 K_\delta\coloneqq
 \left\lceil64M^2\Lambda^2\ln\frac{2B}{\delta}\right\rceil
\end{equation}
iterations suffice to give $e_{K_\delta}\le\delta$. The contraction
constant is independent of $\kappa-1$, including for pure input states.
\end{theorem}
\end{samepage}

\begin{proof}[Proof idea]
The geodesic Hessian in \cref{thm:gconvex} is bounded below by an explicit
positive constant, not merely positive in each direction. More precisely,
\cref{lem:quantitative-curvature} proves strong geodesic convexity with
constant $\mu$. Compare the convex upper function with the point a fraction
$\mu$ of the way along the geometric path from $W_k$ toward the true
optimizer. The decrease furnished by this curvature exceeds the error
introduced by replacing the logarithmic determinant by its tangent.
Because the next iterate minimizes the upper function, it does at least
as well as this comparison point. The optimizer is used only in the
proof, not in the algorithm. \Cref{app:majorization-rate} gives the full
matrix estimates and proves \cref{eq:MM-rate,eq:inexact-rate}.
Iterating \cref{eq:inexact-rate} with the specified error tolerance gives
$e_k\le B e^{-\mu k}+\delta/2$, which proves \cref{eq:iteration-count}.
\end{proof}

\paragraph{Accompanying implementation.}
The accompanying source and ancillary-code package contains
\texttt{code/majorization\_solver.py}, a Python implementation of
\eqref{eq:MM}, the outer iteration underlying \cref{thm:runtime}, using positive graded quadrature and a log-determinant
barrier for the convex subproblems. This file is provided for inclusion
with the arXiv posting. It accepts the full PPT or separable matrix
constraints and records feasibility margins and a subproblem-error
estimate, rather than implementing only a local search in a covariance
ansatz. The demonstration uses floating-point arithmetic: its residuals
are numerical diagnostics, not outward-rounded interval certificates.
Theorem~\ref{prop:MM} applies whenever the stated inner-error and feasibility
conditions are verified; the distributed implementation does not claim
the worst-case bit complexity of the certified solver analyzed in
\cref{thm:runtime}. The code's README states the tested cases and these
implementation limits explicitly.

\Cref{prop:MM} is a finite-accuracy guarantee, not only an asymptotic
convergence statement. Its logarithmic dependence on $1/\delta$ means that
adding a fixed number of accuracy bits requires a bounded additional
number of outer iterations for fixed input data. Small nonzero inner
errors produce a controlled error floor; they are not silently treated
as exact minimizations. An approximate step need not decrease the
objective at every iteration, but the error recurrence still holds. The guarantee concerns the chosen bounded
reference domain. It does not optimize the outer channel-simulation
parameters introduced later in \cref{sec:channels}.

\subsection{Numerical integration, semidefinite subproblems, and total cost}
A convex subproblem is still a numerical task because $\Phi_V$ contains an
integral. A uniform grid is inefficient near $t=1$ when $\kappa$ is close
to one. Instead, divide $[0,1]$ into intervals that shrink geometrically
toward that endpoint and use a positive-weight Gauss--Legendre rule on
each interval. This standard quadrature rule is exact on polynomials up
to a specified degree \cite[Section~3.5(v)]{DLMF}. Our resolvent estimates
in \cref{app:dyadic-quadrature} give a uniform error bound on the entire
reference domain.

Let $\eta>0$ be the allowed integral error, and set
\begin{equation}\label{eq:qbound}
 q_0\coloneqq\frac{M}{1-\kappa^{-1}}.
\end{equation}
The number of quadrature points can be chosen of order
\begin{equation}\label{eq:node-count-main}
 O\!\left(\left[1+\ln\frac{\kappa}{\kappa-1}\right]
           \left[1+\ln^+\frac{\Tr(V)q_0}{\eta}\right]\right),
 \qquad \ln^+x\coloneqq\max\{0,\ln x\}.
\end{equation}
Thus the count depends logarithmically, not as a power, on the inverse
regularization margin and the inverse error tolerance. Positive quadrature
weights preserve ordinary convexity.

For nodes $t_\ell$ and weights $w_\ell>0$, with $\sum_\ell w_\ell=1$,
the quadrature subproblem has the following exact semidefinite
representation before coefficient rounding:
\begin{equation}\label{eq:inverse-SDP}
 \begin{aligned}
 \text{minimize}\quad &\frac12\Tr(W_k^{-1}W)
             +\frac12\sum_\ell w_\ell\Tr(C_VY_\ell)\\
 \text{subject to}\quad & W\in\cK,\qquad
 \begin{pmatrix}W+it_\ell\Omega&I\\I&Y_\ell\end{pmatrix}\succeq0
                      \quad\text{for every }\ell .
 \end{aligned}
\end{equation}
The Schur complement gives $Y_\ell\succeq(W+it_\ell\Omega)^{-1}$;
minimization is exact because $C_V\succeq0$. The auxiliary $Y_\ell$ are
Hermitian matrices, whereas $W$ and the local separability witnesses are
real symmetric. A complex Hermitian inequality can be represented by a
real symmetric one of twice the dimension. The objective retains its
original trace, not twice that trace. A linear objective with such
constraints is called a \emph{semidefinite program (SDP)}
\cite{VandenbergheBoyd1996}.

A complete runtime bound must count the cost of solving these subproblems,
not only their number. The following statement uses rationally specified
covariance data; an arbitrary real number is not automatically available
to a computer at arbitrary precision.

\begin{samepage}
\begin{theorem}[Finite-accuracy computation and bit complexity]
\label{thm:runtime}
Let $V$ be a rational quantum covariance matrix of $N$ modes, with specified
bipartition, and let $b$ denote its total binary encoding length. Choose
the rational bounds $\Lambda,B,M$ by the explicit construction in
\cref{app:bit-cost}, satisfying \eqref{eq:runtime-initial}, \eqref{eq:K} with
$M\ge4$, independently of the requested accuracy. For either
$\cF\in\{\cS,\cP\}$ and $\varepsilon\coloneqq2^{-p}$ with
$p$ a nonnegative integer,
there is a finite-precision implementation returning a feasible reference
covariance matrix $\widehat W\in\cF$ and a rational number $U$ such that
\begin{equation}\label{eq:runtime-output}
 E_{\cF}\le\frac{f_V(\widehat W)}{\ln2}\le U
                         \le E_{\cF}+\varepsilon.
\end{equation}
It uses at most
\begin{equation}\label{eq:runtime-work}
 O\!\left(M^2\Lambda^2\left[1+\ln\frac{B}{\varepsilon}\right]\right)
 \operatorname{poly}\!\left(
 N,b,1+\ln M,1+\ln\Lambda,1+\ln B,p\right)
\end{equation}
elementary bit operations. The polynomial depends on a choice of a
certified bounded-SDP algorithm, not on $V$ or $\varepsilon$.
In particular, for each fixed rational input covariance matrix, the
cost is polynomial in the requested number $p$ of accuracy bits. This is
not a uniform polynomial bound in the encoded input size, because the
prefactor contains the numerical magnitudes $M^2\Lambda^2$.
\end{theorem}
\end{samepage}

The proof in \cref{app:bit-cost} checks the boundedness and strict-feasibility
hypotheses needed for polynomial-time SDP approximation
\cite[Theorem~1.1]{DeKlerkVallentin2016}. It also bounds coefficient
rounding, enforces exact feasibility of the returned rational iterate,
and prevents its encoding length from growing unchecked from one
iteration to the next. These conditions are important near a pure state;
merely calling an SDP solver is not a bit-complexity proof.

For an explicit allocation of errors, take
$\kappa\coloneqq1+\varepsilon/(8N)$, so the regularization error is less
than $\varepsilon/4$ bits, and set $\delta\coloneqq\varepsilon\ln2/2$.
Solve each true surrogate to accuracy at most $\mu\delta/2$, including all
numerical errors. \Cref{eq:iteration-count} then leaves at most
$\varepsilon/2$ bits of optimization error. Finally evaluate the reference
objective from above with at most $\varepsilon/4$ bits of error.
This gives \cref{eq:runtime-output}; it does not identify numerical
quadrature error with entanglement regularization error.

The bounds are conservative. In particular, \cref{lem:compact} allows $M$
to grow exponentially in the initial objective bound $B$. The total-cost
statement therefore gives an explicit dependence on covariance size and
conditioning, rather than claiming that every multimode instance is easy.
Neither the unrestricted Fock-space algorithm nor the outer channel
optimizations inherit \cref{thm:runtime} without further analysis.

\subsection{Projected gradient and practical stopping certificates}
The original projected-gradient method remains an alternative when
semidefinite projections are simpler to implement than full surrogate
minimizations. Its rate can also be quantified, although its conservative
constant deteriorates as the regularization margin becomes small.
On real symmetric matrices the gradient and a sufficient smoothness
constant $L$ are
\begin{equation}\label{eq:gradient}
 G_V(W)\coloneqq\nabla f_V(W)
 =\frac12W^{-1}-\frac12\int_0^1
                  \Rea[Q_t(W)C_VQ_t(W)]\,dt,
 \qquad Q_t(W)\coloneqq(W+it\Omega)^{-1},
\end{equation}
\begin{equation}\label{eq:constants}
 L\coloneqq\frac12M^2+\norm{C_V}q_0^3.
\end{equation}
The derivative bound is proved in \cref{app:derivatives}. Define the
Euclidean projection and the iteration by
\begin{equation}\label{eq:projection}
 \Proj_{\cK}(X)\coloneqq
 \argmin\left\{\tfrac12\norm{Y-X}_{\mathrm F}^2:Y\in\cK\right\},
\end{equation}
\begin{equation}\label{eq:projected-iteration}
 W_{k+1}\coloneqq\Proj_{\cK}(W_k-\alpha G_V(W_k)).
\end{equation}
The projection is a classical nearest-matrix computation with linear
matrix inequalities, not a quantum measurement.

\begin{samepage}
\begin{theorem}[Quantitative projected-gradient convergence]\label{thm:projected}
For a fixed Gaussian input covariance matrix $V$, use $\cK$, $\mu$, and
$L$ from \eqref{eq:K}, \eqref{eq:runtime-gap}, \eqref{eq:constants}. Starting at a feasible reference in $\cK$, exact gradients and
projections in \eqref{eq:projected-iteration}, with constant
$0<\alpha\le1/L$, converge to the unique minimizer on $\cK$ and satisfy
\begin{equation}\label{eq:projected-rate}
 e_{k+1}\le
 \left(1+\frac{\mu\alpha}{4M^2}\right)^{-1}e_k.
\end{equation}
For $\alpha=1/L$, the contraction factor is
$(1+\mu/(4M^2L))^{-1}$.
\end{theorem}
\end{samepage}

For a finite stopping certificate, an outward normal at $W\in\cK$ means
a real symmetric matrix $Z$ with
\begin{equation}\label{eq:normal}
 Z\in N_{\cK}(W)\quad\Longleftrightarrow\quad
       \inner{Z}{Y-W}\le0\quad\text{for every }Y\in\cK.
\end{equation}
It describes a direction pointing away from the feasible set. An exact
projection $Y=\Proj_{\cK}(W-\alpha G_V(W))$ provides
\begin{equation}\label{eq:projection-normal}
 Z\coloneqq(W-Y)/\alpha-G_V(W)\in N_{\cK}(Y).
\end{equation}
The residual must therefore use the gradient at the \emph{new} point $Y$.

\begin{samepage}
\begin{proposition}[Linear and quadratic residual certificates]\label{prop:certificate}
For $W\in\cK$, $Z\in N_{\cK}(W)$, and
$\eta\coloneqq\norm{G_V(W)+Z}_{\mathrm F}$, define
$R_M\coloneqq2\sqrt{2N}M\ln M$. Then
\begin{equation}\label{eq:residual-bound}
 0\le f_V(W)-f_\kappa^\star
 \le\min\left\{R_M\eta,\frac{M^2\eta^2}{2\mu}\right\}.
\end{equation}
Its Gaussian entanglement upper bound consequently has excess at most
\begin{equation}\label{eq:total-error}
 N\log_2\kappa+
 \frac1{\ln2}\min\left\{R_M\eta,\frac{M^2\eta^2}{2\mu}\right\}.
\end{equation}
\end{proposition}
\end{samepage}

\Cref{app:majorization-rate} proves the estimates in
\cref{thm:projected,prop:certificate}. The quadratic
residual term strengthens the earlier linear certificate near an optimum.
In floating-point computation, feasibility, normal-cone membership, and
gradient error must also be bounded. These certificates do not turn an
uncertified local optimizer into a certified implementation. The supplied
numerical examples and their scope are described in \cref{app:numerics}.

\section{Gaussian state entanglement and distillation benchmarks}
\label{sec:operational-states}
The preceding sections explain how to optimize Gaussian reference states.
We now ask what those values tell us about quantum tasks and how tight the
resulting bounds can be. We first define entanglement and secret-key
distillation, then give an analytic two-mode example that can be solved by
one scalar root. We also show why this Gaussian calculation must not be
confused with the unrestricted entanglement measure. Finally, explicit
protocols and two- and three-mode plots compare what can be achieved with
what the converse bounds permit. All rates in this section are per copy
of a shared state, not per channel use.

\subsection{Operational definitions and the converse inequalities}
Entanglement distillation converts many imperfect shared states
into fewer nearly ideal entangled pairs. Secret-key distillation instead
produces uniform shared classical bits hidden from an eavesdropper.
Let $D_{\leftrightarrow}(A;B)_\rho$ be the asymptotic number of ebits per copy
of $\rho_{AB}$ distillable by local operations and unlimited authenticated
public classical communication. For $\ell$ input copies, the output approaches a maximally entangled
state of dimension $d_\ell$, and the rate is measured by
$\ell^{-1}\log_2d_\ell$ as the error tends to zero. We define
\begin{equation}\label{eq:ideal-target-definitions}
 \ket{\Phi^d}\coloneqq\frac1{\sqrt d}\sum_{j=0}^{d-1}\ket{jj},\qquad
 \Phi^d\coloneqq\ket{\Phi^d}\bra{\Phi^d},\qquad
 \overline\Phi^d\coloneqq\frac1d\sum_{j=0}^{d-1}\ket{jj}\bra{jj}.
\end{equation}
For normalized states, fidelity and normalized trace distance are defined
as $F(\rho,\sigma)\coloneqq\norm{\sqrt\rho\sqrt\sigma}_1^2$ and
$\Delta(\rho,\sigma)\coloneqq\tfrac12\norm{\rho-\sigma}_1$.
Entanglement infidelity is $1-F$ with the ideal entangled target. Let
$K_D(A;B)_\rho$ be the corresponding secret bits per copy, with correctness
and secrecy against a purification of the input and the full public transcript.
The operational key error is normalized trace distance from a uniform,
perfectly correlated key independent of Eve, retaining Eve's actual marginal.
For example, if the final state is $\omega_{K_AK_BE}$, this key error
is $\Delta(\omega_{K_AK_BE},\overline\Phi^d_{K_AK_B}\otimes\omega_E)$,
where $E$ includes Eve's systems and the public transcript.
No preshared secret key or entanglement is supplied. All measurement outcomes,
including unsuccessful ones, count in the rate normalization.

The equivalent private-state formulation retains the laboratories' local
shield systems instead of discarding them \cite{HorodeckiEtAl2005,HorodeckiEtAl2009}.
An ideal private state of key dimension $d$ has the form
\begin{equation}\label{eq:private-target}
 \gamma_{K_AA'K_BB'}=U(\Phi^d_{K_AK_B}\otimes\sigma_{A'B'})U^\dagger,
 \qquad
 U\coloneqq\sum_{a,b=0}^{d-1}\ket{ab}\bra{ab}_{K_AK_B}\otimes U^{ab}_{A'B'}.
\end{equation}
The operators $U^{ab}_{A'B'}$ are unitaries, and $\sigma_{A'B'}$ is
a normalized shield state. Measuring the key registers $K_A,K_B$ gives
an ideal secret key, while the shield systems $A',B'$ remain in the trusted
laboratories. The shield is not assumed to have a fixed dimension. The associated privacy
test is $T_\gamma\coloneqq U(\Phi^d\otimes I)U^\dagger$. For every normalized separable
state $\sigma$, $\Tr[T_\gamma\sigma]\le1/d$
\cite[Section~III]{WildeTomamichelBerta2017}. For an entanglement target,
$\Tr[\Phi^d\tau]\le1/d$ for every $\tau\in\mathrm{PPT}'$: partial transposition
of $\Phi^d$ is the swap divided by $d$, and trace-norm duality proves the claim.

\begin{samepage}
\begin{proposition}[State-distillation bounds]\label{prop:state-operational}
Let $\rho_{AB}$ be a Gaussian state with finitely many modes on each
party and a finite covariance matrix. For the unrestricted and
Gaussian-restricted quantities in \eqref{eq:ER}, \eqref{eq:R}, \eqref{eq:ERG}, \eqref{eq:RG} and the
distillation tasks just defined,
\begin{align}
 D_{\leftrightarrow}(A;B)_\rho
 &\le R^\infty(A;B)_\rho\le R(A;B)_\rho\le R_G(A;B)_\rho,
 \label{eq:distillation-chain}\\
 D_{\leftrightarrow}(A;B)_\rho\le K_D(A;B)_\rho
 &\le E_R^\infty(A;B)_\rho\le E_R(A;B)_\rho\le E_R^G(A;B)_\rho.
 \label{eq:key-chain}
\end{align}
Here $E^\infty(\rho)\coloneqq\lim_{k\to\infty}E(\rho^{\otimes k})/k$
for $E\in\{R,E_R\}$. The limits exist by subadditivity and are finite. Both bounds apply
to arbitrary finite mode numbers and bipartitions, not only to two-mode
states.
\end{proposition}
\end{samepage}
\begin{proof}
We recall the testing argument to expose the different reference sets.
If a retained output $\zeta$ passes the appropriate target test with
probability $p$, let $h_2(p)\coloneqq-p\log_2p-(1-p)\log_2(1-p)$ be the binary entropy. Relative-entropy data processing to the binary test gives
\begin{equation}\label{eq:one-test-bound}
 E(\zeta)\ge p\log_2d-h_2(p).
\end{equation}
Use $E=R$ and the maximally entangled test, or $E=E_R$ and the privacy test.
The inequality follows because the reference acceptance is at most $1/d$;
the rejected contribution to its cross entropy is nonnegative. It remains
valid for the subnormalized Rains reference. A vanishing operational key
error permits a retained private-state output whose acceptance tends to
one \cite{HorodeckiEtAl2009,WildeTomamichelBerta2017}.

For an $\ell$-copy input, LOCC monotonicity bounds the output entanglement by
$E(\rho^{\otimes\ell})$. Divide \cref{eq:one-test-bound} by $\ell$, let the
error vanish, and take the regularized limit. Tensor products of feasible
references prove subadditivity. The bound by a local marginal entropy in
\cref{eq:marginal-upper} makes both quantities finite. Distilled ebits yield
secret bits by local measurement. Finally, restricting a reference set to
Gaussian states only increases an infimum. These arguments prove the two
chains, extending the established Rains and REE converses
\cite{Rains1999,Rains2001,HorodeckiEtAl2005} to the finite-energy bosonic setting
without a continuity estimate depending on the shield dimension.
\end{proof}

Rains entanglement is \emph{not} a general secret-key upper bound. It vanishes
on every PPT state, whereas PPT states with distillable key exist
\cite{HorodeckiEtAl2005,HorodeckiEtAl2009}. We do not assume that such a
counterexample is Gaussian. For a $1\times n$ Gaussian partition, however,
\cref{cor:one-many} makes $R_G=E_R^G$, so that common numerical value bounds
both tasks. The same conclusion holds in every mode number for
\cref{prop:phase-symmetry}'s opposite-phase-invariant family.

\subsection{An analytic two-mode family and the Gaussian-reference gap}
\label{sec:symmetric}
An explicit calculation helps interpret the general results. We consider
symmetric squeezed thermal states, for which the reference optimization
reduces to one real parameter. This family also shows a limitation of the
Gaussian restriction: even a faithful Gaussian state can have a smaller
unrestricted entanglement value than its best Gaussian reference gives.

\paragraph{Reduction to a boundary family.}
Consider the symmetric squeezed thermal covariance
\begin{equation}\label{eq:Vvr}
 V(v,r)\coloneqq v\begin{pmatrix}
 \cosh(2r)I_2&\sinh(2r)Z\\
 \sinh(2r)Z&\cosh(2r)I_2
 \end{pmatrix},
 \qquad Z=\diag(1,-1),\quad v\geq 1,\quad r\geq 0.
\end{equation}
Both ordinary symplectic eigenvalues are $v$, whereas the partially
transposed symplectic eigenvalues are
\begin{equation}\label{eq:uw}
 u\coloneqq ve^{-2r},\qquad w\coloneqq ve^{2r}.
\end{equation}
The state is separable exactly when $u\geq 1$. In that case $E_R^G=R_G=0$.
We therefore assume $u<1$ below. The input entropy in natural units is
$s(V(v,r))=2\gent((v-1)/2)$.

Chen's analysis already gives symmetric separable-reference reductions
\cite[Section~V]{Chen2005}. The next proposition recovers the reduction for
\cref{eq:Vvr} from our uniqueness theorem, without assuming in advance that
an arbitrary two-mode optimizer shares an input's standard form.

\begin{samepage}
\begin{proposition}[Optimal symmetric reference]\label{prop:symmetric-ref}
For $v\geq 1$, $r\geq 0$, and $ve^{-2r}<1$, the unique optimal separable
Gaussian covariance matrix for $V(v,r)$ has the form
\begin{equation}\label{eq:Wb}
 W_b\coloneqq\frac12\begin{pmatrix}
 (b^2+1)I_2&(b^2-1)Z\\
 (b^2-1)Z&(b^2+1)I_2
 \end{pmatrix},\qquad b>1.
\end{equation}
It is also the unique optimal PPT Gaussian covariance matrix.
\end{proposition}
\end{samepage}
\begin{proof}
The input is invariant under opposite local phase rotations
$R_\theta\oplus R_{-\theta}$, where
$R_\theta=\left(\begin{smallmatrix}\cos\theta&\sin\theta\\
-\sin\theta&\cos\theta\end{smallmatrix}\right)$. By
\cref{prop:symmetry}, the optimizer is invariant under the same rotations.
Its local blocks must therefore be $aI_2$ and $b'I_2$, and its cross block
must have the form
$\left(\begin{smallmatrix}c&d\\d&-c\end{smallmatrix}\right)$.
For the latter assertion, differentiating rotation invariance gives
$JC+CJ=0$ with $J=\left(\begin{smallmatrix}0&1\\-1&0\end{smallmatrix}\right)$,
whose real solutions have the displayed form. Party-swap symmetry forces
$a=b'$. Full transposition, represented by $Z\oplus Z$, leaves the input
fixed and sends $d$ to $-d$, so $d=0$. Thus
\begin{equation}\label{eq:Wac}
 W=\begin{pmatrix}aI_2&cZ\\cZ&aI_2\end{pmatrix}.
\end{equation}
PPT and separability coincide here and require $a-|c|\geq 1$.
The only nonfaithful covariance matrix in this feasible symmetric family is
$I_4$, which gives infinite relative entropy for the entangled input.
Every candidate of finite objective is therefore faithful.

Set $\nu\coloneqq\sqrt{a^2-c^2}>1$. Its Hamiltonian is
$H(W)=2\nu\arccoth(\nu)W^{-1}$. The only part of its objective that depends
on the sign of $c$ is
\begin{equation}\label{eq:sign-c}
 \frac14\Tr[V(v,r)H(W)]
 =\frac{2v}{\nu}\arccoth(\nu)
      [a\cosh(2r)-c\sinh(2r)].
\end{equation}
For $r>0$, replacing negative $c$ by $|c|$ strictly decreases this expression
and preserves feasibility. Thus an optimizer has $c\geq 0$.
If $a-c>1$, both the physical and PPT inequalities are strict, so the
covariance is interior to the PPT set, contrary to
\cref{prop:symmetry}. Hence $a-c=1$. Setting
$b=\sqrt{a^2-c^2}=\sqrt{a+c}>1$ yields \cref{eq:Wb}.
\end{proof}

\paragraph{A strictly convex scalar objective.}
The ordinary symplectic eigenvalues of $W_b$ are $b,b$ and its partially
transposed eigenvalues are $1,b^2$. Direct substitution into the Gibbs
formula gives the scalar function
\begin{equation}\label{eq:Fb}
 F_b(b)\coloneqq\ln\frac{b^2-1}{4}
        +(ub+w/b)\arccoth b-2\gent\!\left(\frac{v-1}{2}\right).
\end{equation}
A change to the inverse symplectic eigenvalue exposes convexity. Define
$t=1/b\in(0,1)$ and
\begin{equation}\label{eq:Ft}
 F(t)\coloneqq\ln\frac{1-t^2}{4t^2}
       +(u/t+wt)\arctanh t-2\gent\!\left(\frac{v-1}{2}\right).
\end{equation}

\begin{samepage}
\begin{theorem}[Unique scalar root]\label{thm:scalar}
For the entangled symmetric squeezed thermal state in \eqref{eq:Vvr},
let $u,w$ be as in \eqref{eq:uw} and $F$ as in \eqref{eq:Ft}. Then
\begin{equation}\label{eq:scalar-result}
 E_R^G(A;B)_{\rho_{V(v,r)}}=R_G(A;B)_{\rho_{V(v,r)}}
 =\frac{F(t_*)}{\ln2},
\end{equation}
where $F$ is strictly convex on $(0,1)$ and $t_*$ is the unique solution of
\begin{equation}\label{eq:root}
 \frac{(u-2)/t+wt}{1-t^2}
       +\left(w-\frac{u}{t^2}\right)\arctanh t=0.
\end{equation}
The optimal covariance matrix is $W_{1/t_*}$. Bisection on the derivative, with
endpoint bracketing, converges to its global optimum.
\end{theorem}
\end{samepage}
\begin{proof}
\Cref{prop:symmetric-ref,cor:one-many} justify the scalar minimization for
both Gaussian quantities. Differentiation of \cref{eq:Ft} gives the left
side of \cref{eq:root}. To prove strict convexity, expand the two elementary
functions for $0<t<1$:
\begin{equation}\label{eq:positive-series}
 F(t)=u-2\gent((v-1)/2)-\ln4-2\ln t
       +\sum_{k=1}^{\infty}c_kt^{2k},
 \qquad c_k\coloneqq\frac{u}{2k+1}+\frac{w}{2k-1}-\frac1k.
\end{equation}
Every coefficient is strictly positive. Indeed,
\begin{equation}\label{eq:ck-positive}
 k(4k^2-1)c_k
 =2k^2(u+w-2)+k(w-u)+1>0,
\end{equation}
since $u+w=2v\cosh(2r)\geq 2$ and $w\geq  u$. The series and its derivatives
converge uniformly on compact subintervals. Thus
\begin{equation}\label{eq:Fsecond}
 F''(t)=\frac2{t^2}
       +\sum_{k=1}^{\infty}2k(2k-1)c_kt^{2k-2}>0.
\end{equation}
At $t\downarrow0$, $F(t)\to+\infty$ because of $-2\ln t$.
At $t\uparrow1$, the coefficient of $-\ln(1-t)$ in \cref{eq:Ft} tends to
$(u+w)/2-1>0$ for an entangled input. Thus $F(t)\to+\infty$ there as well.
A strictly convex differentiable function with these endpoint limits has
one interior minimizer and precisely one zero of its derivative.
\end{proof}

\begin{table}[t]
\centering
\setlength{\tabcolsep}{10pt}
\begin{tabular}{@{}rrrr@{}}
\toprule
Thermal parameter $v$ & Squeezing $r$ & $E_R^G=R_G$ (bits) & Optimal $b$\\
\midrule
$1$    & $0.3$ & $0.567326784$ & $1.259312812$\\
$1$    & $0.5$ & $1.059118723$ & $1.563795799$\\
$1$    & $1.0$ & $2.392627502$ & $2.660265630$\\
$1.01$ & $0.3$ & $0.509530435$ & $1.268497882$\\
$1.5$  & $1.0$ & $1.319403490$ & $3.286944261$\\
$2$    & $1.0$ & $0.905499583$ & $3.811256658$\\
\bottomrule
\end{tabular}
\caption{Numerical evaluations of the unique root in \cref{thm:scalar}.
The first three input states are pure, and the remaining input states are faithful.
These are Gaussian-restricted values, not unrestricted values. The displayed
decimals are numerical approximations; the strict separation in
\cref{thm:gap} uses a separate analytic lower bound with interval evaluation.}
\label{tab:scalar}
\end{table}

\paragraph{A faithful input for which Gaussian references are suboptimal.}
The distinction between Gaussian-restricted and unrestricted entanglement
can already be seen for a pure two-mode squeezed vacuum. Its unrestricted
REE and Rains entanglement equal its entropy of entanglement
\cite{VedralPlenio1998,Rains1999,Rains2001}:
\begin{equation}\label{eq:pure-unrestricted}
 E_R(A;B)_{\rho_{V(1,r)}}=R(A;B)_{\rho_{V(1,r)}}=g(\sinh^2r).
\end{equation}
An explicit separable reference is obtained by deleting the off-diagonal
terms of the squeezed state's Schmidt expansion. It is
\begin{equation}\label{eq:Schmidt-dephased-reference}
 \tau_r\coloneqq\sum_{k\ge0}p_k\ket{kk}\bra{kk},
 \qquad p_k\coloneqq\frac{(\tanh^2r)^k}{\cosh^2r}.
\end{equation}
Because the input state is pure, its relative entropy with respect to
$\tau_r$ is $-\sum_k p_k\log_2p_k=g(\sinh^2r)$. This gives the upper bound
in \cref{eq:pure-unrestricted}. The known pure-state distillation rate and
the entanglement upper bounds give the opposite inequality. The separation
proof below needs only the explicit separable reference, not an additional
asymptotic converse argument.

To justify the equality in
\cref{eq:pure-unrestricted} in the bosonic setting, one can instead apply
the selective cutoff lower bound of \cref{thm:cutoff} to finite Schmidt
cutoffs of the squeezed vacuum. Each successful state is a finite-dimensional
pure state, for which REE and Rains entanglement equal the entropy of its
Schmidt distribution. Write $a_K\coloneqq\sum_{k=0}^Kp_k$ and
$p^{(K)}\coloneqq(p_k/a_K)_{k=0}^K$ for its success probability and conditional
Schmidt distribution. Explicitly,
\begin{equation}\label{eq:schmidt-entropy-limit}
 a_K H(p^{(K)})
 =-\sum_{k=0}^Kp_k\log_2p_k+a_K\log_2a_K
 \longrightarrow H((p_k)_{k\ge0}).
\end{equation}
The entropy series converges for the geometric distribution and $a_K\to1$.
The explicit dephased reference supplies the matching upper bound.
The proof of \cref{thm:cutoff} does not use \cref{eq:pure-unrestricted},
so this argument is not circular and needs no infinite-dimensional
continuity assumption.

\begin{samepage}
\begin{theorem}[Strict Gaussian-restriction gap for a faithful state]
\label{thm:gap}
Let $\rho_{AB}$ have covariance $V(101/100,3/10)$. Then it is faithful and
\begin{equation}\label{eq:gap-chain}
 \begin{split}
 R(A;B)_\rho\le E_R(A;B)_\rho
 &<0.45795\ \text{bits},\\
 R_G(A;B)_\rho=E_R^G(A;B)_\rho
 &>0.46998\ \text{bits}.
 \end{split}
\end{equation}
In particular, restricting the reference to Gaussian states strictly
increases both values. Numerically, the
Gaussian-restricted value is approximately $0.509530435$ bits.
\end{theorem}
\end{samepage}
\begin{proof}
Write $V_0\coloneqq V(1,3/10)$. Let the parties apply local displacements with a
joint centered classical Gaussian displacement vector $\xi$ of covariance
$\mathbb E[\xi\xi^{\top}]=V_0/200$. This is an operation using shared randomness
and local unitaries. Because our quantum covariance is twice the ordinary
second central moment, it adds $2\mathbb E[\xi\xi^{\top}]=V_0/100$ and produces
$V(101/100,3/10)$. Relative-entropy data processing applied to the input and
its Schmidt-dephased separable reference gives
\begin{equation}\label{eq:unrestricted-upper}
 R(A;B)_\rho\le E_R(A;B)_\rho
 \le g(\sinh^2(3/10))<0.45795.
\end{equation}
The output reference remains separable, as it is a mixture of locally
displaced product states. Thus this bound does not require identifying the
unrestricted optimizer.

For a rigorous Gaussian lower bound, use the first three positive terms
of \cref{eq:positive-series}. With $x=t^2$, define
\begin{equation}\label{eq:Ppoly}
 P(x)\coloneqq u-2\gent((v-1)/2)-\ln4-\ln x
                       +c_1x+c_2x^2+c_3x^3,
 \qquad 0<x<1.
\end{equation}
Then $F(\sqrt x)\geq  P(x)$, and $P$ is convex. The supporting line at
$x_0=27/40$ gives the uniform bound
\begin{equation}\label{eq:scalar-certified}
 F(\sqrt x)\geq 
 P(x_0)+\min\left\{-x_0P'(x_0),(1-x_0)P'(x_0)\right\}.
\end{equation}
For $v=101/100$ and $r=3/10$, direct interval evaluation of these elementary
logarithms and exponentials gives, after division by $\ln2$,
\begin{equation}\label{eq:intervals}
 \begin{aligned}
 \frac{P(x_0)-x_0P'(x_0)}{\ln2}
     &\in(0.46998453525725,\,0.46998453525727),\qquad P'(x_0)>0,\\
 g(\sinh^2(3/10))
     &\in(0.45794979555220,\,0.45794979555221).
 \end{aligned}
\end{equation}
The accompanying short interval-arithmetic script reproduces these
inclusions from rational input parameters; it does not solve an
optimization. \Cref{eq:scalar-result,eq:scalar-certified} prove the
Gaussian lower bound and complete the strict comparison. Finally, both
ordinary symplectic eigenvalues are $101/100>1$, so the input is faithful.
\end{proof}

\begin{figure}[t]
\centering
\begin{tikzpicture}
\begin{axis}[
 width=.86\textwidth,height=6.5cm,
 xmin=1,xmax=1.05,ymin=.34,ymax=.60,
 xlabel={Thermal parameter $v$ at fixed squeezing $r=0.3$},
 ylabel={Entanglement (bits)},
 xtick={1,1.01,1.02,1.03,1.04,1.05},
 xticklabels={$1$,$1.01$,$1.02$,$1.03$,$1.04$,$1.05$},
 ytick={.35,.4,.45,.5,.55,.6},
 scaled ticks=false, tick label style={font=\small},
 legend style={at={(.98,.98)},anchor=north east,font=\small,
               draw=none,fill=white,cells={anchor=west}},
 axis lines=left,clip=false]
\addplot[boundblue,thick,solid,no marks] coordinates {
(1.00000000,0.567326783905)
(1.00062500,0.561233102105)
(1.00125000,0.556386904493)
(1.00187500,0.552010003886)
(1.00250000,0.547936994292)
(1.00312500,0.544089323097)
(1.00375000,0.540420727070)
(1.00437500,0.536900639672)
(1.00500000,0.533507338257)
(1.00562500,0.530224582752)
(1.00625000,0.527039769335)
(1.00687500,0.523942830725)
(1.00750000,0.520925539859)
(1.00812500,0.517981047348)
(1.00875000,0.515103562092)
(1.00937500,0.512288123623)
(1.01000000,0.509530435416)
(1.01062500,0.506826740026)
(1.01125000,0.504173723711)
(1.01187500,0.501568442302)
(1.01250000,0.499008262726)
(1.01312500,0.496490816254)
(1.01375000,0.494013960664)
(1.01437500,0.491575749308)
(1.01500000,0.489174405558)
(1.01562500,0.486808301524)
(1.01625000,0.484475940180)
(1.01687500,0.482175940244)
(1.01750000,0.479907023302)
(1.01812500,0.477668002770)
(1.01875000,0.475457774383)
(1.01937500,0.473275307955)
(1.02000000,0.471119640199)
(1.02062500,0.468989868448)
(1.02125000,0.466885145139)
(1.02187500,0.464804672940)
(1.02250000,0.462747700447)
(1.02312500,0.460713518344)
(1.02375000,0.458701455988)
(1.02437500,0.456710878356)
(1.02500000,0.454741183296)
(1.02562500,0.452791799060)
(1.02625000,0.450862182076)
(1.02687500,0.448951814932)
(1.02750000,0.447060204549)
(1.02812500,0.445186880519)
(1.02875000,0.443331393588)
(1.02937500,0.441493314279)
(1.03000000,0.439672231624)
(1.03062500,0.437867752004)
(1.03125000,0.436079498087)
(1.03187500,0.434307107850)
(1.03250000,0.432550233673)
(1.03312500,0.430808541506)
(1.03375000,0.429081710104)
(1.03437500,0.427369430308)
(1.03500000,0.425671404388)
(1.03562500,0.423987345428)
(1.03625000,0.422316976755)
(1.03687500,0.420660031408)
(1.03750000,0.419016251641)
(1.03812500,0.417385388464)
(1.03875000,0.415767201210)
(1.03937500,0.414161457128)
(1.04000000,0.412567931008)
(1.04062500,0.410986404828)
(1.04125000,0.409416667414)
(1.04187500,0.407858514135)
(1.04250000,0.406311746604)
(1.04312500,0.404776172406)
(1.04375000,0.403251604832)
(1.04437500,0.401737862636)
(1.04500000,0.400234769805)
(1.04562500,0.398742155335)
(1.04625000,0.397259853032)
(1.04687500,0.395787701306)
(1.04750000,0.394325542994)
(1.04812500,0.392873225181)
(1.04875000,0.391430599036)
(1.04937500,0.389997519648)
(1.05000000,0.388573845885)
};
\addlegendentry{$E_R^G=R_G$ (scalar formula)}
\addplot[mutedgray,thick,dashed,no marks] coordinates {(1,0.457949795552) (1.05,0.457949795552)};
\addlegendentry{Upper bound on $E_R$ and $R$}
\addplot[rateorange,only marks,mark=*,mark size=2.2pt] coordinates {(1.01,.46998453525726)};
\addlegendentry{Certified Gaussian lower bound}
\draw[densely dotted] (axis cs:1.01,.35)--(axis cs:1.01,.509530435416);
\end{axis}
\end{tikzpicture}
\caption{A faithful separation at $v=1.01$ within the family $V(v,0.3)$.
The solid curve evaluates the Gaussian optimum from \cref{thm:scalar}.
The dashed line is the entropy of the pure input $V(1,0.3)$ and therefore
an \emph{upper bound}, not an exact curve, for both unrestricted measures
at every $v\geq 1$. The marked point is the analytic lower bound from
\cref{eq:scalar-certified}; it lies above the dashed line. The endpoint
$v=1$ is pure, whereas every $v>1$ is faithful.}
\label{fig:gap}
\end{figure}
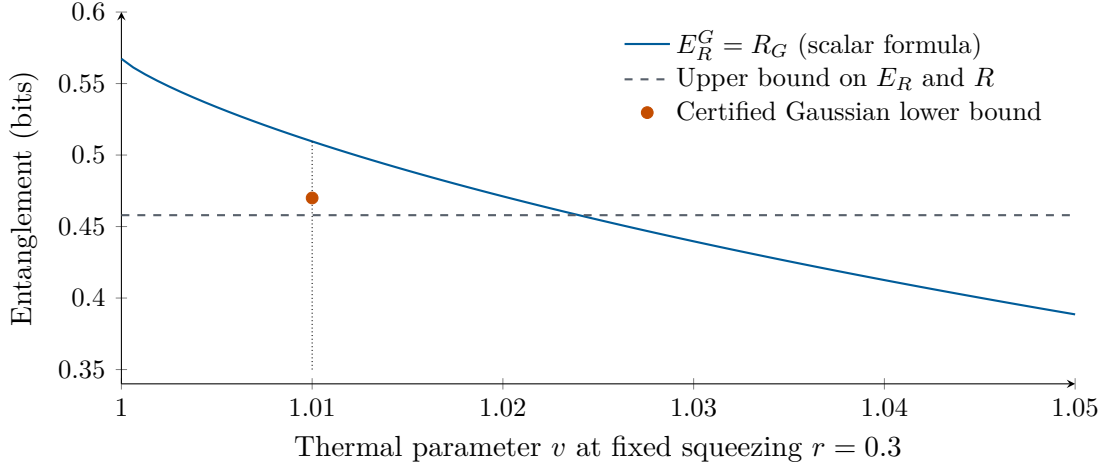


\Cref{fig:gap} compares the scalar Gaussian optimum with the rigorous
unrestricted upper bound. The counterexample rules out a general
Gaussian-optimality statement for
the forward REE and for the Gaussian-reference Rains optimization. It does
not determine the unrestricted values of this mixed state: the first line
of \cref{eq:gap-chain} is an upper bound. It also does not contradict
Gaussian optimality for reverse relative entropy of entanglement
\cite{RippchenEtAl2026}. The position of the variable reference in relative
entropy is essential.

\subsection{Achievable benchmarks: hashing, noisy measurements, and heralding}
To determine how informative a converse is, we must compare it with
rates achieved by actual protocols. The upper bounds should be compared
with operationally attainable rates,
not merely with a lower bound on an entanglement measure. We use the following established protocol families.
For entropy expressions, $S(A)_\rho\coloneqq S(\rho_A)$ and similarly
for joint systems. The one-way hashing inequalities give
\begin{equation}\label{eq:state-hashing}
 L_{\rm hash}(\rho)\coloneqq\max\left\{0,S(A)_\rho-S(AB)_\rho,
                              S(B)_\rho-S(AB)_\rho\right\}
 \le D_{\leftrightarrow}(\rho).
\end{equation}
The finite-dimensional theorem is due to Devetak and Winter
\cite{DevetakWinter2005}; its Gaussian use and reverse-coherent-information
interpretation are discussed in \cite{GarciaPatronEtAl2009}.
Local photon cutoffs followed by the finite-dimensional protocol justify
\cref{eq:state-hashing} here. The cutoffs converge with their mean energies;
entropy convergence follows by comparison with a fixed product thermal
state, whose logarithm is affine in photon number, together with lower
semicontinuity. Fix the cutoff before taking the coding blocklength limit.
The same reasoning justifies finite-energy channel uses below.

Secret-key distillation also permits a trusted noisy measurement at one
party while the other retains quantum memory. Let $X$ denote the measured
party and $Y$ the other party, and purify $\rho_{XY}$ to $XYE$. A Gaussian
measurement with seed covariance $M_X+i\Omega_X\succeq0$ has conditional
covariances
\begin{equation}\label{eq:Gaussian-conditional}
 V_{Y|x}=V_Y-C_{YX}(V_X+M_X)^{-1}C_{YX}^{\top},\qquad
 V_{E|x}=V_E-C_{EX}(V_X+M_X)^{-1}C_{EX}^{\top}.
\end{equation}
Here $C_{YX}$ and $C_{EX}$ are the corresponding off-diagonal
covariance blocks of a Gaussian purification, and $x$ denotes the
classical measurement outcome. The conditional states' covariance
matrices do not depend on $x$; only their means do. Thus their entropies
are outcome independent as well. Devetak--Winter coding therefore achieves
\begin{equation}\label{eq:CQ-key}
 L_{X\to Y}(M_X)\coloneqq S(Y)-S(XY)+S(E|x)-S(Y|x),
\end{equation}
when this quantity is positive. Here $S(E|x)$ and $S(Y|x)$ mean the entropies of the conditional
Gaussian states, not a difference of quantum joint entropies. The rate is
the difference $\chi(x;Y)-\chi(x;E)$ of Holevo informations, where
$\chi(x;Y)\coloneqq S(\rho_Y)-\int p(dx)S(\rho_Y^x)$, and similarly
for Eve. Accordingly, the receiver is not required to measure each copy
separately. Trusted-noise improvements and optimization of Gaussian
measurements have precedents in \cite{PirandolaEtAl2009Key,OttavianiEtAl2016,OrtolanoEtAl2025}.
For a pure seed the last two entropies coincide. A mixed seed may improve
the key rate. Its purification stays in the trusted laboratory, not with
Eve; replacing $S(E|x)$ by $S(Y|x)$ for a mixed seed would incorrectly erase
this improvement. Our numerical purification of $XY$ includes all of Eve's
modes, and \cref{eq:Gaussian-conditional} is used on that purification.

Non-Gaussian local projections can yield positive entanglement rates when
hashing the original Gaussian state does not. If a local filter succeeds
with probability $p$ and produces a two-qubit state $\omega$, then
\begin{equation}\label{eq:heralding-rate}
 D_{\leftrightarrow}(\rho)\ge
 p\max\left\{L_{\rm hash}(\omega),L_{\rm rec}(\omega)\right\}.
\end{equation}
Here $L_{\rm rec}$ consists of a Bell twirl, finitely many successful
recurrence rounds, and final hashing, with the probability and two-copy
cost of every recurrence round included. We use the recurrence protocol
of Deutsch et al.\ \cite{DeutschEtAl1996} and a finite search over local
Bell-label permutations. Fock-state distribution and recurrence protocols
are central to the improved bosonic lower bounds of
\cite{MeleLamiGiovannetti2025,BarberPirandola2026}. Details of the exact Fock
coefficients, searched filters, and recurrence yields are in
\cref{app:operational-numerics}.

For a generic correlated multimode state there is no established closed
formula for the best LOCC lower bound. In the plots, ``achievable lower
bound'' means the maximum of the explicitly evaluated protocol families
above. It is not an assertion of optimization over all Gaussian
measurements, all non-Gaussian filters, or all published recurrence searches.
The pure-loss benchmarks below are different: their operational value is
known exactly, so their lower curve is optimal.

\subsection{Two- and three-mode comparisons}
The examples below separate three pieces of information: a Gaussian
reference upper bound, an achievable protocol rate, and an exact
operational value when one is known. The three-mode example is bipartite,
with two modes in one laboratory; it is not a claim about genuine
tripartite entanglement. Keeping the state partition and the rate
normalization explicit is important when comparing the plots. The plotted
Gaussian upper values are $D(\rho_V\Vert\sigma_{\widehat W})$ for stored
feasible separable reference covariance matrices $\widehat W$. Therefore
$R_G=E_R^G\le D(\rho_V\Vert\sigma_{\widehat W})$ for the families below.
The reference optimization is globally characterized by the earlier theorems,
but the reported floating-point values are not certified global optima.

\begin{figure}[t]
\centering
\begin{tikzpicture}\begin{axis}[
width=.90\textwidth,height=7cm,xmin=.12,xmax=.92,ymin=0,
xlabel={Thermal-loss transmissivity $\eta$ ($E=1$, $n_{\rm th}=0.05$)},
ylabel={Bits per two-mode source copy},axis lines=left,
legend style={at={(.5,-.24)},anchor=north,draw=none,font=\small,legend columns=2,cells={anchor=west}},
tick label style={font=\small}]
\addplot[boundblue,thick,dashed,no marks] coordinates {
(0.12,0.0330524126368)
(0.153333333333,0.0597645644638)
(0.186666666667,0.0903999996044)
(0.22,0.12415153103)
(0.253333333333,0.160587351899)
(0.286666666667,0.199472787108)
(0.32,0.240689723843)
(0.353333333333,0.284196384604)
(0.386666666667,0.330006045921)
(0.42,0.378175643518)
(0.453333333333,0.428800081841)
(0.486666666667,0.482010226056)
(0.52,0.537973620865)
(0.553333333333,0.596897581831)
(0.586666666667,0.659034718085)
(0.62,0.724691318097)
(0.653333333333,0.794239474231)
(0.686666666667,0.868134469325)
(0.72,0.946940018985)
(0.753333333333,1.0313658905)
(0.786666666667,1.12232615978)
(0.82,1.2210342434)
(0.853333333333,1.32916916232)
(0.886666666667,1.44919606429)
(0.92,1.58507903037)
};
\addlegendentry{Gaussian-reference upper bound}
\addplot[accentpurple,very thick,no marks] coordinates {
(0.12,0.0271043252929)
(0.153333333333,0.0501553173191)
(0.186666666667,0.077201319193)
(0.22,0.107506867998)
(0.253333333333,0.140657065028)
(0.286666666667,0.176414942606)
(0.32,0.214653357771)
(0.353333333333,0.255319533782)
(0.386666666667,0.298415673653)
(0.42,0.343988301116)
(0.453333333333,0.392122816692)
(0.486666666667,0.442941526778)
(0.52,0.496604312487)
(0.553333333333,0.5533116389)
(0.586666666667,0.613309988335)
(0.62,0.676900160167)
(0.653333333333,0.744449317163)
(0.686666666667,0.816408302843)
(0.72,0.893336823374)
(0.753333333333,0.975941014156)
(0.786666666667,1.06513165156)
(0.82,1.16211914607)
(0.853333333333,1.26857977224)
(0.886666666667,1.38697616207)
(0.92,1.52127008626)
};
\addlegendentry{Preparation-channel upper bound}
\addplot[rateorange,thick,dash dot,no marks] coordinates {
(0.12,0)
(0.153333333333,5.47894540647e-06)
(0.186666666667,8.79515555779e-05)
(0.22,0.000379656429441)
(0.253333333333,0.0166704127089)
(0.286666666667,0.0625014469937)
(0.32,0.109886323869)
(0.353333333333,0.158938217121)
(0.386666666667,0.209783657854)
(0.42,0.262564815486)
(0.453333333333,0.317442308716)
(0.486666666667,0.374598707606)
(0.52,0.434242950301)
(0.553333333333,0.496615990506)
(0.586666666667,0.561998132551)
(0.62,0.630718730834)
(0.653333333333,0.703169285183)
(0.686666666667,0.77982155728)
(0.72,0.861253370072)
(0.753333333333,0.948186657822)
(0.786666666667,1.04154606082)
(0.82,1.14255422511)
(0.853333333333,1.25289828043)
(0.886666666667,1.37505053568)
(0.92,1.51298142653)
};
\addlegendentry{Achievable entanglement}
\addplot[keygreen,thick,densely dotted,no marks] coordinates {
(0.12,0)
(0.153333333333,5.47894540647e-06)
(0.186666666667,8.79515555779e-05)
(0.22,0.00177898618176)
(0.253333333333,0.0266371718578)
(0.286666666667,0.0656811169536)
(0.32,0.110799164436)
(0.353333333333,0.159164302131)
(0.386666666667,0.209829947128)
(0.42,0.262572293656)
(0.453333333333,0.317443211569)
(0.486666666667,0.374598783771)
(0.52,0.434242954418)
(0.553333333333,0.496615990633)
(0.586666666667,0.561998132551)
(0.62,0.630718730834)
(0.653333333333,0.703169285183)
(0.686666666667,0.77982155728)
(0.72,0.861253370072)
(0.753333333333,0.948186657822)
(0.786666666667,1.04154606082)
(0.82,1.14255422511)
(0.853333333333,1.25289828043)
(0.886666666667,1.37505053568)
(0.92,1.51298142653)
};
\addlegendentry{Achievable secret key}
\end{axis}\end{tikzpicture}
\caption{Thermal-loss variant of the two-mode preparation, with environment
mean photon number $0.05$ and source energy $E=1$. The dashed Gaussian
curve evaluates stored feasible reference states; Gaussian Rains
entanglement and Gaussian REE coincide for this partition. The solid
preparation-channel converse is the scalar sector bound, not a
single-copy Gaussian entanglement value. Lower curves take the maximum
of the evaluated hashing, trusted-measurement, and heralded two-qubit
protocols. No exact mixed-state distillation rate is asserted.}
\label{fig:operational-two}\end{figure}
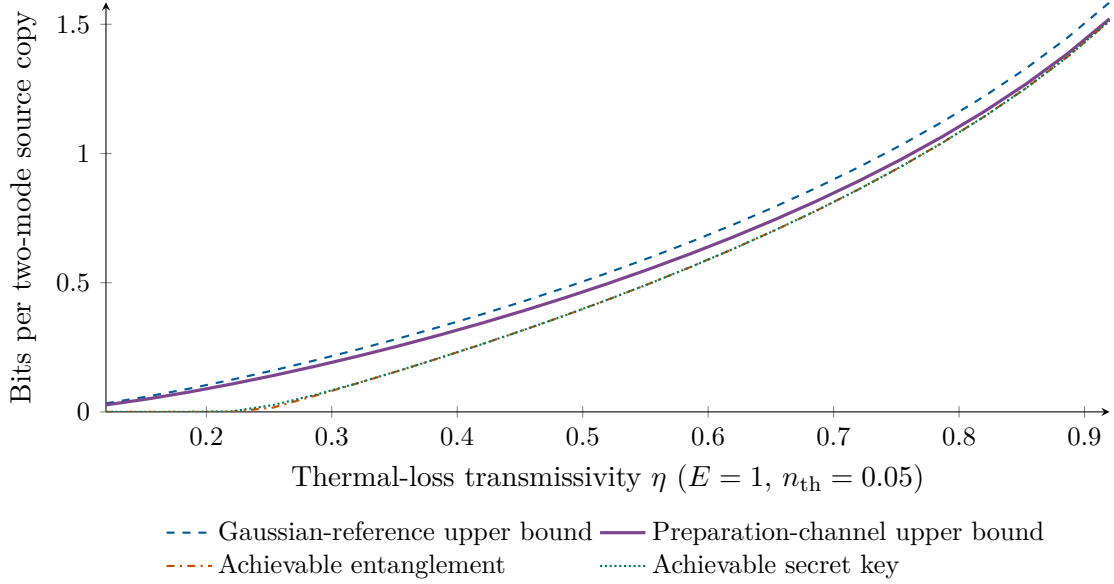

\paragraph{Two modes: an exactly known distillation rate.}
Let one share of a two-mode squeezed vacuum of local mean photon number
$E>0$ pass through pure loss of transmissivity $\eta$. The resulting covariance matrix is
\begin{equation}\label{eq:loss-state}
 V_{\eta,E}\coloneqq\begin{pmatrix}
 (2E+1)I_2&2\sqrt{\eta E(E+1)}Z\\
 2\sqrt{\eta E(E+1)}Z&(2\eta E+1)I_2
 \end{pmatrix},\qquad Z\coloneqq\diag(1,-1).
\end{equation}
Reverse hashing achieves $c_\eta(E)\coloneqq g(E)-g((1-\eta)E)$. The recent pure-loss
capacity converse \cite{PirandolaSector2026,FalcoEtAl2026} bounds the key
obtainable from this state by the same quantity: prepare this state with
one channel use per copy and then apply its state-distillation protocol.
Consequently,
\begin{equation}\label{eq:loss-state-exact}
 D_{\leftrightarrow}(V_{\eta,E})=K_D(V_{\eta,E})=c_\eta(E).
\end{equation}
This equality concerns operational rates, not the single-copy REE or
Rains value. For this $1:1$ partition, Gaussian Rains entanglement and
Gaussian REE coincide by \cref{cor:one-many}; switching between them
cannot lower an already optimized Gaussian-reference bound.
To compare genuinely different converses away from this exactly solved
limit, \cref{fig:operational-two} instead uses a thermal environment of
mean occupation $n_{\rm th}=0.05$, with $E=1$. Its covariance is obtained
from \eqref{eq:loss-state} by replacing Bob's diagonal block with
$[2\eta E+1+2(1-\eta)n_{\rm th}]I_2$.
The preparation-channel converse is
$K_D(\rho)\le\mathcal F_{\eta,(1-\eta)n_{\rm th}}(1)$, where
\eqref{eq:sector-F} recalls the established scalar formula.
This can improve the operational bound without contradicting equality
of the two Gaussian-restricted entanglement quantities. The plotted
lower rates are the explicitly evaluated hashing, trusted-measurement,
and heralded protocols, not a claim of optimal noisy-state distillation.

\paragraph{Three modes with unequal losses and correlated noise.}
A less reducible example starts with a squeezed pair of local mean one on
$AB_1$ and a thermal mode of mean $0.2$ on $B_2$. Bob mixes $B_1B_2$ on the
passive interferometer
\begin{equation}\label{eq:state-beamsplitter}
 O_\theta\coloneqq\begin{pmatrix}\cos\theta&-\sin\theta\\\sin\theta&\cos\theta\end{pmatrix},
 \qquad\theta=0.55,
\end{equation}
then applies losses of transmissivities $(0.85,0.45)$ and adds classical
Gaussian noise with quantum covariance matrix
\begin{equation}\label{eq:state-noise}
 2q\begin{pmatrix}1&0.35\\0.35&1.4\end{pmatrix}\otimes I_2,
 \qquad 0\le q\le0.22.
\end{equation}
The input squeezing is $r=\operatorname{arcsinh}1$. All these specifications
use the interleaved quadrature order. They define the complete covariance matrix
without a Fock cutoff. Unequal losses follow the interferometer, so undoing
that interferometer does not undo the noisy preparation.

\begin{figure}[t]
\centering
\begin{tikzpicture}
\begin{axis}[
name=main,
width=.90\textwidth,
height=7.0cm,
axis lines=left,
enlarge y limits={upper,value=0.05},
ymin=0,
xlabel={Added correlated-noise parameter $q$},
ylabel={Bits per three-mode source copy},
tick label style={font=\small},
label style={font=\small},
legend style={at={(.5,-.24)},anchor=north,draw=none,font=\small,legend columns=2,cells={anchor=west},/tikz/every even column/.append style={column sep=8pt}},
legend cell align=left,
clip=true,
xmin=0.0,
xmax=0.22,
scaled ticks=false,
xticklabel style={/pgf/number format/fixed,/pgf/number format/precision=2},
xtick={0,.04,.08,.12,.16,.20},
ymax=1.16]
\addplot[boundblue,very thick,no marks] coordinates {
(0,1.09401765467)
(0.00785714285714,0.978375218725)
(0.0157142857143,0.898956130528)
(0.0235714285714,0.833112358533)
(0.0314285714286,0.775978467121)
(0.0392857142857,0.725228425431)
(0.0471428571429,0.679485254543)
(0.055,0.637835570334)
(0.0628571428571,0.59962838541)
(0.0707142857143,0.564376087151)
(0.0785714285714,0.531699866187)
(0.0864285714286,0.50129710553)
(0.0942857142857,0.472920664512)
(0.102142857143,0.446365072726)
(0.11,0.421456966493)
(0.117857142857,0.398048250676)
(0.125714285714,0.376011074847)
(0.133571428571,0.355234058632)
(0.141428571429,0.335619401896)
(0.149285714286,0.317080630071)
(0.157142857143,0.29954080977)
(0.165,0.282931122454)
(0.172857142857,0.267189702938)
(0.180714285714,0.252260690083)
(0.188571428571,0.238093440154)
(0.196428571429,0.224641869778)
(0.204285714286,0.211863905717)
(0.212142857143,0.19972101445)
(0.22,0.188177805508)
};
\addlegendentry{Gaussian-reference upper bound}
\addplot[accentpurple,very thick,no marks] coordinates {
(0,1.06327827842)
(0.00785714285714,0.938538585206)
(0.0157142857143,0.855693138188)
(0.0235714285714,0.78852757745)
(0.0314285714286,0.731234277317)
(0.0392857142857,0.681023852464)
(0.0471428571429,0.636254307655)
(0.055,0.595848782652)
(0.0628571428571,0.559050790687)
(0.0707142857143,0.525302627229)
(0.0785714285714,0.494178028801)
(0.0864285714286,0.465341885668)
(0.0942857142857,0.438524701702)
(0.102142857143,0.413505652114)
(0.11,0.390100925647)
(0.117857142857,0.368155456308)
(0.125714285714,0.347536907374)
(0.133571428571,0.32813119721)
(0.141428571429,0.309839107827)
(0.149285714286,0.292573670843)
(0.157142857143,0.276258122556)
(0.165,0.260824282917)
(0.172857142857,0.246211255127)
(0.180714285714,0.232364371149)
(0.188571428571,0.219234328195)
(0.196428571429,0.206776475268)
(0.204285714286,0.194950218801)
(0.212142857143,0.183718523777)
(0.22,0.17304749206)
};
\addlegendentry{Improved operational upper bound}
\addplot[keygreen,dashed,thick,mark=o,mark repeat=3,mark size=1.4pt] coordinates {
(0,0.741986148431)
(0.00785714285714,0.617708141684)
(0.0157142857143,0.529692326649)
(0.0235714285714,0.455395716683)
(0.0314285714286,0.390077597345)
(0.0392857142857,0.331554867378)
(0.0471428571429,0.278611443625)
(0.055,0.230514390653)
(0.0628571428571,0.186816495299)
(0.0707142857143,0.14726386462)
(0.0785714285714,0.111752145258)
(0.0864285714286,0.0803115313511)
(0.0942857142857,0.0531160876844)
(0.102142857143,0.0305253569152)
(0.11,0.0216421468234)
(0.117857142857,0.0189176510783)
(0.125714285714,0.016358243287)
(0.133571428571,0.0149487616338)
(0.141428571429,0.0136755871715)
(0.149285714286,0.0124691256027)
(0.157142857143,0.0113269610209)
(0.165,0.0102466826716)
(0.172857142857,0.00922589185674)
(0.180714285714,0.0082622086832)
(0.188571428571,0.00735327854096)
(0.196428571429,0.00649677822496)
(0.204285714286,0.0056904216319)
(0.212142857143,0.00493196498278)
(0.22,0.00421921153783)
};
\addlegendentry{Achievable key envelope}
\addplot[rateorange,dashdotted,thick,mark=triangle,mark repeat=3,mark size=1.5pt] coordinates {
(0,0.741986148431)
(0.00785714285714,0.617693623517)
(0.0157142857143,0.529580856062)
(0.0235714285714,0.455009200433)
(0.0314285714286,0.389121078419)
(0.0392857142857,0.329597103093)
(0.0471428571429,0.275065961018)
(0.055,0.224619134111)
(0.0628571428571,0.177609840242)
(0.0707142857143,0.133554120301)
(0.0785714285714,0.0920763535586)
(0.0864285714286,0.0528767168661)
(0.0942857142857,0.0280892371768)
(0.102142857143,0.0247372324732)
(0.11,0.0216421468234)
(0.117857142857,0.0189176510783)
(0.125714285714,0.016358243287)
(0.133571428571,0.0149487616338)
(0.141428571429,0.0136755871715)
(0.149285714286,0.0124691256027)
(0.157142857143,0.0113269610209)
(0.165,0.0102466826716)
(0.172857142857,0.00922589185674)
(0.180714285714,0.0082622086832)
(0.188571428571,0.00735327854096)
(0.196428571429,0.00649677822496)
(0.204285714286,0.0056904216319)
(0.212142857143,0.00493196498278)
(0.22,0.00421921153783)
};
\addlegendentry{Achievable entanglement envelope}
\addplot[mutedgray,densely dotted,thick,no marks] coordinates {
(0,0.741986148431)
(0.00785714285714,0.617693623517)
(0.0157142857143,0.529580856062)
(0.0235714285714,0.455009200433)
(0.0314285714286,0.389121078419)
(0.0392857142857,0.329597103093)
(0.0471428571429,0.275065961018)
(0.055,0.224619134111)
(0.0628571428571,0.177609840242)
(0.0707142857143,0.133554120301)
(0.0785714285714,0.0920763535586)
(0.0864285714286,0.0528767168661)
(0.0942857142857,0.0157105256558)
(0.102142857143,0)
(0.11,0)
(0.117857142857,0)
(0.125714285714,0)
(0.133571428571,0)
(0.141428571429,0)
(0.149285714286,0)
(0.157142857143,0)
(0.165,0)
(0.172857142857,0)
(0.180714285714,0)
(0.188571428571,0)
(0.196428571429,0)
(0.204285714286,0)
(0.212142857143,0)
(0.22,0)
};
\addlegendentry{Hashing before filtering}
\addplot[accentpurple,solid,mark=diamond,mark repeat=3,mark size=1.3pt ,thick] coordinates {
(0,0.310464979299)
(0.00785714285714,0.258732371408)
(0.0157142857143,0.216149083193)
(0.0235714285714,0.178884510482)
(0.0314285714286,0.145558685121)
(0.0392857142857,0.115404787363)
(0.0471428571429,0.0879178083304)
(0.055,0.062732396866)
(0.0628571428571,0.0462765979034)
(0.0707142857143,0.0404905130412)
(0.0785714285714,0.0354642580897)
(0.0864285714286,0.0316610850244)
(0.0942857142857,0.0280892371768)
(0.102142857143,0.0247372324732)
(0.11,0.0216421468234)
(0.117857142857,0.0189176510783)
(0.125714285714,0.016358243287)
(0.133571428571,0.0149487616338)
(0.141428571429,0.0136755871715)
(0.149285714286,0.0124691256027)
(0.157142857143,0.0113269610209)
(0.165,0.0102466826716)
(0.172857142857,0.00922589185674)
(0.180714285714,0.0082622086832)
(0.188571428571,0.00735327854096)
(0.196428571429,0.00649677822496)
(0.204285714286,0.0056904216319)
(0.212142857143,0.00493196498278)
(0.22,0.00421921153783)
};
\addlegendentry{Heralded recurrence}
\end{axis}
\begin{axis}[at={(main.north east)},anchor=north east,xshift=-2mm,yshift=-3mm,
width=4.6cm,height=2.9cm,axis lines=left,xmin=.09,xmax=.20,ymin=0,ymax=.04,
xtick={.1,.15,.2},ytick={0,.02,.04},scaled ticks=false,
xticklabel style={/pgf/number format/fixed,/pgf/number format/precision=2},
yticklabel style={/pgf/number format/fixed,/pgf/number format/precision=2},
tick label style={font=\scriptsize},title={Zoom of lower rates},title style={font=\scriptsize},
xlabel={$q$},label style={font=\scriptsize},axis background/.style={fill=white}]
\addplot[mutedgray,densely dotted,thick,no marks] coordinates {
(0,0.741986148431)
(0.00785714285714,0.617693623517)
(0.0157142857143,0.529580856062)
(0.0235714285714,0.455009200433)
(0.0314285714286,0.389121078419)
(0.0392857142857,0.329597103093)
(0.0471428571429,0.275065961018)
(0.055,0.224619134111)
(0.0628571428571,0.177609840242)
(0.0707142857143,0.133554120301)
(0.0785714285714,0.0920763535586)
(0.0864285714286,0.0528767168661)
(0.0942857142857,0.0157105256558)
(0.102142857143,0)
(0.11,0)
(0.117857142857,0)
(0.125714285714,0)
(0.133571428571,0)
(0.141428571429,0)
(0.149285714286,0)
(0.157142857143,0)
(0.165,0)
(0.172857142857,0)
(0.180714285714,0)
(0.188571428571,0)
(0.196428571429,0)
(0.204285714286,0)
(0.212142857143,0)
(0.22,0)
};
\addplot[accentpurple,solid,thick,no marks] coordinates {
(0,0.310464979299)
(0.00785714285714,0.258732371408)
(0.0157142857143,0.216149083193)
(0.0235714285714,0.178884510482)
(0.0314285714286,0.145558685121)
(0.0392857142857,0.115404787363)
(0.0471428571429,0.0879178083304)
(0.055,0.062732396866)
(0.0628571428571,0.0462765979034)
(0.0707142857143,0.0404905130412)
(0.0785714285714,0.0354642580897)
(0.0864285714286,0.0316610850244)
(0.0942857142857,0.0280892371768)
(0.102142857143,0.0247372324732)
(0.11,0.0216421468234)
(0.117857142857,0.0189176510783)
(0.125714285714,0.016358243287)
(0.133571428571,0.0149487616338)
(0.141428571429,0.0136755871715)
(0.149285714286,0.0124691256027)
(0.157142857143,0.0113269610209)
(0.165,0.0102466826716)
(0.172857142857,0.00922589185674)
(0.180714285714,0.0082622086832)
(0.188571428571,0.00735327854096)
(0.196428571429,0.00649677822496)
(0.204285714286,0.0056904216319)
(0.212142857143,0.00493196498278)
(0.22,0.00421921153783)
};
\end{axis}
\end{tikzpicture}
\caption{The $A:B_1B_2$ state defined by \cref{eq:state-beamsplitter,eq:state-noise}. The purple curve additionally uses the analytic predecessor in \cref{prop:scalar-predecessor}; it is an operational bound, not a Gaussian Rains value. Both of Bob\textquotesingle s modes enter the covariance optimization and Gaussian lower bounds. The two lower envelopes include the explicitly searched non-Gaussian protocol. The inset resolves the heralded recurrence rate (solid) after the unfiltered hashing rate (dotted) has vanished. Envelopes refer only to the evaluated protocol families; no globally optimal LOCC lower-bound claim is made.}
\label{fig:operational-three}
\end{figure}
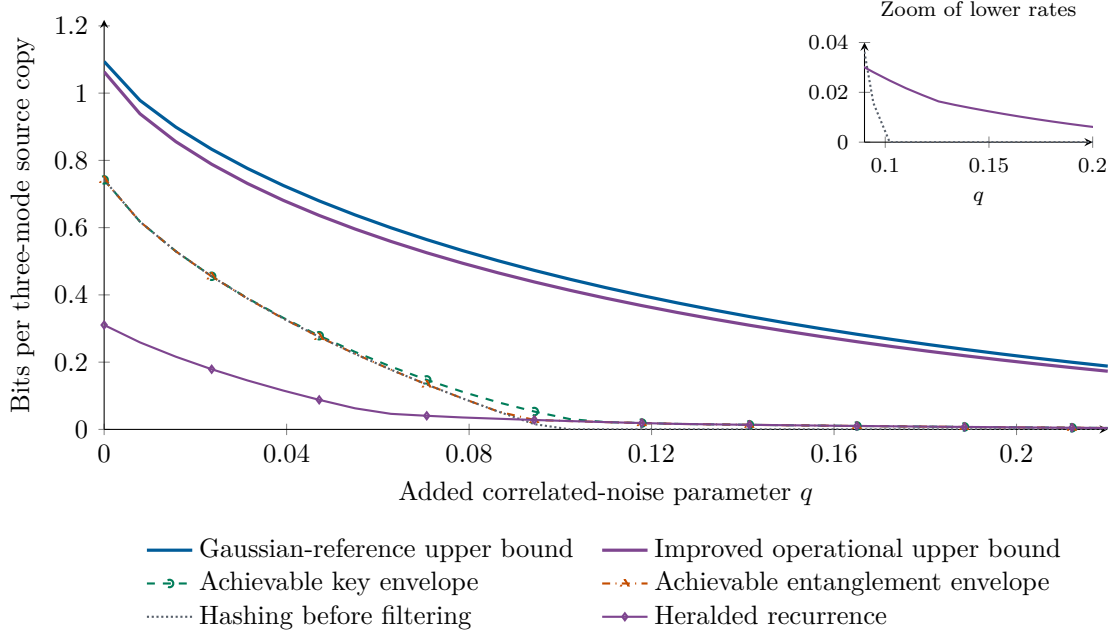

We evaluate the $A:B_1B_2$ entanglement using the full three-mode covariance matrix
and \cref{eq:paired-reduction}. The two Gaussian-restricted quantities coincide, $R_G=E_R^G$; a
computed separable reference supplies an upper bound on that common value.
The lower curves retain both of Bob's modes in hashing and in the noisy
measurement calculation. A further heralded protocol projects Bob onto
vacuum and a searched one-photon superposition of his two modes, and Alice
onto a filtered zero--one-photon subspace. For example, at $q=0.12$ direct
hashing gives zero, whereas this finite recurrence search achieves more
than $0.018$ ebits per input copy. This is a numerical evaluation of an
explicit protocol, not a statement about the optimal distillation rate.
The same state is prepared by a one-input, two-output channel with
\begin{equation}\label{eq:three-preparation}
 \begin{split}
 D_0&\coloneqq\diag(\sqrt{0.85},\sqrt{0.45}),\qquad
 k\coloneqq D_0O_{0.55}(1,0)^{\top},\\
 Q(q)&\coloneqq0.2D_0O_{0.55}(0,1)^{\top}(0,1)O_{0.55}^{\top}D_0
       +q\begin{pmatrix}1&0.35\\0.35&1.4\end{pmatrix}.
 \end{split}
\end{equation}
For $q>0$, \cref{prop:scalar-predecessor} gives the complementary bound
$K_D(\rho(q))\le\mathcal F_{\tau_*(q),v_*(q)}(1)$.
At $q=0$, $Q(0)$ has rank one and $k$ is not in its range.
The rank-one inequalities used in \eqref{eq:predecessor-halfplanes}
then force $b-a=-1$. Setting $\beta^{-1}=0$ in
\eqref{eq:predecessor-parameters} gives the feasible pure-loss
predecessor $\tau_*=\alpha$, $v_*=0$ directly. No continuity
assumption on distillable key is needed at this endpoint.
The new operational curve is the minimum of this preparation bound and
the stored Gaussian-reference bound. It is not labeled as a Gaussian
Rains value.
\Cref{fig:operational-three} shows the upper bounds and the lower-bound
envelopes, including the non-Gaussian contribution.

\FloatBarrier
\section{Energy-constrained communication through Gaussian channels}
\label{sec:channels}
We next convert the state calculations into bounds on adaptive communication.
Two simulation constructions play complementary roles. Photon-sector
simulation preserves the transmitted-energy constraint but uses non-Gaussian
resources. Exact Gaussian-resource teleportation permits the covariance
optimization developed above, but its resource energy is not the transmitted
energy. Combining their bounds, rather than identifying their resources,
gives a converse for every finite-mode Gaussian channel. A converse
is an upper bound on the rate of every allowed communication protocol.
It need not be achieved, and it must keep the sender's energy constraint
separate from the energy used by the mathematical simulator.

\subsection{Channel model, energy constraint, and prior work}
A communication problem specifies a channel, allowed assistance,
and a constraint on transmitted states. We consider
$\mathcal N\colon A\to B$ with $m$ input modes and $n$ output modes. A
centered Gaussian channel is specified by real matrices $X\in\R^{2n\times2m}$
and $Y=Y^{\top}\in\R^{2n\times2n}$, with
\begin{equation}\label{eq:channel-CP}
 V\longmapsto XVX^{\top}+Y,\qquad
 \mathsf H_{X,Y}\coloneqq Y+i(\Omega_n-X\Omega_mX^{\top})\succeq0.
\end{equation}
The inequality in \cref{eq:channel-CP} is the condition for complete
positivity (CP), meaning positivity is preserved even when the channel
acts on part of a larger entangled state. A fixed output displacement is irrelevant to the capacities considered here
and is omitted. We use the standard multimode Gaussian-channel
classification and composition rule \cite{CarusoEtAl2008}.
No square-matrix, invertibility, phase-insensitivity, or independent-noise
assumption is imposed on $X,Y$ unless explicitly stated.

One channel use means one simultaneous use of the entire $m$-input-mode
channel. Alice and Bob start without shared entanglement or secret key and
interleave $\ell$ uses with LOCC. Local memories have no dimension, entropy,
or energy constraint. Each finite code has finitely many local rounds and
finite or countable classical outcome alphabets, with normal states and
instruments on separable Hilbert spaces. Eve holds a channel environment,
its purification, and the whole public transcript. Let $e_{ij}$ be the
unconditional mean photon number in input mode $j$ at use $i$. We impose
\begin{equation}\label{eq:multimode-energy}
 \sum_{i=1}^{\ell}\sum_{j=1}^m e_{ij}\le\ell E,
 \qquad e_{ij}\coloneqq\Tr[\hat n_j\rho_A^{(i)}],\qquad 0\le E<\infty.
\end{equation}
All classical outcomes and failed attempts are included in these means.
No constraint is placed on the simulator's energy or the receiver's output
energy. A bound per physical mode is obtained only after the displayed
rates per $m$-mode use are divided by $m$.

Write $Q_2(\mathcal N,E)$ and $D_2(\mathcal N,E)$ for the vanishing-error
quantum-transmission and entanglement-distribution capacities, and
$P_2(\mathcal N,E)$ and $K_2(\mathcal N,E)$ for private transmission and key
agreement. The errors are the entanglement infidelity for a quantum target
and the joint correctness/secrecy normalized trace distance for a key, as
in \cref{sec:operational-states}. Teleportation and local preparation give
$Q_2=D_2$; a uniformly random private message and a one-time pad over the
public channel give $P_2=K_2$. These conversions preserve every optical input.
Thus
\begin{equation}\label{eq:capacity-orders}
 Q_2=D_2\le P_2=K_2.
\end{equation}
The energy-constrained framework and important earlier upper bounds are
studied in \cite{DavisShirokovWilde2018,GoodenoughEtAl2016}.
Our results concern these vanishing-error capacities. They do not assert an
energy-dependent strong converse under the unconditional mean constraint.

Teleportation simulation bounds LOCC-assisted communication by the
entanglement required for the shared simulation resources
\cite{BennettEtAl1996,MuellerHermes2012,WolfEtAl2007,NisetEtAl2009,WildeTomamichelBerta2017,Wilde2018Simulation}.
The same substitution includes every step of an adaptive protocol in a
final LOCC operation on the resource states.
Finite-resource Gaussian simulations were developed in
\cite{LiuzzoScorpoEtAl2017,Wilde2018Simulation}; a finite resource must reproduce
the channel exactly, or a simulation-error analysis is needed. Squashed
entanglement gives a different energy-dependent converse
\cite{TakeokaGuhaWilde2013,DavisShirokovWilde2018}. More recently, the
sector-teleportation results of Pirandola
\cite{PirandolaSector2026,PirandolaNoisy2026} and Falco et al.
\cite{FalcoEtAl2026} retain the actual input energy in the entanglement
gain. We use those scalar-channel results as imported theorems, with their
specified operational assumptions, rather than claiming their spectral
resource calculations as contributions of the present paper. The earlier works already use multimode photon sectors and arbitrary
correlations among identical channel uses. In particular,
\cite[Appendix~B]{PirandolaSector2026} treats sectors of
$\mathcal L_\eta^{\otimes s}$, and
\cite[Appendix~B.1]{PirandolaNoisy2026} proves additivity for products of
conditional sector resources. We do not claim these constructions as new.
Our mode-resolved allocation theorem is a consequence of their scalar
gain estimates for distinct physical mode parameters; the pure-loss
corollary is presented as the corresponding extension of the scalar
capacity formula, not as the introduction of multimode sectors.
The additional matrix results are the general receiver-postprocessing
criterion, the jointly optimized passive-basis loss cores, the analytic
scalar predecessor, and their combination with optimized Gaussian
resource entanglement.

\subsection{From single-mode photon sectors to multimode energy allocation}
The multimode construction starts from an imported one-mode
entanglement-gain inequality. We state it first, then explain how different
mode types can be combined. Here ``scalar channel'' means a single-mode
phase-insensitive Gaussian channel with scalar gain and noise parameters. The vacuum-noise normalization is
important when composing these channels.
Let $\mathcal G_{\tau,v}$ act on one mode as
\begin{equation}\label{eq:scalar-channel}
 V\longmapsto\tau V+(2v+1-\tau)I_2,
 \qquad \tau>0,\quad v\ge\max\left\{0,\tau-1\right\}.
\end{equation}
The number $v$ is its \emph{output} mean photon number on vacuum input, not
the mean photon number of a thermal environment. Pure loss is $(\tau,v)=(\eta,0)$;
a quantum-limited amplifier is $(\kappa,\kappa-1)$.
For $x\ge0$, define $\varphi(x)\coloneqq x\log_2x$, with $\varphi(0)=0$. Given energy
$e>0$, set
\begin{equation}\label{eq:sector-parameters}
 y\coloneqq\tau e+v,\quad L\coloneqq1+e+y,\quad
 t\coloneqq\frac{L-\sqrt{L^2-4\tau e(e+1)}}2,\quad u\coloneqq e-t,\quad w\coloneqq y-t.
\end{equation}
In the non-entanglement-breaking region $v<\tau$, the scalar sector function is
\begin{equation}\label{eq:sector-F}
 \mathcal F_{\tau,v}(e)\coloneqq
 \varphi(t)+\varphi(L-t)-\varphi(e)-\varphi(y)-g(u)-g(w).
\end{equation}
Set $\mathcal F_{\tau,v}(0)\coloneqq0$ and $\mathcal F_{\tau,v}\coloneqq0$ when $v\ge\tau$.
The quantum-limited boundary values are
\begin{equation}\label{eq:sector-boundaries}
 \begin{split}
 \mathcal F_{\eta,0}(e)&=c_\eta(e)\coloneqq g(e)-g((1-\eta)e),\\
 \mathcal F_{\kappa,\kappa-1}(e)&=
 g(\kappa e+\kappa-1)-g((\kappa-1)(e+1)).
 \end{split}
\end{equation}
The identity-channel boundary is $\mathcal F_{1,0}(e)=g(e)$.

We use the following established scalar gain theorem in a form that makes
its hypotheses visible. For an arbitrary correlated input $\rho_{X_0AY_0}$
of finite $E_R(X_0A;Y_0)$ and finite transmitted mean $e$, let
$\omega=(\id_{X_0Y_0}\otimes\mathcal G_{\tau,v})(\rho)$. Then
\begin{equation}\label{eq:imported-gain}
 E_R^\infty(X_0;BY_0)_\omega
 \le E_R^\infty(X_0A;Y_0)_\rho+\mathcal F_{\tau,v}(e).
\end{equation}
The output also has finite unregularized REE. Moreover $\mathcal F$ is
nonnegative, nondecreasing, concave, and at most $g(e)$.
For positive excess noise these statements are
\cite[Eq.~(18), Appendix~C, Lemmas~4--5]{PirandolaNoisy2026};
the boundary cases follow from
\cite[Appendices~B and~E]{PirandolaSector2026} and, for pure loss including
the identity endpoint, \cite[Lemma~4.1 and Corollary~4.2]{FalcoEtAl2026}.
The formula \cref{eq:sector-F} is Eq.~(9) of \cite{PirandolaNoisy2026}.
An entanglement-breaking channel is an LOCC channel from Alice to Bob
(measure, announce, prepare), and has zero gain. This covers the remaining
parameter region.

\paragraph{Separating the mode types.}
We now explain the finite photon-number spaces behind the
extension before using the gain inequality in an adaptive protocol.
Suppose different physical modes have different scalar parameters. The
replicas of a \emph{fixed} mode can still be mixed passively, but modes with
different parameters cannot in general be mixed as one irreducible sector.
This explains why a vector of photon totals is needed.

For an auxiliary replica number $s$ and vector
$\boldsymbol k=(k_1,\ldots,k_m)$, define
\begin{equation}\label{eq:multisector-space}
 \mathcal H_{s,\boldsymbol k}
 \coloneqq\bigotimes_{j=1}^m\operatorname{Sym}^{k_j}(\mathbb C^s),\qquad
 d_{s,k_j}\coloneqq\binom{s+k_j-1}{k_j}.
\end{equation}
The $s$ modes in factor $j$ are copies of physical input mode $j$ across
replicas, not consecutive adaptive uses within one run. The notation $\operatorname{Sym}^{k_j}(\mathbb C^s)$ denotes the
$k_j$-boson subspace of $s$ modes: its basis consists of occupation vectors
$(n_1,\ldots,n_s)$ with $\sum_a n_a=k_j$. Counting these vectors gives
$d_{s,k_j}$ in \cref{eq:multisector-space}. Passive interferometers mixing
the $s$ replicas independently for each mode type form the group $U(s)^m$.
Each factor carries an irreducible representation of its $U(s)$, meaning
that no proper nonzero subspace is invariant under every such
interferometer, and
the corresponding resource is
\begin{equation}\label{eq:multisector-resource}
 J_{s,\boldsymbol k}
 \coloneqq\bigotimes_{j=1}^m
 (\id\otimes\mathcal G_{\tau_j,v_j}^{\otimes s})
       (\Phi^{d_{s,k_j}}).
\end{equation}
Applying the sector simulator for each factor gives an exact LOCC simulation
on $\mathcal H_{s,\boldsymbol k}$, including arbitrary retained memories and
correlations among the factors. This tensorization organizes the established sector simulations
by physical mode type; it does not introduce sector teleportation anew. Irreducible covariance is the underlying
finite-dimensional channel-programming principle
\cite{GschwendtnerEtAl2021,PirandolaSector2026}.
One can use a finite weighted Bell measurement: Schur orthogonality gives
the required identity resolution for the orbit of a maximally entangled
vector, and a finite convex decomposition of that resolution gives finitely
many effects. The covariance corrections are the passive output unitaries.

The resources in \cref{eq:multisector-resource} are not Gaussian. In
particular, their covariance matrices do not determine their relative
entropy of entanglement. Coherences between different photon totals are
controlled in the scalar proof before the replica limit is taken; simply
measuring and discarding these coherences would change the channel.
The regularized gain theorem \cref{eq:imported-gain} is a convenient way to
retain that control in the adaptive multimode argument.

\begin{samepage}
\begin{theorem}[Mode-resolved sector converse]\label{thm:multisector}
Let $\mathcal N=\bigotimes_{j=1}^m\mathcal G_{\tau_j,v_j}$ be a
product of the single-mode channels in \eqref{eq:scalar-channel}, with
$\tau_j>0$ and $v_j\ge\max\left\{0,\tau_j-1\right\}$. Allow arbitrary adaptive
LOCC-assisted protocols satisfying the unconditional mean-energy
constraint \eqref{eq:multimode-energy}. For $E\ge0$, define
\begin{equation}\label{eq:allocation-B}
 B_{\boldsymbol\tau,\boldsymbol v}(E)
 \coloneqq\max\left\{\sum_{j=1}^m\mathcal F_{\tau_j,v_j}(e_j):
                 e_j\ge0,\ \sum_{j=1}^m e_j\le E\right\}.
\end{equation}
Then
\begin{equation}\label{eq:multisector-bound}
 Q_2\!\left(\bigotimes_j\mathcal G_{\tau_j,v_j},E\right)
 \le K_2\!\left(\bigotimes_j\mathcal G_{\tau_j,v_j},E\right)
 \le B_{\boldsymbol\tau,\boldsymbol v}(E).
\end{equation}
The maximum is attained. The function $B$ is nondecreasing and concave in
$E$, and is at most $m g(E/m)$.
\end{theorem}
\end{samepage}
\begin{proof}
Apply the scalar channels within a fixed use one after another, keeping all
other input modes in Alice's memory and all already received modes in Bob's
memory. The channels act on disjoint inputs, so this ordering does not
change the overall channel. The marginal of an unsent mode is unchanged
by a trace-preserving operation on the other modes. Its energy is therefore
the actual $e_{ij}$ in \cref{eq:multimode-energy}, even for correlated inputs.
Applying \cref{eq:imported-gain} successively gives a gain of at most
$\sum_j\mathcal F_{\tau_j,v_j}(e_{ij})$. The initial REE is zero, intermediate
LOCC cannot increase it, and the scalar unregularized bound guarantees its
finiteness at every step. After all uses, the retained output obeys
\begin{equation}\label{eq:multi-telescope}
 E_R^\infty(\zeta_\ell)
 \le\sum_{i=1}^{\ell}\sum_{j=1}^m\mathcal F_{\tau_j,v_j}(e_{ij})
 \le\ell\sum_{j=1}^m\mathcal F_{\tau_j,v_j}(\bar e_j)
 \le\ell B_{\boldsymbol\tau,\boldsymbol v}(E),
 \qquad \bar e_j\coloneqq\frac1\ell\sum_i e_{ij}.
\end{equation}
The middle inequality is Jensen's inequality applied separately
to each mode type. It does not assume that the transmitted states at
different uses are independent or identically distributed. The averages
$\bar e_j$ satisfy $\sum_j\bar e_j\le E$ by the input constraint.
For a quantum or private-state target, the regularized testing bound in
\cref{lem:regularized-test} below implies
$p_\ell\log_2d_\ell-h_2(p_\ell)\le E_R^\infty(\zeta_\ell)$, with
$p_\ell\to1$. For an operational key code, retain the local shields using
the input-preserving private-state reduction
\cite[Appendix~D.3]{PirandolaSector2026}; it is a statement about purifying
local operations and does not depend on the optical channel being pure loss.
It leaves all marginals in \cref{eq:multimode-energy} unchanged. Divide by
$\ell$ and let the error vanish to obtain the converse.

The feasible energy simplex is compact and $\mathcal F$ is continuous, so
the maximum is attained. Increasing $E$ enlarges that simplex. Mixing two
feasible allocations and using concavity of each summand proves concavity
of $B$. Finally $\sum_j\mathcal F_j(e_j)\le\sum_jg(e_j)\le mg(E/m)$.
\end{proof}

\begin{samepage}
\begin{corollary}[Pure-loss energy allocation from the scalar capacity theorem]\label{cor:loss-capacity}
For $\boldsymbol\eta\in(0,1]^m$,
\begin{equation}\label{eq:exact-multiloss}
 Q_2\!\left(\bigotimes_j\mathcal L_{\eta_j},E\right)
 =K_2\!\left(\bigotimes_j\mathcal L_{\eta_j},E\right)
 =\max\left\{\sum_j c_{\eta_j}(e_j):e_j\ge0,\ \sum_j e_j\le E\right\}.
\end{equation}
The same identity holds after a passive Gaussian unitary before the product
channel and an arbitrary Gaussian unitary after it.
\end{corollary}
\end{samepage}
\begin{proof}
The converse is \cref{thm:multisector}. Independently send one share of a
squeezed pair of local mean $e_j$ through each mode, and reverse-hash the
resulting states. This achieves $\sum_jc_{\eta_j}(e_j)$ by
\cref{eq:loss-state-exact}, or directly by hashing. Independent distillation
errors vanish jointly because there are finitely many modes. Optimize the
allocation. Input passive unitaries preserve total photon number and can
be inverted by Alice; Bob can invert the output unitary without any energy
constraint on his apparatus. Thus neither changes the capacities.
\end{proof}

For $E>0$ and positive transmissivities, the pure-loss allocation is unique.
Indeed,
\begin{equation}\label{eq:c-second}
 c_\eta''(e)=-\frac{\eta}{(\ln2)e(e+1)((1-\eta)e+1)}<0\qquad(e>0).
\end{equation}
Its derivative diverges at $e\downarrow0$, so every mode receives positive
energy. A common derivative $c'_{\eta_j}(e_j)$ determines the allocation.
A zero-transmissivity mode may simply be omitted.

\subsection{Two simulation routes for general multimode channels}
The product-channel converse alone would not cover general
correlations in the noise. We therefore give two ways to reach a general
multimode channel. One places a product of single-mode channels before
additional processing at the receiver. The other implements the target
directly by teleportation through a Gaussian resource. These are separate
constructions, with separate feasible sets and optimization variables.

\paragraph{A product channel followed by receiver processing.}
The product-sector theorem does not require the \emph{target} channel to
have product noise. It suffices to factor the target into a product of
scalar channels followed by a channel local to Bob. The following linear
matrix inequalities (LMIs) provide exactly such a factorization, including for
phase-sensitive and rectangular multimode channels.

Let
\begin{equation}\label{eq:AB-core}
 \mathsf A\coloneqq\bigoplus_{j=1}^m a_jI_2,\qquad
 \mathsf B\coloneqq\bigoplus_{j=1}^m b_jI_2,\qquad
 a_j>0,\quad b_j\ge|a_j-1|.
\end{equation}
Define $\tau_j\coloneqq1/a_j$ and $v_j\coloneqq(b_j+1-a_j)/(2a_j)$, so that each associated
scalar channel is completely positive. Let $\mathcal D(X,Y)$ be the set of
these parameters satisfying
\begin{equation}\label{eq:core-LMI}
 Y-X\mathsf B X^{\top}+i(\Omega_n-X\mathsf A\Omega_mX^{\top})\succeq0.
\end{equation}
The inequalities $b_j\ge|a_j-1|$ are two scalar linear inequalities; the
matrix inequality is affine in $\boldsymbol a,\boldsymbol b$.

\begin{samepage}
\begin{theorem}[Universal Gaussian-core converse]\label{thm:core}
Let $\mathcal N\colon A\to B$ be the $m$-input-mode,
$n$-output-mode Gaussian channel $(X,Y)$ in \eqref{eq:channel-CP}, and
impose \eqref{eq:multimode-energy}. Let $\mathcal D(X,Y)$ be specified by
\eqref{eq:AB-core}, \eqref{eq:core-LMI}, and let $B_{\boldsymbol\tau,\boldsymbol v}$
be defined in \eqref{eq:allocation-B}. For $E\ge0$, define
\begin{equation}\label{eq:Ucore}
 U_{\rm core}(X,Y;E)\coloneqq
 \inf\left\{B_{\boldsymbol\tau,\boldsymbol v}(E):
                    (\boldsymbol a,\boldsymbol b)\in\mathcal D(X,Y)\right\}.
\end{equation}
Then
\begin{equation}\label{eq:core-capacity}
 Q_2(\mathcal N,E)\le K_2(\mathcal N,E)
 \le U_{\rm core}(X,Y;E)\le m g(E/m).
\end{equation}
Every feasible parameter choice supplies a valid upper bound, without
requiring optimality of the outer minimization. The function $U_{\rm core}$
is nondecreasing and concave in $E$ and vanishes at $E=0$.
\end{theorem}
\end{samepage}
\begin{proof}
For feasible parameters set
\begin{equation}\label{eq:postprocessing-C}
 X_{\mathcal C}\coloneqq X\mathsf A^{1/2},\qquad
 Y_{\mathcal C}\coloneqq Y-X\mathsf B X^{\top}.
\end{equation}
Its complete-positivity condition is exactly \cref{eq:core-LMI}.
The product core has matrices
$X_0=\mathsf A^{-1/2}$ and $Y_0=\mathsf A^{-1}\mathsf B$.
The composition rule gives
\begin{equation}\label{eq:core-composition}
 X_{\mathcal C}X_0=X,\qquad
 X_{\mathcal C}Y_0X_{\mathcal C}^{\top}+Y_{\mathcal C}=Y.
\end{equation}
Thus $\mathcal N=\mathcal C\circ\bigotimes_j\mathcal G_{\tau_j,v_j}$ exactly
on arbitrary inputs, not only on Gaussian states. In a communication
protocol using the product core, Bob can apply $\mathcal C$ after each use
and reproduce the target channel. Alice's input, including its mean energy,
is unchanged. \Cref{thm:multisector} therefore proves the first bound after
taking the infimum.

The choice $a_j=1,b_j=0$ is always feasible by \cref{eq:channel-CP}; it is the
identity core with Bob implementing the original channel. Its cost is
$mg(E/m)$, proving finiteness and the last inequality. An infimum of
concave functions is concave, as is seen from the intersection of their
convex hypographs. Here it is also nondecreasing because every member of
the family is nondecreasing. All members vanish at zero.
\end{proof}

This is a universal bound, not a normal-form claim that every Gaussian
channel is unitarily equivalent to independent scalar channels. Its
reference to the input basis matters because \cref{eq:multimode-energy}
uses physical photon number. One may optimize over an additional passive
input interferometer. An active input symplectic transformation would
require transforming the energy constraint and is not silently allowed.
Although the feasible region in \cref{eq:core-LMI} is convex, we do not
claim that its general objective is convex. Compactly truncating the
parameters gives ordinary finite-dimensional global searches; a useful
convex specialization follows next.

\begin{samepage}
\begin{corollary}[Convex pure-loss-core optimization]\label{cor:convex-loss-core}
Restrict to $a_j=\lambda_j\ge1$ and $b_j=\lambda_j-1$, so that the core is
pure loss of transmissivity $1/\lambda_j$. With
$\Lambda=\bigoplus_j\lambda_jI_2$, the feasible condition becomes
\begin{equation}\label{eq:loss-core-LMI}
 \mathsf H_{X,Y}-X(\Lambda-I)(I+i\Omega_m)X^{\top}\succeq0.
\end{equation}
Minimizing
\begin{equation}\label{eq:convex-loss-objective}
 \max\left\{\sum_j\left[g(e_j)-g((1-\lambda_j^{-1})e_j)\right]:
                   e_j\ge0,\ \sum_j e_j\le E\right\}
\end{equation}
over this region, specified by linear matrix inequalities, is a convex minimization in
$\boldsymbol\lambda$. It gives an upper bound on both $Q_2$ and $K_2$.
\end{corollary}
\end{samepage}
\begin{proof}
Substitute the loss parameters into \cref{eq:core-LMI}, using commutation
of $\Lambda$ and $\Omega_m$, to get \cref{eq:loss-core-LMI}. For fixed $e_j$,
$(1-\lambda_j^{-1})e_j$ is concave in $\lambda_j>0$. The function $g$ is
concave and nondecreasing, so the negative of this composition is convex.
The sum is convex in $\boldsymbol\lambda$, and its pointwise maximum over
the energy simplex remains convex. \Cref{thm:core} gives the converse.
No finite-parameter attainment is required when a transmissivity tends to
zero; increasing compact upper bounds on $\lambda_j$ gives the same infimum.
\end{proof}

\paragraph{Optimizing the passive basis together with the losses.}
The preceding diagonal parameterization chooses the input mode basis before
optimizing the losses. For correlated noise, this choice can discard useful
factorizations. A passive change of modes preserves total photon number, so
it can be optimized without changing the communication constraint.
The next result includes this basis choice in a single convex matrix variable.

\begin{samepage}
\begin{theorem}[Passive-basis loss-core bound]\label{thm:passive-core}
Let $(X,Y)$ be an $m$-input-mode, $n$-output-mode Gaussian channel as in
\eqref{eq:channel-CP}. Define
\begin{equation}\label{eq:passive-feasible}
 \mathcal L(X,Y)\coloneqq\left\{L=L^{\top}\in\R^{2m\times2m}:
 \begin{array}{l}L\succeq I,\ L\Omega_m=\Omega_mL,\\
 \mathsf H_{X,Y}-X(L-I)(I+i\Omega_m)X^{\top}\succeq0
 \end{array}\right\}.
\end{equation}
The ordinary eigenvalues of $L$ occur in pairs. Denote one copy of each
pair by $\lambda_1,\ldots,\lambda_m$, and, for $E\ge0$, define
\begin{equation}\label{eq:passive-cost}
 \begin{split}
 \mathcal C_E(L)&\coloneqq\max\left\{
 \sum_{j=1}^m\left[g(e_j)-g((1-\lambda_j^{-1})e_j)\right]:
 e_j\ge0,\ \sum_j e_j\le E\right\},\\
 U_{\rm pass}(X,Y;E)&\coloneqq\inf\left\{\mathcal C_E(L):L\in\mathcal L(X,Y)\right\}.
 \end{split}
\end{equation}
The feasible set and the objective $\mathcal C_E$ are Euclidean convex,
and $Q_2(\mathcal N,E)\le K_2(\mathcal N,E)\le U_{\rm pass}(X,Y;E)$.
The family contains every fixed-basis pure-loss core of
\eqref{eq:loss-core-LMI}. Every feasible $L$ gives a valid upper bound,
without requiring that the infimum be attained.
\end{theorem}
\end{samepage}
\begin{proof}
A real symmetric matrix commuting with $\Omega_m$ is the real representation
of an $m\times m$ Hermitian matrix. Unitary diagonalization in that
representation gives an orthogonal symplectic $O$ with
$L=O(\bigoplus_j\lambda_jI_2)O^{\top}$. Thus the Gaussian channel with
matrices $X_0=L^{-1/2}$ and $Y_0=I-L^{-1}$ is a product of pure-loss
channels in a passive basis. Set
\begin{equation}\label{eq:passive-post}
 X_c\coloneqq XL^{1/2},\qquad Y_c\coloneqq Y-X(L-I)X^{\top}.
\end{equation}
Commutation with $\Omega_m$ shows that the complete-positivity matrix of
$(X_c,Y_c)$ is precisely the final matrix in \eqref{eq:passive-feasible}.
Furthermore $X_cX_0=X$ and $X_cY_0X_c^{\top}+Y_c=Y$.
Bob can implement this postprocessing locally. The input and output
passive changes in the core can be undone locally, and the input change
preserves the total energy. The loss allocation formula
\eqref{eq:exact-multiloss} proves the bound.

For fixed $e\ge0$, the function
$\lambda\mapsto g(e)-g((1-\lambda^{-1})e)$ is convex on $[1,\infty)$:
$1-\lambda^{-1}$ is concave, and $g$ is concave and nondecreasing.
Maximizing the sum over the energy simplex gives a symmetric convex
function of $(\lambda_1,\ldots,\lambda_m)$. The spectral-function theorem
for Hermitian matrices \cite{Lewis1996} then proves convexity in $L$.
The constraints in \eqref{eq:passive-feasible} are affine matrix
inequalities and linear equalities. This proves the two convexity claims.
\end{proof}

For the real phase-insensitive models in the examples, the matrix
optimization has a closed solution. This solution optimizes the entire
passive loss-core family, but not the more general noisy-core or
Gaussian-resource families.
\begin{samepage}
\begin{corollary}[Greatest loss core for real phase-insensitive channels]
\label{cor:greatest-core}
Suppose $X=K\otimes I_2$ and $Y=Y_0\otimes I_2$, where $K$ is a real
invertible $m\times m$ matrix and $(X,Y)$ is completely positive. Set
\begin{equation}\label{eq:Lmax}
 L_{\max}^{(0)}\coloneqq\frac12\left[I_m+
 K^{-1}(Y_0+I_m)K^{-\top}\right],\qquad
 L_{\max}\coloneqq L_{\max}^{(0)}\otimes I_2.
\end{equation}
Then $L_{\max}$ is the greatest element of $\mathcal L(X,Y)$ in matrix
order and
$U_{\rm pass}(X,Y;E)=\mathcal C_E(L_{\max})$.
In particular, if $Y_0=I_m-KK^{\top}+2Q$ with $Q\succeq0$, then
\begin{equation}\label{eq:Lmax-correlated}
 L_{\max}^{(0)}=K^{-1}(I_m+Q)K^{-\top}.
\end{equation}
\end{corollary}
\end{samepage}
\begin{proof}
In the complex Hermitian representation used above, write the core matrix
as $H$. The two conjugate sectors of the postprocessing uncertainty matrix
require
\begin{equation}\label{eq:complex-core-sectors}
 Y_0+KK^{\top}+I_m-2KHK^{\top}\succeq0,\qquad
 Y_0+KK^{\top}-I_m\succeq0.
\end{equation}
The second inequality already follows from complete positivity of the
target. The first is equivalent to $H\preceq L_{\max}^{(0)}$, even when
$H$ has complex off-diagonal entries. The other target complete-positivity
inequality, $Y_0+I_m-KK^{\top}\succeq0$, implies
$L_{\max}^{(0)}\succeq I_m$. Thus the proposed greatest element is feasible.
Ordered Hermitian eigenvalues are monotone in matrix order, and the scalar
cost is nonincreasing in each $\lambda_j$. Hence no smaller feasible
matrix gives a lower cost. Substitution gives \eqref{eq:Lmax-correlated}.
\end{proof}

\paragraph{An analytic scalar predecessor for a one-input preparation.}
For a state obtained by sending one mode of a squeezed pair through a
multiple-output Gaussian channel, a scalar predecessor can produce a
stronger operational converse than the Gaussian-reference value of the
prepared state. The receiver's several output modes are kept together;
the reduction is a channel factorization, not their removal by a measurement.

\begin{samepage}
\begin{proposition}[A least informative scalar predecessor]
\label{prop:scalar-predecessor}
Let a one-input-mode, $n$-output-mode channel have
$X=k\otimes I_2$ and $Y=(I_n-kk^{\top}+2Q)\otimes I_2$, where
$0<\|k\|\le1$, $k\in\R^n$, and $Q\succ0$. Define
\begin{equation}\label{eq:predecessor-parameters}
 \begin{aligned}
 \alpha&\coloneqq k^{\top}(I_n+Q)^{-1}k,&
 \beta&\coloneqq k^{\top}Q^{-1}k,\\
 a_*&\coloneqq\alpha^{-1}-\beta^{-1},&
 b_*&\coloneqq\alpha^{-1}+\beta^{-1}-1.
 \end{aligned}
\end{equation}
Then $a_*>0$, $b_*\ge|a_*-1|$, and the scalar channel
$\mathcal G_{\tau_*,v_*}$ with
$\tau_*\coloneqq1/a_*$ and $v_*\coloneqq(b_*+1-a_*)/(2a_*)$
is a feasible predecessor of $(X,Y)$. It is itself a Gaussian
postprocessing of every other physical scalar predecessor in the family
\eqref{eq:AB-core}--\eqref{eq:core-LMI}. In particular,
\begin{equation}\label{eq:predecessor-bound}
 K_2(\mathcal N,E)\le\mathcal F_{\tau_*,v_*}(E).
\end{equation}
If $\rho$ is prepared by sending the mode of local mean $E$ from a
squeezed pair through $\mathcal N$, then both $K_D(\rho)$ and
$D_{\leftrightarrow}(\rho)$ obey the same upper bound.
\end{proposition}
\end{samepage}
\begin{proof}
For scalar parameters $a,b$, the postprocessing matrix in
\eqref{eq:core-LMI} separates into
$2(I_n+Q)-(1+a+b)kk^{\top}$ and
$2Q+(a-b-1)kk^{\top}$. For $A\succ0$, the rank-one inequality
$A-c kk^{\top}\succeq0$ is equivalent to
$c\le(k^{\top}A^{-1}k)^{-1}$. Consequently feasibility is exactly
\begin{equation}\label{eq:predecessor-halfplanes}
 a+b\le2/\alpha-1,\qquad b-a\le2/\beta-1,
 \qquad a>0,\quad b\ge|a-1|.
\end{equation}
We have $0<\alpha<\beta$ and $\alpha\le1$.
The proposed $(a_*,b_*)$ saturates the two upper bounds and is physical:
$b_*+a_*\ge1$ and $b_*-a_*\ge-1$.
For another feasible pair $(a,b)$, \eqref{eq:predecessor-halfplanes}
implies $b_*-b\ge|a_*-a|$. Therefore the scalar channel of intensity
gain $a/a_*$ and covariance noise $(b_*-b)/a_*$ is completely positive.
Composing it after the $(a,b)$ predecessor gives exactly the
$(a_*,b_*)$ predecessor. This proves the postprocessing-order assertion.
Its capacity bound is the imported scalar converse followed by the
receiver channel. A state-distillation protocol after one preparation
use per copy is a communication protocol with that same energy, proving
the final statement.
\end{proof}
The postprocessing-order assertion does not imply that every chosen
upper-bound formula is monotone under this ordering, or that the bound
is the capacity. We use \eqref{eq:predecessor-bound} as a particular valid
converse and take its minimum with other valid bounds.

\begin{figure}[t]
\centering
\begin{tikzpicture}[>=Latex,font=\small,
box/.style={draw=boundblue,thick,rounded corners=2pt,minimum height=1.15cm,
 text width=2.8cm,align=center,inner sep=6pt}]
\node[box,fill=boundblue!5] (o) at (0,0) {Passive input\\change of modes};
\node[box,fill=keygreen!5] (loss) at (4.0,0) {Independent losses\\$1/\lambda_1,\ldots,1/\lambda_m$};
\node[box,fill=rateorange!5] (post) at (8.0,0) {Receiver's Gaussian\\postprocessing};
\draw[->,thick] (-2.3,0)--(o.west);
\draw[->,thick] (o.east)--(loss.west);
\draw[->,thick] (loss.east)--(post.west);
\draw[->,thick] (post.east)--(10.3,0);
\node[align=center] at (4,-1.45) {Together these reproduce the target channel $(X,Y)$.\\
The matrix $L$ optimizes the passive basis and the losses simultaneously.};
\node[font=\footnotesize,align=center] at (0,1.05) {Total input energy is preserved};
\node[font=\footnotesize,align=center] at (8,1.05) {Local operation at Bob};
\end{tikzpicture}
\caption{The passive-basis factorization of \cref{thm:passive-core}.
Only the middle channel is used to bound communication. The passive input
operation preserves the energy constraint, and the receiver's processing
can only reduce capacity. The scalar predecessor in
\cref{prop:scalar-predecessor} replaces the middle channel by one noisy
scalar channel when there is only one input mode. This is a channel
factorization, not a claim that the target noise is independently diagonalizable.}
\label{fig:core-factorization}
\end{figure}

\paragraph{Exact teleportation with a Gaussian resource.}
We now use the paper's Gaussian entanglement calculation directly. Unlike a
sector resource, the resource in this subsection is an ordinary finite-mode
Gaussian state whose covariance completely specifies the entanglement
optimization. It need not have the same local energy as the code input.

Set $Z_m\coloneqq\bigoplus_{j=1}^m\diag(1,-1)$ and
\begin{equation}\label{eq:LX}
 L_X\coloneqq\begin{pmatrix}-XZ_m&I_{2n}\end{pmatrix}.
\end{equation}
A resource $\sigma_{RB}$ has $m$ modes $R$ on Alice's side and $n$ modes $B$
on Bob's side. Alice performs the continuous-variable Bell measurement
$\hat r_A-Z_m\hat r_R$ and Bob displaces by $X$ times the outcome. These measured quadratures commute, since their commutator matrix
is $i(\Omega_m+Z_m\Omega_mZ_m)=0$. Thus they form a joint quadrature
measurement on Alice's input and her resource modes. The gain-$X$ Braunstein--Kimble construction
\cite{BraunsteinKimble1998,LiuzzoScorpoEtAl2017} has the following exact
multimode covariance condition.

\begin{samepage}
\begin{proposition}[Exact Gaussian simulator]\label{prop:exact-simulator}
Let $\mathcal N=(X,Y)$ be an $m$-input-mode, $n$-output-mode
Gaussian channel as in \eqref{eq:channel-CP}. Let $\sigma_{RB}$ be a
zero-mean Gaussian resource state of quantum covariance matrix $\Sigma$,
with $m$ modes $R$ at Alice and $n$ modes $B$ at Bob. Set
$Z_m\coloneqq\bigoplus_{j=1}^m\diag(1,-1)$ and
$L_X\coloneqq(-XZ_m\ \ I_{2n})$ as in \eqref{eq:LX}.
The Bell measurement of $\hat r_A-Z_m\hat r_R$, followed by Bob's
outcome-dependent displacement with gain $X$, simulates $\mathcal N$
exactly whenever
\begin{equation}\label{eq:simulator-feasible}
 \Sigma+i(\Omega_m\oplus\Omega_n)\succeq0,
 \qquad L_X\Sigma L_X^{\top}=Y.
\end{equation}
The simulation identity holds for arbitrary channel input states and
retained reference systems, including entangled input states.
If $\mathsf H_{X,Y}\succ0$, a finite covariance matrix satisfying
\eqref{eq:simulator-feasible} exists.
\end{proposition}
\end{samepage}
\begin{proof}
For a centered resource, the noise quadratures are
$\hat r_B-XZ_m\hat r_R$. Direct evaluation of the Bell measurement and
displacement in Weyl characteristic functions gives
\begin{equation}\label{eq:simulator-characteristic}
 \chi_{\rm out}(z)=\chi_{\rm in}(X^{\top}z)
                 \exp[-z^{\top}L_X\Sigma L_X^{\top}z/4],
 \qquad\chi_\rho(z)\coloneqq\Tr[e^{iz^{\top}\hat r}\rho].
\end{equation}
This is the characteristic-function definition of the channel $(X,Y)$
under \cref{eq:simulator-feasible}; it establishes equality as channels,
not merely equality on Gaussian input states. Linearity gives the corresponding
identity in the presence of an arbitrary retained reference.

For existence, $L_X$ has full row rank because its right block is the
identity. Complete its rows to an invertible real matrix $T$. Write
\begin{equation}\label{eq:simulator-existence-block}
 T(\Omega_m\oplus\Omega_n)T^{\top}
 =\begin{pmatrix}J&K\\-K^{\top}&J_2\end{pmatrix},
 \qquad J=\Omega_n-X\Omega_mX^{\top}.
\end{equation}
Consider the transformed real covariance matrix $Y\oplus tI_{2m}$. Its uncertainty
matrix has upper-left block $Y+iJ=\mathsf H_{X,Y}\succ0$. The Schur
complement of this block is positive for sufficiently large finite $t$:
for example choose
$t>\norm{J_2}+\norm{K}^2/\lambda_{\min}(\mathsf H_{X,Y})$.
Then $\Sigma=T^{-1}(Y\oplus tI)T^{-\top}$ is physical and has
$L_X\Sigma L_X^{\top}=Y$ by construction.
\end{proof}

\begin{figure}[t]
\centering
\begin{tikzpicture}[>=Latex,font=\small]
\node[anchor=east] at (0,1.4) {Input $A$};
\node[anchor=east] at (0,0) {Resource $R$};
\node[anchor=east] at (0,-2) {Resource $B$};
\draw[thick] (0,1.4)--(3.3,1.4);
\draw[thick] (0,0)--(3.3,0);
\draw[thick,draw=boundblue] (3.3,-.4) rectangle (5.7,1.8);
\node[align=center] at (4.5,.7) {Alice's Bell\\measurement};
\draw[->,thick,dashed,draw=rateorange] (5.7,.7)--(8.2,-1.55);
\node[align=center,font=\footnotesize,fill=white,inner sep=3pt] at (7.7,.45)
{Classical outcome $z$};
\draw[thick] (0,-2)--(7.9,-2);
\draw[thick,draw=keygreen] (7.9,-2.45) rectangle (10.3,-1.55);
\node at (9.1,-2) {$D_B(Xz)$};
\draw[->,thick] (10.3,-2)--(12,-2);
\node[above,font=\footnotesize] at (11.1,-2) {Output};
\draw[dotted,thick,draw=boundblue] (.75,0)--(.75,-2);
\node[fill=white,font=\footnotesize,align=center] at (.75,-1)
{Shared\\state $\sigma_\Sigma$};
\node[align=center] at (6,-3.2)
{Exact channel identity: $L_X\Sigma L_X^{\top}=Y$.\\
The shared resource energy is not the code's transmitted energy.};
\end{tikzpicture}
\caption{Exact Gaussian-resource teleportation in
\cref{prop:exact-simulator}. Each solid wire is a separate quantum
system. The dashed diagonal arrow carries the Bell-measurement outcome
from Alice to Bob, who displaces his resource modes with gain $X$.
The dotted line marks the initially shared state, not a transmission
through the channel. The equality in \eqref{eq:simulator-feasible}
is essential: an arbitrary finite-energy channel-output state need not
satisfy it.}
\label{fig:resource-simulation}
\end{figure}

The Bell measurement is an ideal quadrature measurement, allowed in the
unconstrained local apparatus of this converse. ``Finite resource'' refers
to the shared state's covariance and energy, not to a claim that ideal
homodyne detection has a bounded-energy experimental realization.
Finite-resource simulation and this distinction are discussed in
\cite{Wilde2018Simulation}. At a singular CP boundary a finite exact resource
may or may not exist; \cref{thm:core} remains applicable regardless.

Let $\mathcal T(X,Y)$ be the set in \cref{eq:simulator-feasible}, and define
\begin{equation}\label{eq:UG-def}
 \begin{split}
 U_G^R(X,Y)&\coloneqq\inf\left\{R_G(R;B)_{\sigma_\Sigma}:\Sigma\in\mathcal T(X,Y)\right\},\\
 U_G^E(X,Y)&\coloneqq\inf\left\{E_R^G(R;B)_{\sigma_\Sigma}:\Sigma\in\mathcal T(X,Y)\right\}.
 \end{split}
\end{equation}
An empty infimum is $+\infty$. For each fixed resource, the inner reference
optimization is the globally solvable problem of \cref{sec:algorithms}.
The additional optimization over resource covariances is not covered by
that convexity theorem and is not asserted to be geodesically convex.
A single feasible resource and a single feasible reference already give
a valid, possibly nonoptimal, capacity upper bound.

Ideal Bell measurements have continuous outcomes, whereas the finite codes
above use finite or countable alphabets. This does not enlarge the capacities
being bounded. For a fixed normal input and fixed finite protocol, truncate
the Bell outcome range and replace the correction on each bin by a fixed
representative displacement. Strong continuity of displacements and the
vanishing tail probability imply trace-norm convergence to the ideal
simulator. The approximating channels are finite-outcome LOCC. Lower
semicontinuity of the unrestricted REE and Rains entanglement
\cite{LamiShirokov2023} passes their monotonicity bounds to the limiting
output. No uniform approximation over an unbounded-energy input family is
needed, because exact simulation is established before fixing each finite
code. The same finite-outcome approximation justifies Gaussian random
postprocessing when it is included in the local simulator.

\begin{samepage}
\begin{theorem}[Combined universal communication bounds]\label{thm:combined}
Let $\mathcal N=(X,Y)$ be an $m$-input-mode, $n$-output-mode
Gaussian channel satisfying \eqref{eq:channel-CP}, with the adaptive
communication model and finite input-energy constraint
\eqref{eq:multimode-energy}. Let $U_{\rm core},U_{\rm pass},U_G^R,U_G^E$ be defined
in \eqref{eq:Ucore}, \eqref{eq:passive-cost}, and \eqref{eq:UG-def}. Then
\begin{align}
 Q_2(\mathcal N,E)&\le
 \min\left\{U_{\rm core}(X,Y;E),U_{\rm pass}(X,Y;E),U_G^R(X,Y)\right\},\label{eq:combined-Q}\\
 P_2(\mathcal N,E)=K_2(\mathcal N,E)&\le
 \min\left\{U_{\rm core}(X,Y;E),U_{\rm pass}(X,Y;E),U_G^E(X,Y)\right\}.\label{eq:combined-K}
\end{align}
Both right sides are finite after the minimum is taken, even when the
exact-Gaussian-resource set is empty.
\end{theorem}
\end{samepage}
\begin{proof}
Fix a resource satisfying \cref{eq:simulator-feasible}. Replace every channel
use in a fixed adaptive protocol by its exact LOCC simulator. The final
state is an LOCC output of $\sigma_\Sigma^{\otimes\ell}$; all adaptive
classical processing can be included in that final LOCC channel. For the
quantum task,
\begin{equation}\label{eq:Gaussian-simulation-Q}
 R(\zeta_\ell)\le R(\sigma_\Sigma^{\otimes\ell})
 \le R_G(\sigma_\Sigma^{\otimes\ell})=\ell R_G(\sigma_\Sigma).
\end{equation}
For a retained private-state output, replace $R$ by $E_R$ and $R_G$ by
$E_R^G$. The last equalities are \cref{thm:additivity}. The target tests in
\cref{eq:one-test-bound} give the respective vanishing-error converses.
Infimize over exact resources and combine with \cref{thm:core,thm:passive-core}.
Finiteness follows from $U_{\rm core}\le mg(E/m)$.
\end{proof}

\Cref{thm:combined} separates two optimizations that are sometimes inadvertently
identified. The code energy $E$ is enforced in $U_{\rm core}$. The resource
in $U_G$ is optimized under an \emph{exact channel identity}, not under a
local-energy equality to $E$. Substituting a finite-energy Choi state of
energy $E$ into a teleportation-simulation bound without checking that
identity generally simulates a noisier channel and does not prove a
converse for $\mathcal N$.

A convenient explicit exact-resource family illustrates the distinction.
For $\mu>1$, put
\begin{equation}\label{eq:mu-resource}
 \begin{split}
 \xi_\mu&\coloneqq2(\mu-\sqrt{\mu^2-1}),\qquad Y_0\coloneqq Y-\xi_\mu XX^{\top},\\
 \Sigma_\mu&\coloneqq\begin{pmatrix}
 \mu I_{2m}&\sqrt{\mu^2-1}Z_mX^{\top}\\
 \sqrt{\mu^2-1}XZ_m&\mu XX^{\top}+Y_0
 \end{pmatrix}.
 \end{split}
\end{equation}
If $Y_0+i(\Omega_n-X\Omega_mX^{\top})\succeq0$, this is the state produced by
sending the channel $(X,Y_0)$ one share of $m$ squeezed pairs of local
covariance $\mu I$. It is therefore physical. Substitution shows
$L_X\Sigma_\mu L_X^{\top}=Y$ exactly. For a strictly positive CP matrix,
$\xi_\mu\to0$ guarantees feasibility at some finite $\mu$. In our plots
we test $\mu\in\{20,50,120\}$ and keep the lowest feasible Gaussian
entanglement upper bound. The simulator parameter $\mu$ is independent of
$E$ throughout this calculation.

\subsection{Two- and three-mode channel comparisons}
We give both exactly solvable benchmarks and genuinely correlated-noise
examples. All horizontal energy axes refer to total photons per multimode
use. Two-mode and three-mode \emph{channels} below have two or three input
and output modes, respectively; their Gaussian simulator resources have
four or six modes, unlike the three-mode \emph{state} in
\cref{sec:operational-states}.

\begin{figure}[t]
\centering
\begin{tikzpicture}
\begin{axis}[
name=main,
width=.90\textwidth,
height=7.0cm,
axis lines=left,
enlarge y limits={upper,value=0.05},
ymin=0,
xlabel={Total input photons $E$ per 3-mode use},
ylabel={Qubits or secret bits per use},
tick label style={font=\small},
label style={font=\small},
legend style={at={(.5,-.24)},anchor=north,draw=none,font=\small,legend columns=2,cells={anchor=west},/tikz/every even column/.append style={column sep=8pt}},
legend cell align=left,
clip=true,
xmode=log,
log basis x=10,
xmin=.025,
xmax=15.0,
xtick={.03,.1,.3,1,3,10,30},
xticklabels={$0.03$,$0.1$,$0.3$,$1$,$3$,$10$,$30$}]
\addplot[boundblue,very thick,no marks] coordinates {
(0.025,0.132911840442)
(0.0289121114823,0.148949914589)
(0.0334364076145,0.166775045161)
(0.0386686857807,0.186559230056)
(0.0447197341666,0.208485893461)
(0.0517176775873,0.232749601523)
(0.0598106904003,0.259555504496)
(0.0691701339475,0.289118453401)
(0.0799941849573,0.321661736824)
(0.0925120317368,0.357415383148)
(0.106988727001,0.396613975901)
(0.123730800096,0.439493935661)
(0.143092747447,0.486290231509)
(0.165484538659,0.537232498393)
(0.191380297212,0.592540553418)
(0.221328339541,0.652419322724)
(0.255962785079,0.717053209763)
(0.296016983101,0.786599953453)
(0.342339040643,0.86118404044)
(0.395909780312,0.940889750517)
(0.457863508212,1.02575393221)
(0.529512031723,1.11575863316)
(0.612372435696,1.21082375456)
(0.70819920518,1.31079996653)
(0.819021374873,1.41546221059)
(0.947185491868,1.52450421989)
(1.09540530141,1.63753458837)
(1.26681920771,1.75407499514)
(1.46505672644,1.87356121254)
(1.69431533611,1.99534747392)
(1.95944935535,2.11871463608)
(2.26607272823,2.24288233935)
(2.62067789382,2.36702506017)
(3.03077325701,2.49029159622)
(3.50504217137,2.61182716425)
(4.05352680035,2.73079697494)
(4.68784074993,2.84640992506)
(5.42141497492,2.95794095208)
(6.26978216586,3.06475065106)
(7.25090563796,3.16630095555)
(8.38555968608,3.26216600988)
(9.69776945941,3.35203776854)
(11.2153196696,3.43572629501)
(12.9703429039,3.51315514502)
(15,3.58435255971)
};
\addlegendentry{Exact $Q_2=K_2$ / sector bound}
\addplot[mutedgray,dashed,thick,no marks] coordinates {
(0.025,0.139345430385)
(0.0289121114823,0.156396744556)
(0.0334364076145,0.17539391503)
(0.0386686857807,0.196535825559)
(0.0447197341666,0.220028664351)
(0.0517176775873,0.24610096973)
(0.0598106904003,0.274998161216)
(0.0691701339475,0.30697580164)
(0.0799941849573,0.342303831765)
(0.0925120317368,0.381269276817)
(0.106988727001,0.42416648774)
(0.123730800096,0.471303090644)
(0.143092747447,0.522993141799)
(0.165484538659,0.579557867656)
(0.191380297212,0.641304575228)
(0.221328339541,0.708558019666)
(0.255962785079,0.781620665841)
(0.296016983101,0.860777706803)
(0.342339040643,0.946298959419)
(0.395909780312,1.03842369516)
(0.457863508212,1.13735343731)
(0.529512031723,1.24324629269)
(0.612372435696,1.35620421032)
(0.70819920518,1.47627077599)
(0.819021374873,1.6034141231)
(0.947185491868,1.7375228943)
(1.09540530141,1.87839374881)
(1.26681920771,2.02572498189)
(1.46505672644,2.17910681692)
(1.69431533611,2.33801905687)
(1.95944935535,2.50182604178)
(2.26607272823,2.66978005748)
(2.62067789382,2.84102490833)
(3.03077325701,3.01461011489)
(3.50504217137,3.18949711119)
(4.05352680035,3.36459398225)
(4.68784074993,3.53876613912)
(5.42141497492,3.71086994077)
(6.26978216586,3.87978120453)
(7.25090563796,4.04442339405)
(8.38555968608,4.16991345993)
(9.69776945941,4.16991345993)
(11.2153196696,4.16991345993)
(12.9703429039,4.16991345993)
(15,4.16991345993)
};
\addlegendentry{Earlier squashing/loss envelope}
\addplot[accentpurple,dotted,thick,no marks] coordinates {
(0.025,4.16991345993)
(0.0289121114823,4.16991345993)
(0.0334364076145,4.16991345993)
(0.0386686857807,4.16991345993)
(0.0447197341666,4.16991345993)
(0.0517176775873,4.16991345993)
(0.0598106904003,4.16991345993)
(0.0691701339475,4.16991345993)
(0.0799941849573,4.16991345993)
(0.0925120317368,4.16991345993)
(0.106988727001,4.16991345993)
(0.123730800096,4.16991345993)
(0.143092747447,4.16991345993)
(0.165484538659,4.16991345993)
(0.191380297212,4.16991345993)
(0.221328339541,4.16991345993)
(0.255962785079,4.16991345993)
(0.296016983101,4.16991345993)
(0.342339040643,4.16991345993)
(0.395909780312,4.16991345993)
(0.457863508212,4.16991345993)
(0.529512031723,4.16991345993)
(0.612372435696,4.16991345993)
(0.70819920518,4.16991345993)
(0.819021374873,4.16991345993)
(0.947185491868,4.16991345993)
(1.09540530141,4.16991345993)
(1.26681920771,4.16991345993)
(1.46505672644,4.16991345993)
(1.69431533611,4.16991345993)
(1.95944935535,4.16991345993)
(2.26607272823,4.16991345993)
(2.62067789382,4.16991345993)
(3.03077325701,4.16991345993)
(3.50504217137,4.16991345993)
(4.05352680035,4.16991345993)
(4.68784074993,4.16991345993)
(5.42141497492,4.16991345993)
(6.26978216586,4.16991345993)
(7.25090563796,4.16991345993)
(8.38555968608,4.16991345993)
(9.69776945941,4.16991345993)
(11.2153196696,4.16991345993)
(12.9703429039,4.16991345993)
(15,4.16991345993)
};
\addlegendentry{Unconstrained pure-loss bound}
\end{axis}
\end{tikzpicture}
\caption{Pure-loss transmissivities $\boldsymbol\eta=(0.83,0.57,0.24)$. The solid line is the exact allocation formula \cref{eq:exact-multiloss}, attained by independent squeezed input states and reverse hashing. It also applies to passively mixed input modes and local output interferometers. The dashed line is the minimum of the balanced-squashing allocation \cref{eq:older-loss-upper} and the energy-unconstrained pure-loss sum. The capacity and the energy-dependent envelope vanish at $E=0$, which is omitted on the logarithmic axis; the energy-independent pure-loss benchmark does not.}
\label{fig:channel-loss-three}
\end{figure}
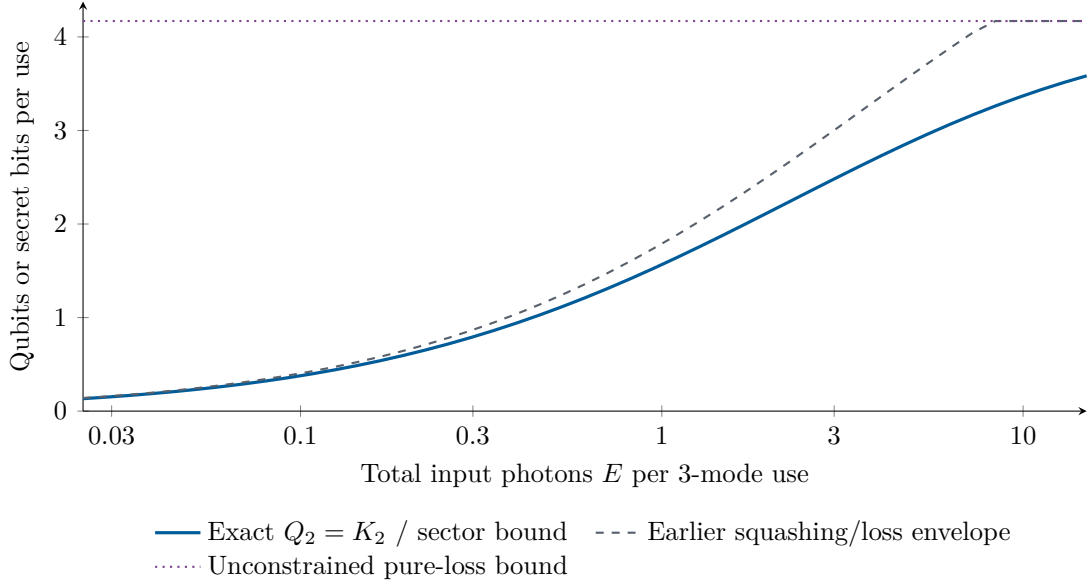

First take pure-loss transmissivities $(0.72,0.31)$ or $(0.83,0.57,0.24)$.
Arbitrary passive mixing of the input modes and local output interferometers
may be added without changing the answer in \cref{eq:exact-multiloss}.
The exact curve is simultaneously the multimode sector upper bound and
an attainable entanglement/key rate. As a prior benchmark, we plot the
minimum of the energy-unconstrained pure-loss sum $-\sum_j\log_2(1-\eta_j)$ and the optimized balanced
squashing upper bound
\begin{equation}\label{eq:older-loss-upper}
 \max\left\{\sum_j\left[
 g\!\left(\frac{1+\eta_j}{2}e_j\right)
 -g\!\left(\frac{1-\eta_j}{2}e_j\right)\right]:
              e_j\ge0,\ \sum_j e_j\le E\right\}.
\end{equation}
These two established upper-bound constructions are from
\cite{TakeokaGuhaWilde2013,DavisShirokovWilde2018,PLOB2017}.
\Cref{fig:channel-loss-two,fig:channel-loss-three} show how the exact
finite-energy formula closes the comparison for these benchmarks.

\begin{figure}[t]
\centering
\begin{tikzpicture}
\begin{axis}[
name=main,
width=.90\textwidth,
height=7.0cm,
axis lines=left,
enlarge y limits={upper,value=0.05},
ymin=0,
xlabel={Total input photons $E$ per 2-mode use},
ylabel={Qubits or secret bits per use},
tick label style={font=\small},
label style={font=\small},
legend style={at={(.5,-.24)},anchor=north,draw=none,font=\small,legend columns=2,cells={anchor=west},/tikz/every even column/.append style={column sep=8pt}},
legend cell align=left,
clip=true,
xmode=log,
log basis x=10,
xmin=.025,
xmax=15.0,
xtick={.03,.1,.3,1,3,10,30},
xticklabels={$0.03$,$0.1$,$0.3$,$1$,$3$,$10$,$30$}]
\addplot[boundblue,very thick,no marks] coordinates {
(0.025,0.10942766634)
(0.0289121114823,0.122283465003)
(0.0334364076145,0.136501332587)
(0.0386686857807,0.15219875006)
(0.0447197341666,0.169498495142)
(0.0517176775873,0.188527963819)
(0.0598106904003,0.209418281826)
(0.0691701339475,0.232303179915)
(0.0799941849573,0.257317604805)
(0.0925120317368,0.284596035436)
(0.106988727001,0.314270471896)
(0.123730800096,0.34646806279)
(0.143092747447,0.381308337206)
(0.165484538659,0.418900011386)
(0.191380297212,0.459337349895)
(0.221328339541,0.502696078616)
(0.255962785079,0.549028874212)
(0.296016983101,0.598360493048)
(0.342339040643,0.65068265208)
(0.395909780312,0.705948833477)
(0.457863508212,0.764069250539)
(0.529512031723,0.824906279366)
(0.612372435696,0.888270720906)
(0.70819920518,0.95391930171)
(0.819021374873,1.02155383758)
(0.947185491868,1.0908224609)
(1.09540530141,1.16132324033)
(1.26681920771,1.23261039573)
(1.46505672644,1.30420313417)
(1.69431533611,1.37559691522)
(1.95944935535,1.44627671832)
(2.26607272823,1.51573165832)
(2.62067789382,1.58347011332)
(3.03077325701,1.64903441998)
(3.50504217137,1.71201418018)
(4.05352680035,1.77205731578)
(4.68784074993,1.8288781983)
(5.42141497492,1.88226244204)
(6.26978216586,1.93206824603)
(7.25090563796,1.97822446287)
(8.38555968608,2.02072582157)
(9.69776945941,2.05962591076)
(11.2153196696,2.09502862393)
(12.9703429039,2.12707878007)
(15,2.15595257452)
};
\addlegendentry{Exact $Q_2=K_2$ / sector bound}
\addplot[mutedgray,dashed,thick,no marks] coordinates {
(0.025,0.116931111244)
(0.0289121114823,0.130942602511)
(0.0334364076145,0.146491022451)
(0.0386686857807,0.163719392078)
(0.0447197341666,0.182777986163)
(0.0517176775873,0.203828694811)
(0.0598106904003,0.227038814039)
(0.0691701339475,0.252583655616)
(0.0799941849573,0.280643702936)
(0.0925120317368,0.311407066534)
(0.106988727001,0.345063605349)
(0.123730800096,0.38180539298)
(0.143092747447,0.42182424234)
(0.165484538659,0.465307344215)
(0.191380297212,0.512438439792)
(0.221328339541,0.563388738707)
(0.255962785079,0.618315465985)
(0.296016983101,0.677356132995)
(0.342339040643,0.740623406811)
(0.395909780312,0.808195549588)
(0.457863508212,0.880116128074)
(0.529512031723,0.956380078531)
(0.612372435696,1.03692980197)
(0.70819920518,1.1216474774)
(0.819021374873,1.21034828565)
(0.947185491868,1.3027759439)
(1.09540530141,1.39859838625)
(1.26681920771,1.49740691778)
(1.46505672644,1.5987172811)
(1.69431533611,1.70197455668)
(1.95944935535,1.80656123961)
(2.26607272823,1.91180846692)
(2.62067789382,2.01701059049)
(3.03077325701,2.12144245287)
(3.50504217137,2.22437854656)
(4.05352680035,2.32511297708)
(4.68784074993,2.37183300071)
(5.42141497492,2.37183300071)
(6.26978216586,2.37183300071)
(7.25090563796,2.37183300071)
(8.38555968608,2.37183300071)
(9.69776945941,2.37183300071)
(11.2153196696,2.37183300071)
(12.9703429039,2.37183300071)
(15,2.37183300071)
};
\addlegendentry{Earlier squashing/loss envelope}
\addplot[accentpurple,dotted,thick,no marks] coordinates {
(0.025,2.37183300071)
(0.0289121114823,2.37183300071)
(0.0334364076145,2.37183300071)
(0.0386686857807,2.37183300071)
(0.0447197341666,2.37183300071)
(0.0517176775873,2.37183300071)
(0.0598106904003,2.37183300071)
(0.0691701339475,2.37183300071)
(0.0799941849573,2.37183300071)
(0.0925120317368,2.37183300071)
(0.106988727001,2.37183300071)
(0.123730800096,2.37183300071)
(0.143092747447,2.37183300071)
(0.165484538659,2.37183300071)
(0.191380297212,2.37183300071)
(0.221328339541,2.37183300071)
(0.255962785079,2.37183300071)
(0.296016983101,2.37183300071)
(0.342339040643,2.37183300071)
(0.395909780312,2.37183300071)
(0.457863508212,2.37183300071)
(0.529512031723,2.37183300071)
(0.612372435696,2.37183300071)
(0.70819920518,2.37183300071)
(0.819021374873,2.37183300071)
(0.947185491868,2.37183300071)
(1.09540530141,2.37183300071)
(1.26681920771,2.37183300071)
(1.46505672644,2.37183300071)
(1.69431533611,2.37183300071)
(1.95944935535,2.37183300071)
(2.26607272823,2.37183300071)
(2.62067789382,2.37183300071)
(3.03077325701,2.37183300071)
(3.50504217137,2.37183300071)
(4.05352680035,2.37183300071)
(4.68784074993,2.37183300071)
(5.42141497492,2.37183300071)
(6.26978216586,2.37183300071)
(7.25090563796,2.37183300071)
(8.38555968608,2.37183300071)
(9.69776945941,2.37183300071)
(11.2153196696,2.37183300071)
(12.9703429039,2.37183300071)
(15,2.37183300071)
};
\addlegendentry{Unconstrained pure-loss bound}
\end{axis}
\end{tikzpicture}
\caption{Pure-loss transmissivities $\boldsymbol\eta=(0.72,0.31)$. The solid line is the exact allocation formula \cref{eq:exact-multiloss}, attained by independent squeezed input states and reverse hashing. It also applies to passively mixed input modes and local output interferometers. The dashed line is the minimum of the balanced-squashing allocation \cref{eq:older-loss-upper} and the energy-unconstrained pure-loss sum. The capacity and the energy-dependent envelope vanish at $E=0$, which is omitted on the logarithmic axis; the energy-independent pure-loss benchmark does not.}
\label{fig:channel-loss-two}
\end{figure}
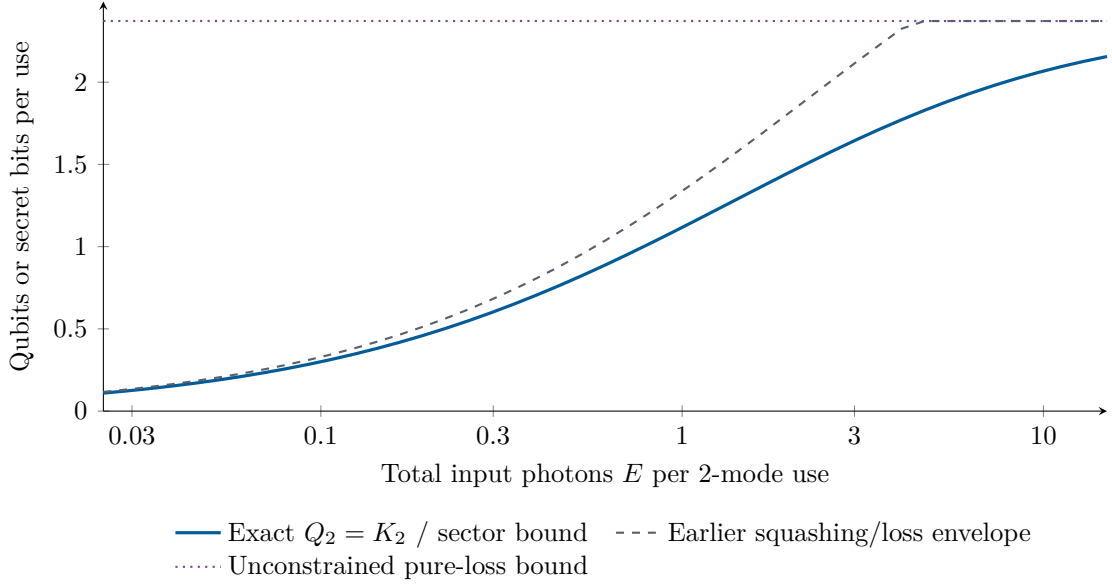

For the correlated examples let
\begin{equation}\label{eq:correlated-channel}
 X\coloneqq \diag(\sqrt{\tau_1},\ldots,\sqrt{\tau_m})\otimes I_2,\qquad
 Y\coloneqq (I_m-\diag\boldsymbol\tau+2Q)\otimes I_2.
\end{equation}
These are independent attenuations followed by correlated classical
Gaussian displacement noise. We use
\begin{equation}\label{eq:channel-two-parameters}
 \boldsymbol\tau\coloneqq (0.68,0.34),\qquad
 Q\coloneqq \begin{pmatrix}0.035&0.018\\0.018&0.065\end{pmatrix}
\end{equation}
and
\begin{equation}\label{eq:channel-three-parameters}
 \boldsymbol\tau\coloneqq (0.78,0.53,0.27),\qquad
 Q\coloneqq \begin{pmatrix}0.040&0.014&0.006\\0.014&0.065&0.020\\0.006&0.020&0.045\end{pmatrix}.
\end{equation}
Both $Q$ matrices are positive definite and do not commute with
$\diag\boldsymbol\tau$. Hence the attenuation and noise cannot be
simultaneously diagonalized by the same output passive change of modes;
these are not merely independent thermal channels written in a rotated basis.

We evaluate two core families. Pure-loss cores use
\begin{equation}\label{eq:example-loss-LMI}
 \diag(\tau_j\lambda_j)\preceq I_m+Q,\qquad\lambda_j\ge1.
\end{equation}
A noisy-core family keeps the attenuation fixed and chooses vacuum-output
noise $r_j\ge0$ subject to
\begin{equation}\label{eq:example-noisy-LMI}
 Q-\diag(r_1,\ldots,r_m)\succeq0.
\end{equation}
The remaining correlated displacement noise is implemented locally by Bob.
The cost is the energy allocation of $\mathcal F_{\tau_j,r_j}$. Every
feasible vector gives an upper bound; we report the best value found by
the documented numerical searches, not the unverified global infimum over
all core parameters. In the inner energy maximization, a concave supporting
hyperplane supplies an upper bound on the maximum, rather than using only
a solver's feasible allocation value; see \cref{app:operational-numerics}.

The additional passive-basis loss-core curve evaluates
$L_{\max}^{(0)}=K^{-1}(I+Q)K^{-\top}$ from
\cref{cor:greatest-core}. It optimizes the loss family over passive bases,
unlike the diagonal restriction in \eqref{eq:example-loss-LMI}.
The resulting curve need not improve the noisy-core or resource bounds;
we display it separately rather than implying universal superiority.
The overall converse may take the minimum of all displayed upper bounds.

For the Gaussian-resource bound, \cref{eq:mu-resource} and
\cref{prop:phase-symmetry} give a common $R_G=E_R^G$ upper bound. Optimizing
the full four- or six-mode reference covariance matrix gives respectively
$1.49639$ and $2.40926$ bits per use as upward-rounded numerical resource
bounds. The underlying evaluations are approximately $1.496382114$ and
$2.409250252$. These values come from explicitly feasible reference
covariances, with the floating-point qualifications in
\cref{app:operational-numerics}; they are not interval-certified optimal values.

The achievable curves include optimized thermal input allocations with
hashing, product noisy Gaussian measurement seeds in both reconciliation
directions, and non-Gaussian Fock-pair distribution followed by heralded
recurrence and hashing. We also take achievable time-sharing envelopes
under the same mean-energy constraint. For diagonal $X$ in
\cref{eq:correlated-channel}, each input--corresponding-output marginal is
a scalar thermal-loss channel even when the other inputs are used.
Independent marginal entanglement-distillation protocols can therefore
be run simultaneously; their errors vanish jointly by the union bound on
the commuting maximally entangled target tests. This justifies adding their
achieved rates despite correlated output noise.

The non-Gaussian comparisons draw on the protocol families of
\cite{MeleLamiGiovannetti2025,BarberPirandola2026}; the trusted Gaussian
measurement calculation uses the finite-energy multimode form of
\cref{eq:CQ-key}, related to \cite{OrtolanoEtAl2025}. Interpolation and
lookahead optimizations provide additional possibilities
\cite{BarberPirandolaInterpolation2026,BarberPirandolaLookahead2026}.
The finite filter and pairing search here does not exhaust those
optimizations. It retains every preparation and recurrence success factor,
and does not replace finite transmitted energy by an infinite-energy
limit. The pure-loss lower curves are optimal; the correlated-noise curves
are the best \emph{evaluated} attainable envelopes specified here.

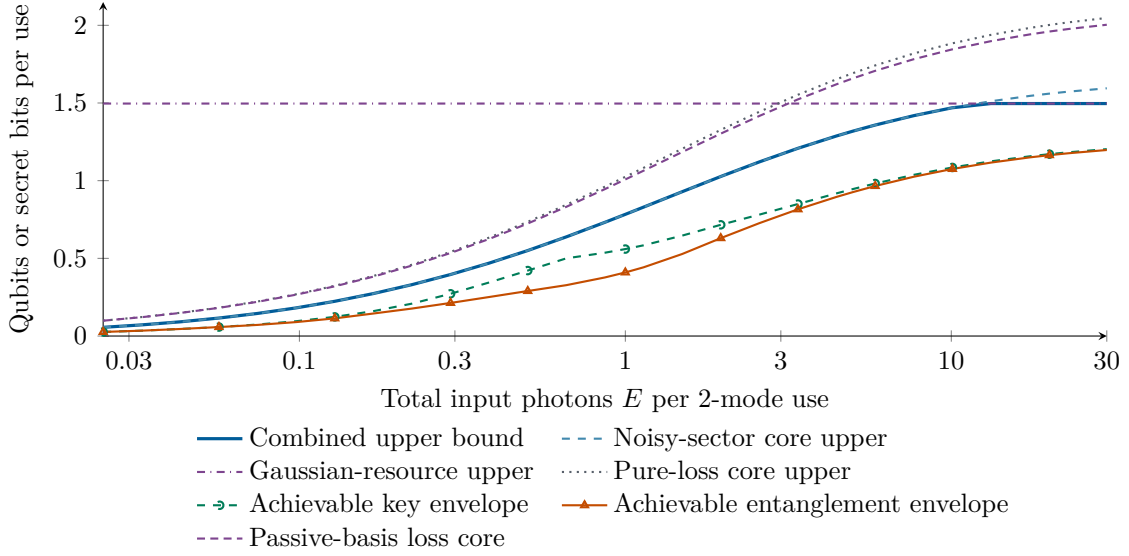
\begin{figure}[t]
\centering
\begin{tikzpicture}
\begin{axis}[
name=main,
width=.90\textwidth,
height=6.0cm,
axis lines=left,
enlarge y limits={upper,value=0.05},
ymin=0,
xlabel={Total input photons $E$ per 2-mode use},
ylabel={Qubits or secret bits per use},
tick label style={font=\small},
label style={font=\small},
legend style={at={(.5,-.24)},anchor=north,draw=none,font=\small,legend columns=2,cells={anchor=west},/tikz/every even column/.append style={column sep=8pt}},
legend cell align=left,
clip=true,
xmode=log,
log basis x=10,
xmin=.025,
xmax=30.0,
xtick={.03,.1,.3,1,3,10,30},
xticklabels={$0.03$,$0.1$,$0.3$,$1$,$3$,$10$,$30$}]
\addplot[boundblue,very thick,no marks] coordinates {
(0.025,0.0560713391876)
(0.0328374977924,0.0718313594948)
(0.0431320504506,0.0914952467265)
(0.056653944458,0.115774546943)
(0.0744149510428,0.145403755159)
(0.0977440316234,0.181105657198)
(0.128386776906,0.223540448227)
(0.168636020129,0.27326149859)
(0.221503397548,0.330676031078)
(0.290944693119,0.396018774315)
(0.38215582872,0.469319478144)
(0.5,0.549105792112)
(0.501961647277,0.550320788856)
(0.659326579373,0.638335923203)
(0.866025403784,0.732107847876)
(1,0.783277531706)
(1.1375242914,0.829764893144)
(1.4941380563,0.928910966004)
(1.96255020501,1.02684135014)
(2,1.03351407599)
(2.57780952098,1.12083913877)
(3.38595257817,1.20849182024)
(4.44744841243,1.2879498794)
(5,1.31924731552)
(5.84172309699,1.35807364416)
(7.67310277204,1.41845398392)
(10,1.46797820081)
(10.0786198135,1.46931136917)
(13.238266235,1.49638211409)
(17.3884615307,1.49638211409)
(20,1.49638211409)
(22.8397426851,1.49638211409)
(30,1.49638211409)
};
\addlegendentry{Combined upper bound}
\addplot[auxblue,dashed,thick,no marks] coordinates {
(0.025,0.0560713391876)
(0.0328374977924,0.0718313594948)
(0.0431320504506,0.0914952467265)
(0.056653944458,0.115774546943)
(0.0744149510428,0.145403755159)
(0.0977440316234,0.181105657198)
(0.128386776906,0.223540448227)
(0.168636020129,0.27326149859)
(0.221503397548,0.330676031078)
(0.290944693119,0.396018774315)
(0.38215582872,0.469319478144)
(0.5,0.549105792112)
(0.501961647277,0.550320788856)
(0.659326579373,0.638335923203)
(0.866025403784,0.732107847876)
(1,0.783277531706)
(1.1375242914,0.829764893144)
(1.4941380563,0.928910966004)
(1.96255020501,1.02684135014)
(2,1.03351407599)
(2.57780952098,1.12083913877)
(3.38595257817,1.20849182024)
(4.44744841243,1.2879498794)
(5,1.31924731552)
(5.84172309699,1.35807364416)
(7.67310277204,1.41845398392)
(10,1.46797820081)
(10.0786198135,1.46931136917)
(13.238266235,1.51133404024)
(17.3884615307,1.54549486678)
(20,1.56032640092)
(22.8397426851,1.57288995308)
(30,1.59461651991)
};
\addlegendentry{Noisy-sector core upper}
\addplot[accentpurple,dashdotted,thick,no marks] coordinates {
(0.025,1.49638211409)
(0.0328374977924,1.49638211409)
(0.0431320504506,1.49638211409)
(0.056653944458,1.49638211409)
(0.0744149510428,1.49638211409)
(0.0977440316234,1.49638211409)
(0.128386776906,1.49638211409)
(0.168636020129,1.49638211409)
(0.221503397548,1.49638211409)
(0.290944693119,1.49638211409)
(0.38215582872,1.49638211409)
(0.5,1.49638211409)
(0.501961647277,1.49638211409)
(0.659326579373,1.49638211409)
(0.866025403784,1.49638211409)
(1,1.49638211409)
(1.1375242914,1.49638211409)
(1.4941380563,1.49638211409)
(1.96255020501,1.49638211409)
(2,1.49638211409)
(2.57780952098,1.49638211409)
(3.38595257817,1.49638211409)
(4.44744841243,1.49638211409)
(5,1.49638211409)
(5.84172309699,1.49638211409)
(7.67310277204,1.49638211409)
(10,1.49638211409)
(10.0786198135,1.49638211409)
(13.238266235,1.49638211409)
(17.3884615307,1.49638211409)
(20,1.49638211409)
(22.8397426851,1.49638211409)
(30,1.49638211409)
};
\addlegendentry{Gaussian-resource upper}
\addplot[mutedgray,dotted,thick,no marks] coordinates {
(0.025,0.0992076223224)
(0.0328374977924,0.122124090571)
(0.0431320504506,0.149757265909)
(0.056653944458,0.182872876586)
(0.0744149510428,0.222286717462)
(0.0977440316234,0.268842568682)
(0.128386776906,0.32336936819)
(0.168636020129,0.386627992157)
(0.221503397548,0.459235726283)
(0.290944693119,0.541576097578)
(0.38215582872,0.633685615448)
(0.5,0.733621256049)
(0.501961647277,0.735140269549)
(0.659326579373,0.844951652262)
(0.866025403784,0.961501191841)
(1,1.02495416402)
(1.1375242914,1.08254365517)
(1.4941380563,1.20530598788)
(1.96255020501,1.32669110457)
(2,1.33497307553)
(2.57780952098,1.44356699304)
(3.38595257817,1.5530908397)
(4.44744841243,1.65299924656)
(5,1.69255251934)
(5.84172309699,1.74180008305)
(7.67310277204,1.81883021406)
(10,1.88246512518)
(10.0786198135,1.88418435083)
(13.238266235,1.93855210874)
(17.3884615307,1.98301760652)
(20,2.00240392177)
(22.8397426851,2.01886478432)
(30,2.04742162016)
};
\addlegendentry{Pure-loss core upper}
\addplot[keygreen,dashed,mark=o,mark repeat=3,mark size=1.4pt ,thick] coordinates {
(0.025,0.026302199306)
(0.0328374977924,0.0345479364658)
(0.0431320504506,0.0453787114971)
(0.056653944458,0.0577828455067)
(0.0744149510428,0.0740226953629)
(0.0977440316234,0.0953537367151)
(0.128386776906,0.123372057647)
(0.168636020129,0.160174119718)
(0.221503397548,0.208513624998)
(0.290944693119,0.272007560914)
(0.38215582872,0.344730844494)
(0.5,0.420222253154)
(0.501961647277,0.421342282503)
(0.659326579373,0.500142062326)
(0.866025403784,0.536345739469)
(1,0.559811635033)
(1.1375242914,0.5838992662)
(1.4941380563,0.646360819361)
(1.96255020501,0.715938792223)
(2,0.720627442037)
(2.57780952098,0.781500536854)
(3.38595257817,0.849709935028)
(4.44744841243,0.918360051349)
(5,0.946411574577)
(5.84172309699,0.981744709723)
(7.67310277204,1.03752398326)
(10,1.0837730571)
(10.0786198135,1.08502258098)
(13.238266235,1.12448845684)
(17.3884615307,1.15666104316)
(20,1.17064567012)
(22.8397426851,1.18249628829)
(30,1.20299801371)
};
\addlegendentry{Achievable key envelope}
\addplot[rateorange,solid,mark=triangle,mark repeat=3,mark size=1.5pt ,thick] coordinates {
(0.025,0.026302199306)
(0.0328374977924,0.0345479364658)
(0.0431320504506,0.0453787114971)
(0.056653944458,0.0577828455067)
(0.0744149510428,0.0716053879062)
(0.0977440316234,0.0897612961274)
(0.128386776906,0.113609079973)
(0.168636020129,0.144933141948)
(0.221503397548,0.176863804111)
(0.290944693119,0.213029791295)
(0.38215582872,0.251174172381)
(0.5,0.289235894216)
(0.501961647277,0.289719467446)
(0.659326579373,0.327505939179)
(0.866025403784,0.377138466463)
(1,0.409308450115)
(1.1375242914,0.442330786668)
(1.4941380563,0.527960893501)
(1.96255020501,0.629976875447)
(2,0.636877305654)
(2.57780952098,0.726620247645)
(3.38595257817,0.815694389832)
(4.44744841243,0.895621472434)
(5,0.92689934782)
(5.84172309699,0.965544448625)
(7.67310277204,1.02530660075)
(10,1.0740252192)
(10.0786198135,1.07533304107)
(13.238266235,1.1164594461)
(17.3884615307,1.1497542218)
(20,1.16417214258)
(22.8397426851,1.17636689981)
(30,1.19741787942)
};
\addlegendentry{Achievable entanglement envelope}
\addplot[accentpurple,thick,densely dashed,no marks] coordinates {
(0,0)
(0.025,0.098623915848)
(0.0328374977924,0.121325419864)
(0.0431320504506,0.148669641856)
(0.056653944458,0.181402972909)
(0.0744149510428,0.220318269307)
(0.0977440316234,0.266230784257)
(0.128386776906,0.319943392653)
(0.168636020129,0.38218928465)
(0.221503397548,0.453561674712)
(0.290944693119,0.534420484953)
(0.38215582872,0.624791217716)
(0.5,0.722756768776)
(0.501961647277,0.724245222502)
(0.659326579373,0.831802095563)
(0.866025403784,0.945865887263)
(1,1.00792681875)
(1.1375242914,1.06422956194)
(1.4941380563,1.18417406868)
(1.96255020501,1.30266990033)
(2,1.31075087506)
(2.57780952098,1.41666170505)
(3.38595257817,1.52338559015)
(4.44744841243,1.62065140549)
(5,1.65913371452)
(5.84172309699,1.70702708462)
(7.67310277204,1.78189094343)
(10,1.8436895486)
(10.0786198135,1.84535857482)
(13.238266235,1.89812062798)
(17.3884615307,1.94124730539)
(20,1.96004241196)
(22.8397426851,1.97599761079)
(30,2.00366867206)
};
\addlegendentry{Passive-basis loss core}
\end{axis}
\end{tikzpicture}
\caption{Correlated-noise two-mode channel in \cref{eq:channel-two-parameters}. The solid blue bound is the original minimum of the feasible noisy-core and Gaussian-resource bounds; the purple dashed curve adds the passive-basis loss-core comparison. The blue curve coincides with the noisy-core bound at low energy and the resource bound at high energy. Lower envelopes combine the evaluated Gaussian and heralded non-Gaussian protocols with energy-preserving time sharing; they are not known to be optimal. The simulator resource has four modes and $\mu=120$, independently of $E$.}
\label{fig:channel-noisy-two}
\end{figure}

\begin{figure}[t]
\centering
\begin{tikzpicture}
\begin{axis}[
name=main,
width=.90\textwidth,
height=6.0cm,
axis lines=left,
enlarge y limits={upper,value=0.05},
ymin=0,
xlabel={Total input photons $E$ per 3-mode use},
ylabel={Qubits or secret bits per use},
tick label style={font=\small},
label style={font=\small},
legend style={at={(.5,-.24)},anchor=north,draw=none,font=\small,legend columns=2,cells={anchor=west},/tikz/every even column/.append style={column sep=8pt}},
legend cell align=left,
clip=true,
xmode=log,
log basis x=10,
xmin=.025,
xmax=30.0,
xtick={.03,.1,.3,1,3,10,30},
xticklabels={$0.03$,$0.1$,$0.3$,$1$,$3$,$10$,$30$}]
\addplot[boundblue,very thick,no marks] coordinates {
(0.025,0.065085962148)
(0.0328374977924,0.0835374243612)
(0.0431320504506,0.10665381973)
(0.056653944458,0.135344467936)
(0.0744149510428,0.170596131146)
(0.0977440316234,0.213461989415)
(0.128386776906,0.265059596369)
(0.168636020129,0.326565800432)
(0.221503397548,0.399157549252)
(0.290944693119,0.483860303723)
(0.38215582872,0.581349778436)
(0.5,0.690115078018)
(0.501961647277,0.691791311954)
(0.659326579373,0.814777957263)
(0.866025403784,0.949371088628)
(1,1.02456603385)
(1.1375242914,1.0940856169)
(1.4941380563,1.24661964792)
(1.96255020501,1.40370413585)
(2,1.41466100778)
(2.57780952098,1.5612982328)
(3.38595257817,1.7150412407)
(4.44744841243,1.8607715297)
(5,1.9199959754)
(5.84172309699,1.99500645803)
(7.67310277204,2.11528423078)
(10,2.21750093501)
(10.0786198135,2.22030152467)
(13.238266235,2.30986075417)
(17.3884615307,2.38467143691)
(20,2.40925025243)
(22.8397426851,2.40925025243)
(30,2.40925025243)
};
\addlegendentry{Combined upper bound}
\addplot[auxblue,dashed,thick,no marks] coordinates {
(0.025,0.065085962148)
(0.0328374977924,0.0835374243612)
(0.0431320504506,0.10665381973)
(0.056653944458,0.135344467936)
(0.0744149510428,0.170596131146)
(0.0977440316234,0.213461989415)
(0.128386776906,0.265059596369)
(0.168636020129,0.326565800432)
(0.221503397548,0.399157549252)
(0.290944693119,0.483860303723)
(0.38215582872,0.581349778436)
(0.5,0.690115078018)
(0.501961647277,0.691791311954)
(0.659326579373,0.814777957263)
(0.866025403784,0.949371088628)
(1,1.02456603385)
(1.1375242914,1.0940856169)
(1.4941380563,1.24661964792)
(1.96255020501,1.40370413585)
(2,1.41466100778)
(2.57780952098,1.5612982328)
(3.38595257817,1.7150412407)
(4.44744841243,1.8607715297)
(5,1.9199959754)
(5.84172309699,1.99500645803)
(7.67310277204,2.11528423078)
(10,2.21750093501)
(10.0786198135,2.22030152467)
(13.238266235,2.30986075417)
(17.3884615307,2.38467143691)
(20,2.41774320482)
(22.8397426851,2.44604860176)
(30,2.49564912224)
};
\addlegendentry{Noisy-sector core upper}
\addplot[accentpurple,dashdotted,thick,no marks] coordinates {
(0.025,2.40925025243)
(0.0328374977924,2.40925025243)
(0.0431320504506,2.40925025243)
(0.056653944458,2.40925025243)
(0.0744149510428,2.40925025243)
(0.0977440316234,2.40925025243)
(0.128386776906,2.40925025243)
(0.168636020129,2.40925025243)
(0.221503397548,2.40925025243)
(0.290944693119,2.40925025243)
(0.38215582872,2.40925025243)
(0.5,2.40925025243)
(0.501961647277,2.40925025243)
(0.659326579373,2.40925025243)
(0.866025403784,2.40925025243)
(1,2.40925025243)
(1.1375242914,2.40925025243)
(1.4941380563,2.40925025243)
(1.96255020501,2.40925025243)
(2,2.40925025243)
(2.57780952098,2.40925025243)
(3.38595257817,2.40925025243)
(4.44744841243,2.40925025243)
(5,2.40925025243)
(5.84172309699,2.40925025243)
(7.67310277204,2.40925025243)
(10,2.40925025243)
(10.0786198135,2.40925025243)
(13.238266235,2.40925025243)
(17.3884615307,2.40925025243)
(20,2.40925025243)
(22.8397426851,2.40925025243)
(30,2.40925025243)
};
\addlegendentry{Gaussian-resource upper}
\addplot[mutedgray,dotted,thick,no marks] coordinates {
(0.025,0.118348465346)
(0.0328374977924,0.146301719059)
(0.0431320504506,0.180254092006)
(0.056653944458,0.221275095435)
(0.0744149510428,0.270544329073)
(0.0977440316234,0.329332211284)
(0.128386776906,0.398963760999)
(0.168636020129,0.480770325371)
(0.221503397548,0.576007972282)
(0.290944693119,0.685760753033)
(0.38215582872,0.810814046223)
(0.5,0.949374107465)
(0.501961647277,0.951504502176)
(0.659326579373,1.10755364306)
(0.866025403784,1.27790032496)
(1,1.37288461129)
(1.1375242914,1.4605679658)
(1.4941380563,1.65259650288)
(1.96255020501,1.85010228023)
(2,1.86388129973)
(2.57780952098,2.04847719165)
(3.38595257817,2.24276442329)
(4.44744841243,2.42811641296)
(5,2.50388694105)
(5.84172309699,2.60030329083)
(7.67310277204,2.75611406277)
(10,2.88991892009)
(10.0786198135,2.8936031279)
(13.238266235,3.0121088278)
(17.3884615307,3.11210193476)
(20,3.15662646475)
(22.8397426851,3.19490039394)
(30,3.2623585213)
};
\addlegendentry{Pure-loss core upper}
\addplot[keygreen,dashed,mark=o,mark repeat=3,mark size=1.4pt ,thick] coordinates {
(0.025,0.0351336379582)
(0.0328374977924,0.0461480303556)
(0.0431320504506,0.060615433797)
(0.056653944458,0.0772245652037)
(0.0744149510428,0.0967793962334)
(0.0977440316234,0.122464665064)
(0.128386776906,0.156202263405)
(0.168636020129,0.200516595847)
(0.221503397548,0.258723467596)
(0.290944693119,0.335178188498)
(0.38215582872,0.42383993334)
(0.5,0.51661544413)
(0.501961647277,0.517997897355)
(0.659326579373,0.615727182651)
(0.866025403784,0.714684833232)
(1,0.766506080793)
(1.1375242914,0.812294972587)
(1.4941380563,0.90599723396)
(1.96255020501,0.97725024217)
(2,0.982946957247)
(2.57780952098,1.07084106216)
(3.38595257817,1.19377259596)
(4.44744841243,1.31633866509)
(5,1.36889034127)
(5.84172309699,1.43742842909)
(7.67310277204,1.55010404067)
(10,1.64730423224)
(10.0786198135,1.64997603124)
(13.238266235,1.73564601309)
(17.3884615307,1.80732752617)
(20,1.8390197727)
(22.8397426851,1.86613544307)
(30,1.91362212642)
};
\addlegendentry{Achievable key envelope}
\addplot[rateorange,solid,mark=triangle,mark repeat=3,mark size=1.5pt ,thick] coordinates {
(0.025,0.0351336379582)
(0.0328374977924,0.0461480303556)
(0.0431320504506,0.060615433797)
(0.056653944458,0.0772245652037)
(0.0744149510428,0.0957952733939)
(0.0977440316234,0.120187896962)
(0.128386776906,0.152227605865)
(0.168636020129,0.194311760679)
(0.221503397548,0.237860738249)
(0.290944693119,0.287710960718)
(0.38215582872,0.341373901406)
(0.5,0.401189119521)
(0.501961647277,0.402119856451)
(0.659326579373,0.468789808166)
(0.866025403784,0.537950789824)
(1,0.574790258528)
(1.1375242914,0.609716971312)
(1.4941380563,0.70028530957)
(1.96255020501,0.819246813874)
(2,0.828757848035)
(2.57780952098,0.975502739272)
(3.38595257817,1.13394831956)
(4.44744841243,1.28226654736)
(5,1.34205814296)
(5.84172309699,1.41740441111)
(7.67310277204,1.53736242645)
(10,1.6385153511)
(10.0786198135,1.6412751114)
(13.238266235,1.72931074718)
(17.3884615307,1.8024444737)
(20,1.83466136601)
(22.8397426851,1.86217902066)
(30,1.91027850699)
};
\addlegendentry{Achievable entanglement envelope}
\addplot[accentpurple,thick,densely dashed,no marks] coordinates {
(0,0)
(0.025,0.117242821242)
(0.0328374977924,0.144862348524)
(0.0431320504506,0.178387977346)
(0.056653944458,0.218866061592)
(0.0744149510428,0.267449278235)
(0.0977440316234,0.325376756826)
(0.128386776906,0.393938938165)
(0.168636020129,0.474426434134)
(0.221503397548,0.568055357646)
(0.290944693119,0.675867213508)
(0.38215582872,0.798614014249)
(0.5,0.93451393997)
(0.501961647277,0.936602643151)
(0.659326579373,1.08953791356)
(0.866025403784,1.25635593084)
(1,1.34931740736)
(1.1375242914,1.43509901224)
(1.4941380563,1.62284823948)
(1.96255020501,1.81578417838)
(2,1.82923769569)
(2.57780952098,2.00939107793)
(3.38595257817,2.19882044487)
(4.44744841243,2.37935304062)
(5,2.45309804865)
(5.84172309699,2.54688398898)
(7.67310277204,2.69832309782)
(10,2.82823855692)
(10.0786198135,2.83181355611)
(13.238266235,2.94676078719)
(17.3884615307,3.04366329792)
(20,3.08678548869)
(22.8397426851,3.12383931096)
(30,3.1891162061)
};
\addlegendentry{Passive-basis loss core}
\end{axis}
\end{tikzpicture}
\caption{Correlated-noise three-mode channel in \cref{eq:channel-three-parameters}. Curve conventions and evaluated protocol families are as in \cref{fig:channel-noisy-two}. The simulator resource now has six modes and $\mu=120$, independently of the input energy $E$. Both correlations between modes and the total energy allocation enter the bounds.}
\label{fig:channel-noisy-three}
\end{figure}
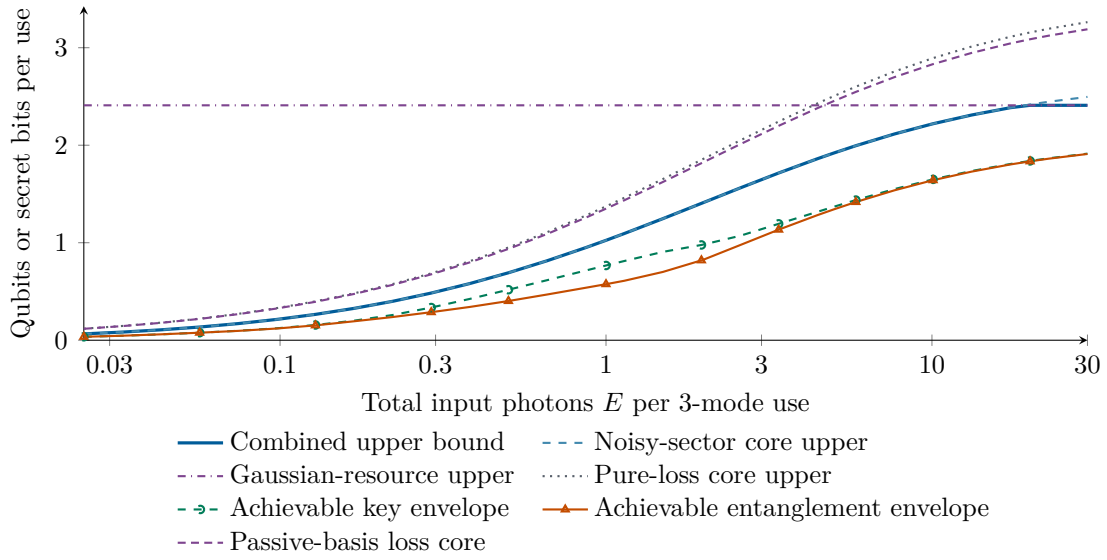

\Cref{fig:channel-noisy-two,fig:channel-noisy-three} show the complementary
roles of the two converses. The sector curves retain the small-energy
behavior. At larger energy, the Gaussian resource calculation gives a
lower ceiling for these examples. Taking their minimum supplies the
stronger displayed upper bound. No equality of the quantum and secret-key
capacities is asserted for these noisy channels, even though their
Gaussian upper bounds coincide by symmetry.

\clearpage
\section{Controlled computation of the unrestricted quantities}
\label{sec:unrestricted}
\Cref{sec:symmetric} shows that Gaussian restriction cannot be used as an
exact replacement for the unrestricted problems. This section gives a
different route: finite-dimensional optimization after local photon
cutoffs, with a quantitative error estimate. The result applies to an
arbitrary finite-energy state, not only a Gaussian state. The cutoff is a
local measurement with a success flag: we keep track of both its success
probability and the possible entanglement outside the retained subspace. Energy-constrained
continuity estimates for these entanglement quantities are developed more
generally in \cite[Section~5]{LamiShirokov2023}. Here the local measurement
outcomes give a direct, success-weighted bound tailored to the cutoff.

\subsection{A two-sided photon-cutoff estimate}
Let $\hat N_A$ and $\hat N_B$ be the total photon number operators in the
$m$ and $n$ local modes. Set
\begin{equation}\label{eq:energies}
 E_A\coloneqq\Tr[\hat N_A\rho_{AB}]<\infty,
 \qquad E_B\coloneqq\Tr[\hat N_B\rho_{AB}]<\infty.
\end{equation}
For nonnegative integers $K_A,K_B$, let $P_A$ and $P_B$ be the spectral
projectors onto total photon numbers at most $K_A$ and $K_B$, respectively.
Define
\begin{equation}\label{eq:cutoff-state}
 P\coloneqq P_A\otimes P_B,\qquad p\coloneqq\Tr[P\rho],\qquad q\coloneqq1-p,
 \qquad \rho_{00}\coloneqq P\rho P/p\quad(p>0).
\end{equation}
The subscript $00$ denotes success of both local cutoff tests. It is an
outcome label, not a pair of photon occupations.

\begin{samepage}
\begin{theorem}[Finite-energy cutoff bound]\label{thm:cutoff}
Let $\rho_{AB}$ be a state on $m$ modes of $A$ and $n$ modes of $B$,
with finite mean photon numbers $E_A,E_B$ as in \eqref{eq:energies}.
For the local photon cutoffs in \eqref{eq:cutoff-state}, let $p>0$ be their
joint success probability, $q=1-p$ the failure probability, and $\rho_{00}$
the normalized successful state. For either $E=E_R$ or $E=R$,
\begin{equation}\label{eq:cutoff-main}
 pE(A;B)_{\rho_{00}}\le E(A;B)_\rho
 \le pE(A;B)_{\rho_{00}}+\epsilon(q),
\end{equation}
where, in bits,
\begin{equation}\label{eq:cutoff-error}
 \epsilon(q)\coloneqq h_2(q)+q\log_2 3
 +q\min\left\{m g\!\left(\frac{E_A}{mq}\right),
              n g\!\left(\frac{E_B}{nq}\right)\right\}.
\end{equation}
The error is defined to be zero at $q=0$ and tends to zero as $q\downarrow0$.
Here $h_2$ is the binary entropy. The successful local Hilbert-space
dimensions are $\binom{K_A+m}{m}$ and $\binom{K_B+n}{n}$.
\end{theorem}
\end{samepage}

\begin{proof}
There are two effects to control: the entanglement in the discarded sectors
and the coherences destroyed by measuring the sectors. Their energy and probability bound the first effect; the entropy of four
classical outcomes bounds the second effect.

Let $P_{A,0}\coloneqq P_A$, $P_{A,1}\coloneqq I-P_A$, and similarly for $B$. Define
$P_{ij}\coloneqq P_{A,i}\otimes P_{B,j}$ and the local pinching channel
\begin{equation}\label{eq:pinching}
 \Pi(X)\coloneqq\sum_{i,j=0}^1P_{ij}XP_{ij},\qquad
 p_{ij}\coloneqq\Tr[P_{ij}\rho],\qquad
 \rho_{ij}\coloneqq P_{ij}\rho P_{ij}/p_{ij}
\end{equation}
for positive-probability outcomes. These outcomes are locally
distinguishable. Selective monotonicity gives
\begin{equation}\label{eq:selective}
 E(A;B)_\rho\geq \sum_{ij}p_{ij}E(A;B)_{\rho_{ij}}
                    \geq  pE(A;B)_{\rho_{00}}.
\end{equation}
For REE this follows from data processing and preservation of separability.
For Rains entanglement, selective monotonicity is established in
\cite{EisertWilde2022}. A direct proof for this particular pinching, including
subnormalized Rains references, is given in \cref{app:cutoff-details}.

For the reverse estimate, choose nearly optimal reference operators in
each sector, supported on that sector. Such support may be imposed by
local compression, which preserves feasibility and does not increase the
relative entropy. A block-diagonal direct sum, weighted by $p_{ij}$, is
then a feasible reference for the whole state: for REE it is separable and
normalized; for Rains its partial-transpose trace norm is at most
$\sum_{ij}p_{ij}=1$. The logarithm of the weighted direct-sum reference is block diagonal.
Taking the approximation error to zero gives
\begin{equation}\label{eq:pinching-upper}
 E(A;B)_\rho
 \le\sum_{ij}p_{ij}E(A;B)_{\rho_{ij}}
                       +S(\Pi(\rho))-S(\rho).
\end{equation}
The entropy increase of a projective pinching is at most the Shannon
entropy of its outcome distribution. One way to see this is to coherently
record the four projectors in an auxiliary register and apply the entropy
triangle inequality; the register marginal has distribution $p_{ij}$.
Since the success probability is $p$ and there are three other outcomes,
\begin{equation}\label{eq:pinching-entropy}
 S(\Pi(\rho))-S(\rho)\le H(\{p_{ij}\})
                       \le h_2(q)+q\log_2 3.
\end{equation}
All entropies here are finite: pinching commutes with the number operators
and preserves the finite energy.

It remains to bound the bad-outcome entanglement. For a finite-entropy
bipartite state $\omega$,
\begin{equation}\label{eq:marginal-upper}
 R(A;B)_\omega\le E_R(A;B)_\omega
                  \le\min\left\{S(\omega_A),S(\omega_B)\right\}.
\end{equation}
For the last inequality, dephase $A$ in an eigenbasis of $\omega_A$.
The resulting classical--quantum state is separable, and the
relative-entropy cost is the entropy increase, at most $S(\omega_A)$.
Interchanging parties gives the other bound. Finite-entropy limits justify
the countable eigenbasis argument.

An $m$-mode state with total mean photon number $e$ has entropy at most
$m g(e/m)$. Compare it in relative entropy with a product of $m$ identical
thermal states of that mean, or maximize the sum of single-mode entropies
using concavity of $g$. Let $e_{A,ij}\coloneqq\Tr[\hat N_A\rho_{ij}]$. Because the
projectors commute with $\hat N_A$,
\begin{equation}\label{eq:bad-energy}
 \sum_{(i,j)\ne(0,0)}p_{ij}e_{A,ij}\le E_A.
\end{equation}
Applying the entropy bound to each bad outcome, and then concavity and
monotonicity of $g$, gives
\begin{equation}\label{eq:bad-ent}
 \sum_{(i,j)\ne(0,0)}p_{ij}E(A;B)_{\rho_{ij}}
 \le q\,m g\!\left(\frac{E_A}{mq}\right).
\end{equation}
The same reasoning on $B$ gives the other term in the minimum.
Combining \cref{eq:pinching-upper,eq:pinching-entropy,eq:bad-ent} proves the
upper bound. For fixed finite energy, $qg(E/(mq))$ tends to zero (it is of
order $q\log(1/q)$ when $E>0$). Finally, counting occupation vectors of
$m$ modes with sum at most $K_A$ gives $\binom{K_A+m}{m}$, and similarly
for $B$.
\end{proof}

The error bound concerns the true projected state $P\rho P/p$. It does not
replace $\rho$ by the exponential of a truncated Hamiltonian. Those two
constructions are generally different.

\subsection{Gaussian tail estimates and finite-dimensional optimization}
The cutoff theorem becomes a computational procedure once its failure
probability and its finite matrix entries can be evaluated. For an arbitrary
finite-energy state, Markov's inequality and a union bound
give
\begin{equation}\label{eq:markov}
 q\le\frac{E_A}{K_A+1}+\frac{E_B}{K_B+1}.
\end{equation}
This already ensures convergence of \cref{eq:cutoff-main}. For Gaussian
inputs, local Williamson unitaries give substantially better estimates.
After these local unitaries, each local marginal is a product of thermal
states. Its total photon number has generating function
\begin{equation}\label{eq:generating}
 \mathbb E[z^{\hat N_A}]
 =\prod_{j=1}^m[1-(z-1)\bar n_{A,j}]^{-1},
 \qquad \bar n_{A,j}\coloneqq\frac{\nu_j(V_A)-1}{2}.
\end{equation}
For $z>1$ in the convergence interval, multiplying by $z^{-(K_A+1)}$
bounds the local tail probability. The analogous bound on $B$ and a union
bound give an exponentially decreasing bound on $q$. Energies in
\cref{eq:cutoff-error} are then those of the locally transformed state;
entanglement is unchanged by these local unitaries. If only an upper bound
$\delta\le1/2$ on $q$ is known, each term in \cref{eq:cutoff-error} is
nondecreasing on that interval, so $\epsilon(q)\le\epsilon(\delta)$.

The matrix elements of the true cutoff state are computable from the
Gaussian coherent-state generating function. \Cref{app:fock} gives the
formula, including the factors associated with our covariance convention.
For its finite-dimensional Rains entanglement, the optimization is
\begin{equation}\label{eq:finite-Rains}
 \begin{aligned}
 \text{minimize}\quad &D(\rho_{00}\Vert\tau)\\
 \text{subject to}\quad&\tau\succeq0,\quad X,Y\succeq0,\\
 &\tau^{\Gamma_B}=X-Y,\qquad\Tr(X+Y)\le1.
 \end{aligned}
\end{equation}
The constraints are semidefinite. The relative-entropy objective is convex
but is not itself an exact finite linear SDP objective. Semidefinite
matrix-logarithm approximations provide controlled finite approximations
\cite{FawziSaundersonParrilo2019}. A small maximally mixed reference component
keeps all logarithms uniformly away from the singular boundary, as detailed
in \cref{app:finite-optimization}.

For finite-dimensional REE, the complete symmetric-extension hierarchy of
Doherty, Parrilo, and Spedalieri gives convergent lower bounds
\cite{DohertyParriloSpedalieri2004}. Convex hulls of successively refined finite
nets of product pure states give convergent upper bounds. Compactness and
faithful reference regularization justify convergence of the objective
values, not only convergence of the sets; we include that argument in
\cref{app:finite-optimization}. If $L\le E(\rho_{00})\le U$ are finite-dimensional
bounds, the resulting bosonic interval is
\begin{equation}\label{eq:final-interval}
 pL\le E(\rho)\le pU+\epsilon(q).
\end{equation}
Thus the cutoff error and the optimization error remain distinct and
explicit.

The successful cutoff state $\rho_{00}$ is generally non-Gaussian.
Its finite-dimensional optimizations permit arbitrary reference operators
of the required type; they do not reimpose the Gaussian restriction.
This procedure is a convergent computational characterization, not an
efficient covariance formula for the unrestricted measures. The cutoff
dimensions and the separability approximations can be large. The Gaussian
results of earlier sections therefore retain a different computational
role even when the unrestricted algorithm is available.

\section{Conclusion}\label{sec:conclusion}
\subsection{Summary}
Combining the Gaussian Gibbs formula \eqref{eq:gibbs-f} with the known
Hamiltonian inverse integral \eqref{eq:H-integral} yields the positive
relative-entropy representation \eqref{eq:resolvent} (\cref{thm:resolvent}).
The uncertainty relation then becomes a tool for optimization, not only a
physicality constraint on the covariance matrix. It gives a
reference-noise inequality whose determinant cost matches the Gaussian
partial-transpose trace-norm penalty, reducing Gaussian Rains entanglement
to Gaussian PPT relative entropy (\cref{thm:rains}). Its geodesic curvature gives unique
optimal covariance matrices, Gaussian additivity, symmetry reductions, and globally
convergent covariance-space algorithms. The improper-integral formulation
includes nonfaithful states rather than excluding them from attainment.

The optimization analysis gives more than eventual convergence. An explicit
strong-geodesic-convexity constant yields geometric convergence of the
convex-majorization method, uniformly in its faithful regularization
parameter. The proof accounts for approximate subproblem minimization and
also strengthens projected-gradient rates and stopping certificates.
Geometrically graded positive quadrature and bounded semidefinite
optimization give a finite-precision implementation with polynomial cost
in the requested accuracy bits for fixed rational input data. Covariance
size enters explicitly in the prefactor, so this is not a uniform
polynomial-time theorem in the input encoding size. These guarantees apply
to the Gaussian reference problem, not to the outer communication-channel
optimizations or the unrestricted entanglement hierarchy.

These structural statements support operational upper bounds but do not
remove the Gaussian restriction from the underlying entanglement measures.
Rains entanglement bounds distillable entanglement; separable-reference REE
also bounds distillable key. Opposite-phase symmetry gives a useful
multimode family in which the Gaussian bounds coincide. The two- and
three-mode state comparisons show both exactly known operational rates
and noisy cases where explicit non-Gaussian heralding improves on raw
hashing. A faithful-state separation (\cref{thm:gap}) and the controlled unrestricted
photon-cutoff theorem (\cref{thm:cutoff}) identify the distinction between covariance
optimization and unrestricted entanglement computation.

For channels, mode-resolved sector teleportation combines the recently
established scalar energy gains without assuming independent adaptive
inputs. Linear matrix inequalities then identify scalar Gaussian cores
followed by arbitrary Gaussian processing at the receiver. This gives a
finite energy-dependent upper bound for every finite-mode Gaussian
channel. The loss-only specialization is a convex optimization over the
core parameters (\cref{cor:convex-loss-core}). More strongly,
\cref{thm:passive-core} includes the passive basis in that convex
optimization. The greatest feasible core \eqref{eq:Lmax} solves the
real phase-insensitive case, and the scalar predecessor
\eqref{eq:predecessor-parameters} gives an additional operational bound
for the three-mode preparation (\cref{prop:scalar-predecessor}). A complementary exact Gaussian-resource construction
uses the covariance entanglement optimization directly. In the correlated
examples, the sector bound is stronger at lower transmitted energies,
whereas the optimized Gaussian resource becomes useful at higher energies.
The two constructions have different resources and different constraints;
only their upper bounds, not their states or energy parameters, are combined.

\subsection{Future directions}
The first algorithmic question is a practical, fully certified implementation
of the finite-precision method analyzed here. Sharper bounds on optimal
reference covariance matrices and input-adapted curvature estimates could
substantially improve the conservative worst-case constants. In particular,
a uniform polynomial bound in the encoded input size remains to be
established; the present polynomial accuracy dependence does not settle
that separate question. The new numerical references are explicitly
feasible upper-bound evaluations, but their objective values are not
interval-certified global optima. Efficient outer optimization over exact
Gaussian simulator resources, or over the general noisy-core family, is
another problem: the inner reference convexity does not automatically
extend to these outer variables.

A sharper energy-dependent converse for a general phase-sensitive channel
might retain correlations beyond a scalar-core factorization. A direct
sector argument would have to treat the channel's actual representation
and the coherences between its sectors; covariance formulas cannot simply
be applied to the non-Gaussian sector resources. Establishing such a
reduction is different from choosing an input-matched finite-energy Choi
state, which generally does not exactly simulate the channel.

On the achievable side, larger non-Gaussian filter families, improved
recurrence searches, and genuinely joint multimode Gaussian measurements
may narrow the remaining gaps. The correlated examples do not establish
optimal noisy-channel capacities or universal superiority over every
known converse. They provide reproducible comparisons among precise
operational constructions. Finally, the Gaussian-restriction gap motivates
systematic non-Gaussian comparison states and quantitative estimates using
the cutoff theorem, rather than an unproved Gaussian-optimality assumption.

\section*{Acknowledgements}
We thank Ludovico Lami and Hami Mehrabi for helpful discussions.

\section*{Use of artificial intelligence}
OpenAI's ChatGPT was used extensively in developing and revising this
manuscript. The assistance was substantive rather than limited to
copyediting: it included proposing and developing mathematical arguments,
examining counterexamples and boundary cases, deriving optimization and
communication bounds, searching and checking literature, preparing
numerical tests and illustrations, and writing and revising the LaTeX and
BibTeX sources. The research questions, author-supplied notes, local
manuscript edits, and detailed comments guided this work.

The assistance also included repeated internal review-and-revision cycles
with separate mathematical, expository, and presentation checks. These
were role-separated self-audits within the same AI-assisted workflow, not
independent external peer review, human referee endorsements, or formal
verification. Numerical tests supplement the analytical proofs and do not
establish their universal claims. The authors are responsible for assessing
and approving the mathematical arguments, citations, interpretations, and
final presentation; AI assistance does not replace independent expert
assessment or journal peer review.

\clearpage
\begingroup
\small
\setlength{\parskip}{0pt}
\interlinepenalty=10000
\bibliographystyle{alphaurl}
\bibliography{gaussian_entanglement}
\endgroup

\clearpage
\appendix
\crefalias{section}{appendix}
\crefalias{subsection}{appendix}
\crefname{appendix}{Appendix}{Appendices}
\Crefname{appendix}{Appendix}{Appendices}

\section{Quantitative convergence and finite-precision complexity}
\label{app:runtime}
This appendix supplies the optimization details behind
\cref{sec:algorithms}. All matrix dimensions are finite. We first turn the
strict curvature proved earlier into a uniform lower bound, then compare
that curvature with the error in the convex surrogate. Finally, we
control numerical integration and check the hypotheses needed to solve
the resulting semidefinite programs in bounded bit complexity. The
arguments concern Gaussian \emph{reference-state} optimization only.

\subsection{An explicit strong-geodesic-convexity constant}
\label{app:quantitative-curvature}
For positive definite real matrices, the affine-invariant distance is
\begin{equation}\label{eq:affine-distance}
 d_{\rm aff}(W_0,W_1)\coloneqq
 \norm{\ln(W_0^{-1/2}W_1W_0^{-1/2})}_{\mathrm F}.
\end{equation}
It is the length distance for the metric
$\inner{X}{Y}_W\coloneqq\Tr(W^{-1}XW^{-1}Y)$
\cite{Bhatia2007,LawsonLim2021}. A differentiable function $f$ is strongly
geodesically convex with constant $\mu>0$ on a set containing its
geometric-mean paths if
\begin{equation}\label{eq:strong-geodesic}
 \begin{split}
 f(W_0\#_sW_1)\le{}&(1-s)f(W_0)+sf(W_1)\\
 &-\frac\mu2s(1-s)d_{\rm aff}(W_0,W_1)^2
 \qquad(0\le s\le1).
 \end{split}
\end{equation}
Unlike strict convexity alone, this inequality fixes a quantitative
amount by which the function lies below each chord. Strong geodesic
convexity and its use in rate analysis are standard notions
\cite{ZhangSra2016}; the constant below is derived for the present
relative-entropy objective.

\begin{samepage}
\begin{lemma}[Uniform Gaussian curvature]\label{lem:quantitative-curvature}
Let $V$ be a quantum covariance matrix and let $\Lambda\ge\norm V$, with
$\Lambda\ge1$. On a geodesically convex set of faithful reference matrices
satisfying $M^{-1}I\preceq W\preceq MI$, with $M\ge1$, the function $f_V$
satisfies \eqref{eq:strong-geodesic} with
$\mu=1/(64M^2\Lambda^2)$. A sharper valid curvature constant is
\begin{equation}\label{eq:sharp-curvature}
 \widetilde\mu\coloneqq
 \frac{\lambda_{\min}(V)^{3/2}}
      {32M^2\sqrt{\norm{V+i\Omega}}}.
\end{equation}
The sharper constant in \eqref{eq:sharp-curvature} is a curvature bound.
We retain the smaller $\mu$ in the contraction theorem, because the
surrogate comparison also requires control of the chosen step along the
geodesic; a larger curvature constant cannot be substituted there without
checking that comparison again.

Neither estimate requires $V+i\Omega$ to be positive definite, and
neither depends on a lower bound for $\nu_j(W)-1$.
\end{lemma}
\end{samepage}

\begin{proof}
The integrand of the geodesic Hessian in
\cref{eq:geodesic-hessian} is nonnegative for every $t$. We obtain a uniform
positive contribution by using only a short interval near $t=0$, which
avoids the singular endpoint at $t=1$.
Fix a point $W$ and tangent $X=\dot W$ on a reference geodesic, and define
\begin{equation}\label{eq:normalized-hessian-data}
 \begin{aligned}
 A&\coloneqq iW^{-1/2}\Omega W^{-1/2},&
 D&\coloneqq W^{-1/2}VW^{-1/2},& C&\coloneqq D+A,\\
 X_0&\coloneqq W^{-1/2}XW^{-1/2},&
 P_t&\coloneqq(I+tA)^{-1}.&&
 \end{aligned}
\end{equation}
The matrix $A$ is Hermitian and purely imaginary with antisymmetric
transpose; physicality implies $\norm A\le1$. The matrix $C\succeq0$ is
the congruence of $V+i\Omega$, while $X_0$ is real symmetric. Set
\begin{equation}\label{eq:curvature-ab}
 a\coloneqq\lambda_{\min}(V)/M,\qquad
 b_0\coloneqq M\norm{V+i\Omega},\qquad
 t_0\coloneqq\tfrac14\sqrt{a/b_0}.
\end{equation}
Then $D\succeq aI$ and $\norm C\le b_0$. Also $a\le b_0$: since
$\Tr A=0$, we have $\Tr C=\Tr D\ge2Na$, so $\norm C\ge a$.
In particular, $0<t_0\le1/4$.

Cyclically rearranging the Hessian integrand, using
$Q_t=W^{-1/2}P_tW^{-1/2}$, gives
\begin{equation}\label{eq:normalized-hessian}
 \Tr[C_VQ_tX(2Q_t-W^{-1})XQ_t]
   =\Tr[CP_tX_0(2P_t-I)X_0P_t].
\end{equation}
For $0\le t\le t_0$, the spectrum of $A$ is in $[-1,1]$, hence
\begin{equation}\label{eq:P-bounds}
 2P_t-I\succeq\frac35I,\qquad
 \norm{P_t-I}\le\frac{t}{1-t}\le\frac43t.
\end{equation}
Furthermore,
\begin{equation}\label{eq:real-direction-bound}
 \norm{X_0C^{1/2}}_{\mathrm F}^2
 =\Tr(CX_0^2)=\Tr(DX_0^2)
 \ge a\norm{X_0}_{\mathrm F}^2.
\end{equation}
The term $\Tr(AX_0^2)$ vanishes because $X_0^2$ is real symmetric and
$A^{\top}=-A$. Therefore the triangle inequality gives
\begin{equation}\label{eq:perturbed-direction-bound}
 \begin{split}
 \norm{X_0P_tC^{1/2}}_{\mathrm F}
 &\ge\norm{X_0C^{1/2}}_{\mathrm F}
       -\norm{X_0(P_t-I)C^{1/2}}_{\mathrm F}\\
 &\ge\left(\sqrt a-\tfrac43t\sqrt{b_0}\right)
                   \norm{X_0}_{\mathrm F}
 \ge\tfrac23\sqrt a\norm{X_0}_{\mathrm F}.
 \end{split}
\end{equation}
Combining \cref{eq:normalized-hessian,eq:P-bounds,eq:perturbed-direction-bound}
yields a lower bound $(4a/15)\norm{X_0}_{\mathrm F}^2$ on the Hessian
integrand. Including its factor $1/2$ and integrating over $[0,t_0]$ gives
\begin{equation}\label{eq:curvature-integrated}
 \frac{d^2}{ds^2}f_V(W(s))
 \ge\frac{a^{3/2}}{30\sqrt{b_0}}
                 \norm{W^{-1/2}\dot W W^{-1/2}}_{\mathrm F}^2.
\end{equation}
The slightly smaller constant in \cref{eq:sharp-curvature} is thus valid.
The inverse uncertainty inequality gives
\begin{equation}\label{eq:V-norm-lower}
 V\succeq\Omega V^{-1}\Omega^{\top}\succeq\Lambda^{-1}I,\qquad
 \norm{V+i\Omega}\le\Lambda+1\le2\Lambda.
\end{equation}
Substitution shows that $\widetilde\mu\ge1/(64M^2\Lambda^2)$.

Finally, write a geometric-mean path as $W(s)=B e^{sH}B$, where
$B=W_0^{1/2}$ and $H=\ln(W_0^{-1/2}W_1W_0^{-1/2})$.
Cyclicity of the trace gives
\begin{equation}\label{eq:constant-speed}
 \norm{W^{-1/2}\dot W W^{-1/2}}_{\mathrm F}^2
 =\Tr(W^{-1}\dot W W^{-1}\dot W)=\Tr H^2
 =d_{\rm aff}(W_0,W_1)^2.
\end{equation}
Subtracting $(\mu/2)s^2d_{\rm aff}^2$ from the function along the path
leaves a convex scalar function. Its chord inequality is
\cref{eq:strong-geodesic}. Pure input states cause no difficulty:
\cref{eq:real-direction-bound} uses $V\succ0$, not $V+i\Omega\succ0$.
\end{proof}

\subsection{Contraction, approximate minimization, and gradient certificates}
\label{app:majorization-rate}
We first prove the rate of the convex-majorization method. We then use the
same curvature estimate to strengthen the residual certificate and the
projected-gradient rate. Throughout, $\cK$ and the constants are those
specified in \cref{sec:algorithms}.

Let $W_*$ be the unique optimizer on $\cK$ and put
$Y_s\coloneqq W_k\#_sW_*$. The algorithm does not require $W_*$; $Y_s$
is only a feasible comparison point. With
$H\coloneqq\ln(W_k^{-1/2}W_*W_k^{-1/2})$, the surrogate error equals
\begin{equation}\label{eq:surrogate-geodesic-error}
 \mathcal B(Y_s,W_k)=\tfrac12\Tr(e^{sH}-I-sH).
\end{equation}
For real $x$, the identity
$e^x-1-x=x^2\int_0^1(1-u)e^{ux}du$ implies
$0\le e^x-1-x\le x^2e^{|x|}/2$.
Since $M^{-1}I\preceq W_k,W_*\preceq MI$, the eigenvalues of
$W_k^{-1/2}W_*W_k^{-1/2}$ lie in $[M^{-2},M^2]$, so
\begin{equation}\label{eq:surrogate-geodesic-upper}
 \mathcal B(Y_s,W_k)
 \le\frac{s^2e^{2s\ln M}}4d_{\rm aff}(W_k,W_*)^2.
\end{equation}
Strong geodesic convexity gives
\begin{equation}\label{eq:curvature-comparison}
 f_V(Y_s)\le f_V(W_k)-s e_k
     -\frac\mu2s(1-s)d_{\rm aff}(W_k,W_*)^2.
\end{equation}
Take $s=\mu$. The numerical inequality needed to absorb the positive
surrogate error is $e^{2\mu\ln M}\le2(1-\mu)$.
Indeed, $\mu\le1/64$ and
$2\mu\ln M\le1/(64e)$, since
$\sup_{M\ge1}(\ln M)/M^2=1/(2e)$. Thus the inequality holds with slack.
Combining \cref{eq:surrogate-geodesic-upper,eq:curvature-comparison} yields
\begin{equation}\label{eq:surrogate-comparison}
 \min_{W\in\cK}q_k(W)\le q_k(Y_\mu)
                   \le f_V(W_k)-\mu e_k.
\end{equation}
Since $f_V\le q_k$, exact minimization proves \cref{eq:MM-rate} and
an inner error $\zeta_k$ proves \cref{eq:inexact-rate}. More explicitly,
\begin{equation}\label{eq:accumulated-inner-errors}
 e_k\le(1-\mu)^ke_0+
            \sum_{j=0}^{k-1}(1-\mu)^{k-1-j}\zeta_j.
\end{equation}
For $\zeta_j\le\mu\delta/2$, the sum is at most $\delta/2$.
Exact iterations have $e_k\to0$, so compactness and uniqueness imply
$W_k\to W_*$. For completeness, the Hessian of the convex inverse part is
\begin{equation}\label{eq:phi-hessian}
 \nabla^2\Phi_V(W)[X,X]
   =\int_0^1\Tr(C_VQ_tXQ_tXQ_t)dt>0\qquad(X\ne0).
\end{equation}
The integrand is nonnegative because
$Q_tXQ_tXQ_t=Q_t^{1/2}(Q_t^{1/2}XQ_t^{1/2})^2Q_t^{1/2}\succeq0$.
At $t=0$ its trace is strictly positive by $V\succ0$ and the reality of
$X,W$. Continuity gives strict positivity of the integral. Each surrogate
therefore has a unique minimizer on the compact ordinary-convex set.

We next establish the residual estimates. The tangent from $W$ toward
$Y$ is
\begin{equation}\label{eq:geodesic-stationary}
 \Log_W(Y)\coloneqq W^{1/2}
       \ln(W^{-1/2}YW^{-1/2})W^{1/2}.
\end{equation}
Differentiating the chord inequality \cref{eq:strong-geodesic} at its
initial endpoint gives
\begin{equation}\label{eq:strong-first-order}
 f_V(Y)\ge f_V(W)+\inner{G_V(W)}{\Log_W(Y)}
                         +\tfrac\mu2d_{\rm aff}(W,Y)^2.
\end{equation}
An outward normal $Z$ in \cref{eq:normal} satisfies
$\inner{Z}{\Log_W(Y)}\le0$, by applying its defining inequality to
$W\#_sY$ and taking $s\downarrow0$. Also
$\norm{\Log_W(Y)}_{\mathrm F}\le M d_{\rm aff}(W,Y)$.
Writing $\eta=\norm{G_V(W)+Z}_{\mathrm F}$ and $d=d_{\rm aff}(W,W_*)$,
\begin{equation}\label{eq:quadratic-residual-proof}
 f_V(W)-f_\kappa^\star
 \le M\eta d-\tfrac\mu2d^2\le\frac{M^2\eta^2}{2\mu}.
\end{equation}
Dropping the curvature term and using $d\le2\sqrt{2N}\ln M$ instead
recovers the linear bound. This proves \cref{prop:certificate}.

For projected gradient, write $D_k\coloneqq W_{k+1}-W_k$.
Projection optimality and $L$-smoothness give
\begin{equation}\label{eq:projected-descent}
 f_V(W_k)-f_V(W_{k+1})
 \ge(\alpha^{-1}-L/2)\norm{D_k}_{\mathrm F}^2
 \ge\frac1{2\alpha}\norm{D_k}_{\mathrm F}^2.
\end{equation}
The normal
$Z_k\coloneqq-D_k/\alpha-G_V(W_k)$ belongs to $N_{\cK}(W_{k+1})$,
and its residual there obeys
\begin{equation}\label{eq:projected-residual-size}
 \norm{G_V(W_{k+1})+Z_k}_{\mathrm F}
 \le(L+\alpha^{-1})\norm{D_k}_{\mathrm F}
 \le2\alpha^{-1}\norm{D_k}_{\mathrm F}.
\end{equation}
Apply \cref{eq:quadratic-residual-proof} at $W_{k+1}$ and combine with
\cref{eq:projected-descent} to obtain
$e_{k+1}\le[4M^2/(\mu\alpha)](e_k-e_{k+1})$.
This is \cref{eq:projected-rate}; it implies convergence of values and,
by compactness and uniqueness, convergence of the iterates. The argument
uses exact projections. Inexact projections need a certified feasibility
and normal-residual error, not just a small iterate difference.

\subsection{Geometrically graded quadrature and bounded semidefinite programs}
\label{app:dyadic-quadrature}
The quadrature below handles the nearly singular endpoint without a
uniformly fine grid. Define interval endpoints recursively by
\begin{equation}\label{eq:dyadic-partition}
 a_0\coloneqq0,\qquad
 a_{j+1}\coloneqq\min\{1,(a_j+\kappa)/2\},
\end{equation}
until $a_J=1$. There are
$J=\lceil\log_2(\kappa/(\kappa-1))\rceil$ intervals.
On each interval use the $r$-point Gauss--Legendre rule. Its nodes are
interior, its weights are positive and sum to the interval length, and
it integrates polynomials of degree at most $2r-1$ exactly
\cite[Section~3.5(v)]{DLMF}. Let $\Phi_{V,J,r}$ denote the resulting
approximation to \cref{eq:phi}.

\begin{samepage}
\begin{lemma}[Uniform geometric quadrature error]\label{lem:graded-quadrature}
For every $W\in\cK$, $r\ge1$, and the partition \eqref{eq:dyadic-partition},
\begin{equation}\label{eq:graded-error}
 |\Phi_{V,J,r}(W)-\Phi_V(W)|
 \le\frac32\Tr(V)q_0\,3^{-2r},
 \qquad q_0=\frac{M}{1-\kappa^{-1}}.
\end{equation}
The same bound holds for pure Gaussian input states.
\end{lemma}
\end{samepage}
\begin{proof}
Let an interval have center $c$ and length $h$.
The construction ensures $h/[2(\kappa-c)]\le1/3$.
Set $A_c\coloneqq W+ic\Omega$ and
$B_c\coloneqq A_c^{-1/2}i\Omega A_c^{-1/2}$.
Both $A_c+(\kappa-c)i\Omega$ and
$A_c-(\kappa-c)i\Omega$ are positive semidefinite. The second assertion
uses $|2c-\kappa|\le\kappa$ and the conjugate uncertainty inequality
$W-i\kappa\Omega\succeq0$. Thus
$\norm{B_c}\le(\kappa-c)^{-1}$, and on the interval
\begin{equation}\label{eq:resolvent-neumann}
 (W+it\Omega)^{-1}
 =A_c^{-1/2}\sum_{j=0}^{\infty}[-(t-c)B_c]^jA_c^{-1/2}.
\end{equation}
Truncating at degree $2r-1$ leaves operator-norm error at most
$\frac32q_0\,3^{-2r}$, because $\norm{A_c^{-1}}\le q_0$ and
$\norm{(t-c)B_c}\le1/3$. The corresponding scalar integrand
$\Tr[C_V(W+it\Omega)^{-1}]/2$ has polynomial-approximation error at most
$\frac34\Tr(V)q_0\,3^{-2r}$, using $C_V\succeq0$ and $\Tr C_V=\Tr V$.
The quadrature integrates this polynomial exactly. Positivity and total
weight $h$ bound the sum of the integral remainder and the quadrature
remainder by twice $h$ times that supremum. Summing $h=1$ proves
\cref{eq:graded-error}.
\end{proof}

For uniform error at most $\eta$, choose
\begin{equation}\label{eq:quadrature-degree}
 r\ge\max\left\{1,\left\lceil
 \frac{\ln^+(3\Tr(V)q_0/(2\eta))}{2\ln3}
 \right\rceil\right\}.
\end{equation}
The total number $n_q=Jr$ of nodes obeys
\cref{eq:node-count-main}. This proof uses only a convergent inverse
series and polynomial exactness. It does not assume that a fixed coarse
quadrature is accurate for every $\kappa$.

There is a square-root-free SDP representation, \cref{eq:inverse-SDP},
whose objective coefficients are the entries of $C_V$ itself.
It can be made bounded by adding
\begin{equation}\label{eq:auxiliary-caps}
 0\preceq Y_\ell\preceq2q_0I,
 \qquad 0\preceq G_A\preceq MI_A,\quad
        0\preceq G_B\preceq MI_B
\end{equation}
when local witnesses are present. These caps do not change the optimum.
For each fixed feasible $W$, one may take
$Y_\ell=(W+it_\ell\Omega)^{-1}\preceq q_0I$; every local witness is
already bounded above by the corresponding principal block of $W$.

For $M\ge4$ and $1<\kappa\le2$, an explicit strictly feasible point is
\begin{equation}\label{eq:strict-SDP-point}
 W^{\rm c}\coloneqq3I,
 \qquad G_A^{\rm c}\coloneqq\tfrac32I_A,\quad
        G_B^{\rm c}\coloneqq\tfrac32I_B,
 \qquad Y_\ell^{\rm c}\coloneqq I.
\end{equation}
The local witnesses are omitted for the PPT formulation.
The local uncertainty margins are $1/2$, the scaled global uncertainty
margin is at least one, and $W^{\rm c}-(G_A^{\rm c}\oplus G_B^{\rm c})
\succeq(3/2)I$. The inverse block has minimum eigenvalue at least
$(3-\sqrt5)/2>1/3$, because $3I+it_\ell\Omega\succeq2I$.
The upper caps also have positive slack, since $q_0\ge8$.
These uniform margins hold for every quadrature node, regardless of
its proximity to one. They will supply an explicit interior ball for
finite-precision optimization.

\subsection{Bit complexity, feasibility repair, and full error accounting}
\label{app:bit-cost}
We prove \cref{thm:runtime} for rational covariance data. The computational
model counts elementary operations on bits, not unit-cost arithmetic on
arbitrary real numbers. The SDP result used here is the bounded,
strictly feasible approximation theorem originating in
\cite{GroetschelLovaszSchrijver1993}, in the formulation and
interior-point treatment of \cite[Theorem~1.1]{DeKlerkVallentin2016}.
It gives rational feasible
solutions of prescribed objective accuracy with cost polynomial in the
data length, the logarithm of an outer-to-inner radius ratio, and the
number of requested accuracy bits. Its rounding analysis also explains
why the lengths of intermediate iterates must be controlled. We verify
the requisite radii below rather than assuming a complexity guarantee
for an arbitrary semidefinite feasibility problem.

First, the covariance bound can be fixed before an accuracy is requested.
A rational $\Lambda\ge\norm V$ can be obtained, for example, from the
maximum absolute row sum of the rational symmetric matrix $V$.
By \cref{prop:noise}, the product-reference relative entropy and entropy
subadditivity give
\begin{equation}\label{eq:initial-B-bound}
 \begin{split}
 f_V(2(V_A\oplus V_B))
 &\le f_V(V_A\oplus V_B)+N\ln2\\
 &=s(V_A)+s(V_B)-s(V)+N\ln2\\
 &\le N[1+\ln(\Lambda+1)].
 \end{split}
\end{equation}
For the last step, every marginal symplectic eigenvalue is at most
$\Lambda$, and
$g_{\rm e}(x)=\ln(x+1)+x\ln(1+1/x)\le\ln(x+1)+1$.
The limiting value at $x=0$ is included. Thus a rational $B\ge1$ strictly
larger than the last expression is computable without optimizing a
reference. An explicit integer choice is
\begin{equation}\label{eq:bit-B}
 B\coloneqq N\left[3+\left\lceil\log_2(\Lambda+1)\right\rceil\right].
\end{equation}
The ceiling is computed by exact comparisons of the rational number
$\Lambda+1$ with integer powers of two, not by rounding a numerical
logarithm.
If $a_+(B)$ is defined in \cref{eq:aplus}, then
\begin{equation}\label{eq:exponential-M}
 e^{2B}\le a_+(B)<e^{2B+1}.
\end{equation}
For example, the integer
\begin{equation}\label{eq:bit-M}
 M\coloneqq 2^{4B+2}\lceil\Lambda+1\rceil
       \ge e^{2B+1}(\Lambda+1)
\end{equation}
is valid in \cref{eq:K} and has $M\ge4$. Its bit length is polynomial
in $N$ and the input length,
but its \emph{magnitude} can be exponentially large. The time to construct
these bounds is covered by the total-cost estimate; their numerical
magnitudes are not suppressed.

Next, fix $\varepsilon=2^{-p}$ and take the rational margin
$\kappa=1+\varepsilon/(8N)$. This lies in $(1,2]$ and
\begin{equation}\label{eq:bit-margin}
 N\log_2\kappa\le\frac{\varepsilon}{8\ln2}<\frac\varepsilon4.
\end{equation}
Set $\delta=\varepsilon\ln2/2$ and choose a rational inner accuracy
$0<\zeta\le\mu\delta/2$. For example,
$\zeta=\mu\varepsilon/8$ is sufficient. The quadrature is chosen so that
its uniform error, together with coefficient-rounding error, is at most
$\zeta/4$. Solving the rounded SDP to error $\zeta/4$ and allowing another
$\zeta/4$ for iterate rounding and feasibility repair then gives a true
surrogate gap at most $\zeta$: the uniform approximation contributes
twice its error when comparing a minimizer with a returned point.
A sufficient iteration count requiring only rational arithmetic is
\begin{equation}\label{eq:bit-iteration-count}
 K\coloneqq\left\lceil\mu^{-1}\left[1+
          \left\lceil\log_2(8B/\varepsilon)\right\rceil\right]\right\rceil.
\end{equation}
Indeed, $\delta>\varepsilon/4$, so the exponent in this count exceeds
$\ln(2B/\delta)$. As in \cref{eq:bit-B}, comparisons with powers of
two compute the inner ceiling exactly. The looser count changes only
the constant in the stated asymptotic bound.

We now detail these finite-precision assertions.
The $r$-point Gauss--Legendre nodes are the roots of a degree-$r$
polynomial with rational coefficients of bit length polynomial in $r$.
Its roots are real and simple. Rational root isolation followed by
refinement computes each node to prescribed precision in polynomial time
\cite[Theorems~2 and~3]{SagraloffMehlhorn2015}; its weights are obtained
from the polynomial derivative. Positivity of
the weights and their sum can be preserved by positive rational
approximation followed by normalization. The interval endpoints in
\cref{eq:dyadic-partition} are rational. Node errors $\theta$ and total
absolute weight error $\theta$ change the objective uniformly by at most
\begin{equation}\label{eq:quadrature-rounding}
 \tfrac12\Tr(V)(q_0^2+q_0)\theta,
\end{equation}
because $\norm{\partial_t Q_t}\le q_0^2$ and $\norm{Q_t}\le q_0$.
All rounded nodes are kept inside $[0,1]$. Thus polynomially many bits in
$\ln q_0$, the input length, $r$, and $\ln(1/\zeta)$ suffice for the
coefficient accuracy. Polynomial root separation also controls the
precision needed to preserve positive weights. These claims concern
certified rational arithmetic, not a double-precision quadrature call.

Let $d\coloneqq2N$. Parameterize the real symmetric and complex Hermitian
variables in \cref{eq:inverse-SDP,eq:auxiliary-caps} by their independent
real entries. Their number $D_0$ is at most $(n_q+3)d^2$.
The strict point \cref{eq:strict-SDP-point} has a Euclidean ball of radius
$r_{\rm in}=1/(64D_0)$ contained in the feasible set. Indeed, the Frobenius norm of a symmetric or Hermitian matrix
perturbation is at most $\sqrt2$ times the Euclidean norm of its
independent-entry vector. Each constraint contains either one variable,
a block diagonal pair, or $W-(G_A\oplus G_B)$; its operator-norm
perturbation is therefore at most three times the norm of the full
independent-entry perturbation. Realification preserves the operator
norm. Every margin at the center is greater than $1/3$, so the stated
radius is sufficient.
The caps imply containment in a ball about the same point of radius
\begin{equation}\label{eq:SDP-radius}
 R_{\rm out}\coloneqq8D_0(M+q_0+1).
\end{equation}
A smaller positive rational inner radius or a larger rational outer
radius may be used. Passing to the
block-diagonal standard SDP form is an injective affine transformation
on these coordinates. Its norm and inverse norm on its image are bounded
by explicit polynomial factors in $D_0$; each variable is visible in its
own capped diagonal block. Thus the relative affine-space radii required
by the bounded-SDP theorem remain of the same logarithmic order.
All coefficients are rational after the quadrature rounding.
Their lengths and $\ln(R_{\rm out}/r_{\rm in})$ are polynomial in
$N,b,\ln M,\ln\Lambda,p$, and $\ln B$.

It remains to prevent arithmetic growth across outer iterations.
Suppose a rational feasible SDP point $x$ has been computed. Let $x^{\rm c}$
be the strict point above and set
$x_\lambda\coloneqq(1-\lambda)x+\lambda x^{\rm c}$.
Every matrix inequality at $x_\lambda$ has slack at least $\lambda/3$.
Round its independent entries to a rational grid fine enough that all
matrix perturbations have norm at most $\lambda/6$. The resulting point
is still feasible for the \emph{exact} covariance constraints.
The linear objective is bounded on the capped domain by
\begin{equation}\label{eq:SDP-objective-bound}
 F_{\max}\coloneqq\tfrac d2M^2+q_0\Tr V.
\end{equation}
Choose a rational $\lambda$ of order $\zeta/(1+F_{\max})$ and make the
rounding-grid error small enough that its objective effect is at most
$\zeta/8$. Decreasing its constant if necessary makes the combined mixing
and rounding loss at most $\zeta/4$. The required grid precision is
polynomial in the listed logarithmic parameters and is fixed independently
of the iteration number. This procedure bounds the encoding length of
$W_{k+1}$ and of all witnesses uniformly across outer iterations.
Inversion of the resulting positive definite rational $W_{k+1}$ is
polynomial-time rational linear algebra
\cite{DeKlerkVallentin2016}.

Consequently, each true surrogate can be minimized to accuracy $\zeta$
with a feasible output in a number of bit operations polynomial in
$N,b,\ln M,\ln\Lambda,\ln B,p$. Here
$\ln(1/\zeta)=O(p+\ln M+\ln\Lambda+1)$ and
$\ln q_0=O(\ln M+\ln N+p+1)$, so \cref{eq:node-count-main} gives a
polynomial number of nodes. Multiplication by the outer-iteration count
\cref{eq:iteration-count} proves \cref{eq:runtime-work}.

Finally, evaluate $f_V(\widehat W)/\ln2$ from above with error at most
$\varepsilon/4$. The determinant and integral terms are controlled by
rational linear algebra and \cref{lem:graded-quadrature}. The symplectic
eigenvalues needed for $s(V)$ are the positive eigenvalues of $i\Omega V$;
for rational $V$, they are real algebraic numbers described by a
polynomial of controlled degree and coefficient length. Exact square-free
factorization records multiplicities before root isolation; it avoids
assuming a simple symplectic spectrum. Certified root refinement
\cite{SagraloffMehlhorn2015} and elementary logarithm evaluation
\cite[Chapter~4]{BrentZimmermann2010} give the desired entropy accuracy.
A root exactly equal to one is detected algebraically and contributes
$g(0)=0$. For roots close to one, the continuity modulus
$g(x)=O(x\log(1/x))$ as $x\downarrow0$ ensures that polynomially many
additional accuracy bits suffice. The support singularity of the
reference is absent because $\widehat W+i\kappa\Omega\succeq0$.
The resulting rational upper enclosure is $U$ in
\cref{eq:runtime-output}. The sum of regularization, optimization, and
one-sided evaluation errors is less than $\varepsilon$.

This establishes a finite-precision existence and complexity theorem,
not a claim that the accompanying floating-point example is a certified
SDP implementation. Nor does it cover an arbitrary unspecified real
covariance oracle. Extending the input class requires a stated efficient
procedure for evaluating its entries and maintaining valid covariance
constraints. The prefactor $M^2\Lambda^2$ remains the obstruction in this
bound to a uniform polynomial-time statement in the input encoding size.

\section{Derivative and quadrature bounds}\label{app:derivatives}
This appendix supplies the gradient smoothness bounds used by the
projected-gradient alternative and a simple midpoint quadrature estimate.
The sharper geometrically graded quadrature for the runtime theorem is
proved in \cref{app:dyadic-quadrature}; the midpoint estimate below remains
useful as a transparent alternative. All norms and derivatives are
finite-dimensional.

\subsection{Gradient and smoothness constants}
Differentiating \cref{eq:resolvent} in a real symmetric direction $X$ gives
\begin{equation}\label{eq:direction-gradient}
 df_V(W)[X]=\frac12\Tr[W^{-1}X]
 -\frac12\int_0^1\Tr[C_VQ_tXQ_t]dt,
\end{equation}
which proves \cref{eq:gradient}. On $\cK$, the scaled uncertainty relation
implies
\begin{equation}\label{eq:Q-norm}
 \norm{itW^{-1/2}\Omega W^{-1/2}}\le t/\kappa,
 \qquad \norm{Q_t(W)}\le\frac{M}{1-t/\kappa}\le q_0.
\end{equation}
Here $\norm{W^{-1}}\le M$ and the eigenvalues of the first Hermitian matrix
are $\pm t/\nu_j(W)$. Differentiating the gradient yields
\begin{equation}\label{eq:grad-derivative}
 \begin{split}
 dG_V(W)[X]={}&-\frac12W^{-1}XW^{-1}\\
 &+\frac12\int_0^1\Rea[
       Q_tXQ_tC_VQ_t+Q_tC_VQ_tXQ_t]dt.
 \end{split}
\end{equation}
Using $\norm{AXB}_{\mathrm F}\le\norm{A}\norm{X}_{\mathrm F}\norm{B}$
in each term gives
\begin{equation}\label{eq:L-proof}
 \norm{dG_V(W)[X]}_{\mathrm F}
 \le\left(\frac12M^2+\norm{C_V}q_0^3\right)\norm{X}_{\mathrm F}.
\end{equation}
Integrating along line segments inside the Euclidean-convex set $\cK$
proves the stated Lipschitz bound.

\subsection{Explicit midpoint quadrature errors}
Consider the composite midpoint rule on $J$ equal subintervals of $[0,1]$,
with nodes $t_\ell\coloneqq(\ell-1/2)/J$. Write $\Phi_{V,J}$ and $G_{V,J}$ for
\cref{eq:phi,eq:gradient} with their integrals replaced by this rule.
Let $d\coloneqq2N$. Then, uniformly over $W\in\cK$,
\begin{align}
 |\Phi_{V,J}(W)-\Phi_V(W)|
 &\le\frac{\Tr(C_V)q_0^3}{24J^2},\label{eq:quad-value}\\
 \norm{G_{V,J}(W)-G_V(W)}_{\mathrm F}
 &\le\frac{\sqrt d\,\norm{C_V}q_0^4}{8J^2}.
 \label{eq:quad-gradient}
\end{align}
Although $C_V$ may have complex entries, its trace is real and equals
$\Tr V$.

To verify these estimates, set $B_0\coloneqq i\Omega$, so $\norm{B_0}=1$.
Differentiation with respect to $t$ gives
\begin{equation}\label{eq:Qt-derivatives}
 Q_t'=-Q_tB_0Q_t,\qquad
 Q_t''=2Q_tB_0Q_tB_0Q_t.
\end{equation}
For $\psi(t)\coloneqq\Tr[C_VQ_t]/2$, positivity of $C_V$ and
\cref{eq:Q-norm} imply
$|\psi''(t)|\le\Tr(C_V)q_0^3$.
The second derivative of $Q_tC_VQ_t$ is the sum of
$Q_t''C_VQ_t$, $2Q_t'C_VQ_t'$, and $Q_tC_VQ_t''$.
After the factor $1/2$ in the gradient, its Frobenius norm is at most
$3\sqrt d\,\norm{C_V}q_0^4$. The scalar and Banach-space midpoint error
bound $\sup\norm{h''}/(24J^2)$ gives
\cref{eq:quad-value,eq:quad-gradient}. It follows, for example, by the
second-order integral remainder on each centered subinterval and the
triangle inequality.

These bounds can be used to certify a computed gradient in a residual
estimate. They do not by themselves certify an approximate projection or
a semidefinite normal: those numerical errors need their own bounds.
Increasing $J$ reduces the stated quadrature error, while small
$\kappa-1$ can make the conservative constants large.

\section{Details of the unrestricted cutoff algorithm}\label{app:cutoff-details}
The cutoff argument uses finite-dimensional optimization only after the
quantum projection has been defined. This appendix checks the
subnormalized-reference step and explains how the finite matrices and
objective bounds are obtained.

\subsection{Selective monotonicity for the cutoff pinching}
We prove the lower bound in \cref{eq:selective} directly for the projectors
used here. They are real in the Fock basis, so partial transposition
commutes with the local block pinching. For an arbitrary
$\tau\in\mathrm{PPT}'$, set $\tau_{ij}\coloneqq P_{ij}\tau P_{ij}$ and
$r_{ij}\coloneqq\norm{\tau_{ij}^{\Gamma_B}}_1$. Trace-norm contractivity of pinching on
Hermitian trace-class operators gives
\begin{equation}\label{eq:block-norms}
 \sum_{ij}r_{ij}=\norm{\Pi(\tau)^{\Gamma_B}}_1
 =\norm{\Pi(\tau^{\Gamma_B})}_1\le\norm{\tau^{\Gamma_B}}_1\le1.
\end{equation}
Every nonzero $\tau_{ij}/r_{ij}$ is a feasible Rains reference in its sector.
Relative-entropy data processing followed by the direct-sum formula gives
\begin{align}
 D(\rho\Vert\tau)
 &\geq  D(\Pi(\rho)\Vert\Pi(\tau))\nonumber\\
 &=\sum_{ij:p_{ij}>0}p_{ij}
       D(\rho_{ij}\Vert\tau_{ij}/r_{ij})
     +\sum_{ij:p_{ij}>0}p_{ij}\log_2\frac{p_{ij}}{r_{ij}}\nonumber\\
 &\geq \sum_{ij}p_{ij}R(A;B)_{\rho_{ij}}.
 \label{eq:block-rains-proof}
\end{align}
If a positive-probability sector has $r_{ij}=0$, the relative entropy is
infinite and the inequality is automatic. Otherwise the last classical
term is nonnegative because $\sum r_{ij}\le1$: normalize this subprobability
distribution and use classical relative-entropy nonnegativity. Taking the
infimum over $\tau$ proves the Rains statement. For REE, use normalized
separable references, their sector traces in place of $r_{ij}$, and
normalized separable conditional references. This proves the corresponding
statement without an infinite-dimensional continuity assumption on the
entanglement measure itself.

\subsection{Exact Fock matrix elements from the covariance}\label{app:fock}
For a centered Gaussian state, write
$\xi=(\sqrt2\Rea\alpha_1,\sqrt2\operatorname{Im}\alpha_1,\ldots)^{\top}$.
Its coherent-state diagonal is
\begin{equation}\label{eq:coherent}
 \bra{\boldsymbol\alpha}\rho_V\ket{\boldsymbol\alpha}
 =\frac{2^N}{\sqrt{\det(V+I)}}
      \exp[-\xi^{\top}(V+I)^{-1}\xi].
\end{equation}
For the vacuum this gives $e^{-\norm{\boldsymbol\alpha}^2}$, checking the
normalization. The formula follows from the overlap of the Gaussian
Wigner function with a coherent-state Wigner function
\cite{Serafini2023}. Expanding coherent states in the Fock basis gives
\begin{equation}\label{eq:Fock-generate}
 e^{\norm{\boldsymbol\alpha}^2}
 \bra{\boldsymbol\alpha}\rho_V\ket{\boldsymbol\alpha}
 =\sum_{\boldsymbol j,\boldsymbol k}
 \frac{\overline{\boldsymbol\alpha}^{\boldsymbol j}
       \boldsymbol\alpha^{\boldsymbol k}}
      {\sqrt{\boldsymbol j!\boldsymbol k!}}
 \bra{\boldsymbol j}\rho_V\ket{\boldsymbol k}.
\end{equation}
Treating $\boldsymbol\alpha$ and its conjugate as independent variables,
\begin{equation}\label{eq:Fock-derivative}
 \bra{\boldsymbol j}\rho_V\ket{\boldsymbol k}
 =\left.
 \frac{\partial_{\overline{\boldsymbol\alpha}}^{\boldsymbol j}
       \partial_{\boldsymbol\alpha}^{\boldsymbol k}}
      {\sqrt{\boldsymbol j!\boldsymbol k!}}
 \left[e^{\norm{\boldsymbol\alpha}^2}
       \bra{\boldsymbol\alpha}\rho_V\ket{\boldsymbol\alpha}\right]
 \right|_{\boldsymbol\alpha=0}.
\end{equation}
The differentiated expression is an exponential quadratic, so a finite
recursion or polynomial expansion gives the finite set of entries needed
for $P\rho_VP$. Its trace is $p$; division by this trace gives $\rho_{00}$.
Displacements can first be removed locally without changing entanglement.
For reference, the total photon number of a covariance $V$ with mean $d$ is
$(\Tr V-2N)/4+\norm{d}^2/2$.

\subsection{Finite-dimensional reference approximations}
\label{app:finite-optimization}
Let $\omega$ be a normalized state in a finite bipartite Hilbert space of
total dimension $d$. Both separable and $\mathrm{PPT}'$ references permit
the regularization
\begin{equation}\label{eq:finite-reg}
 \tau_\delta\coloneqq(1-\delta)\tau+\delta I_d/d,
 \qquad 0<\delta<1.
\end{equation}
For a separable reference this preserves unit trace and separability. For
a Rains reference, convexity of the partial-transpose trace norm preserves
$\mathrm{PPT}'$; also $\Tr\tau_\delta\le1$. Since
$\tau_\delta\succeq(1-\delta)\tau$, operator monotonicity of the logarithm,
with a limiting interpretation on the support of $\omega$, yields
\begin{equation}\label{eq:finite-reg-error}
 D(\omega\Vert\tau_\delta)
 \le D(\omega\Vert\tau)-\log_2(1-\delta).
\end{equation}
Every eigenvalue of $\tau_\delta$ is at least $\delta/d$, and at most one.
Consequently uniform matrix-logarithm approximations on $[\delta/d,1]$
control the relative-entropy objective, with an explicit regularization
error tending to zero. The regularization in \cref{eq:finite-reg} supplies the bounded-spectrum setting needed
for the approximations of \cite{FawziSaundersonParrilo2019}.

For REE, let $\mathcal S_k$ be a nested complete family of normalized
symmetric-extension outer approximations to separability, optionally
including their PPT constraints \cite{DohertyParriloSpedalieri2004}. These
are compact sets satisfying
$\mathcal S_{k+1}\subseteq\mathcal S_k$ and
$\bigcap_k\mathcal S_k=\mathrm{SEP}$. Let
$L_k\coloneqq\min_{\sigma\in\mathcal S_k}D(\omega\Vert\sigma)$.
The values increase and are bounded above by $E_R(\omega)$. To check that
the limit is exactly the desired value, choose minimizers $\sigma_k$ and
extract a convergent subsequence. For every fixed $\ell$, all sufficiently
late members belong to $\mathcal S_\ell$, hence their limit $\sigma_*$ is
separable. Lower semicontinuity gives
\begin{equation}\label{eq:DPS-limit}
 E_R(\omega)\le D(\omega\Vert\sigma_*)
          \le\liminf_k L_k\le E_R(\omega).
\end{equation}
A faithful feasible product reference ensures the values are finite.

For upper bounds, take nested finite sets of product pure states with
union dense in the product-state set, and include a fixed product basis
so that their convex hulls contain $I_d/d$. Denote these hulls by
$\mathcal T_k\subseteq\mathrm{SEP}$. The minima over $\mathcal T_k$ decrease.
To prove convergence, first replace a nearly optimal separable reference
by its faithful regularization \cref{eq:finite-reg}. Approximate its
unregularized part by a finite product-state mixture from some
$\mathcal T_k$, and retain the same $\delta I_d/d$ component. The resulting
references remain bounded below by $\delta I_d/d$ and converge in operator
norm. Their relative entropies converge because the logarithm is continuous
on that spectral interval. \Cref{eq:finite-reg-error}, followed by
$\delta\downarrow0$, proves that the upper values tend to $E_R(\omega)$.
This argument explains why approximation of the reference sets alone would
not suffice without controlling the logarithmic singularity.

All these are finite-dimensional optimization problems at each fixed
cutoff and approximation level. Their computational cost can be large.
For computable Gaussian input state parameters, certified evaluation of finite
matrix elements and optimization bounds, together with
\cref{eq:final-interval}, gives arbitrary-precision two-sided approximations.
The present numerical package does not implement this full unrestricted
hierarchy.

\section{Numerical checks and reproducibility}\label{app:numerics}
The analytical proofs establish the statements for all specified inputs.
The accompanying computations are consistency checks and evaluations of
those formulas, not replacements for proofs or claims of external review.
The scripts use the vacuum-covariance-$I$ convention throughout. This appendix describes the structural tests; \cref{app:operational-details} describes the additional operational calculations.

The function \texttt{symmetric\_gaussian\_entanglement} implements
\cref{thm:scalar} by bracketing the zero of \cref{eq:root}. For small $t$, it
uses the positive power series to avoid cancellation in the derivative.
It returns a numerical value and the optimal reference covariance matrix.
\texttt{faithful\_gap\_check.py} instead evaluates the explicit expressions
in \cref{eq:scalar-certified,eq:intervals} with interval arithmetic and
rational parameter initialization. The latter is a bound on the entire
scalar domain, not a check at a putative optimizer.

The finite-matrix tests compare the Gibbs and resolvent formulas, the
analytic gradient with directional finite differences, and the geodesic
Hessian with finite differences along the actual matrix-geometric-mean
path. Random tests also check the order $W_+\succeq W$, PPT of the
replacement, and the determinant identity \cref{eq:volume-neg}. Tests
include pure input states; separate boundary examples exercise the improper
integral and strict convexity at nonfaithful references. Seeds, numerical
outputs, and the generation script for the plot data are retained in the
source package.

The general-purpose covariance optimizer supplied as a numerical example
uses multistart local optimization and explicit feasibility repair. Its
outputs are feasible upper-bound estimates, not certified global optima.
The globally convergent methods of \cref{sec:algorithms} are specified and
proved here, but the accompanying code is not a verified projection solver
or a certified implementation of their normal-residual tests. Likewise,
the unrestricted cutoff algorithm is proved but not supplied as a complete
end-to-end solver. These implementation distinctions are necessary when
interpreting numerical output.

The runtime revision adds \texttt{runtime\_checks.py}. It independently
checks the normalized geodesic Hessian and the quantitative lower-curvature
estimate, the surrogate remainder on geometric paths, the positive-weight
quadrature estimate, and the strict SDP center. A scalar symmetric-state
implementation illustrates majorization against the independently computed
one-variable optimum. Its floating-point values are consistency checks;
the scalar demonstration is not an arbitrary-mode certified SDP solver.
The code and test records distinguish inherited numerical data from checks
executed for this revision.

\section{Regularized target tests and numerical operational protocols}
\label{app:operational-details}
This appendix makes explicit two supporting points that are easy to miss:
a single-copy testing bound cannot simply be divided by a regularization
parameter, and a heralded protocol's successful conditional rate must be
weighted by its success probability. Neither point requires Gaussian
optimality for an unrestricted entanglement measure.

\subsection{A target test remains useful after regularization}
The scalar sector converse propagates $E_R^\infty$, rather than $E_R$.
The following argument verifies the test used in \cref{thm:multisector}.
It is consistent with the regularized privacy-test treatments in
\cite[Appendices~C--D]{PirandolaSector2026} and
\cite[Section~5.1]{FalcoEtAl2026}. Here $E_R^\infty$ is the regularized
relative entropy of entanglement defined in \cref{prop:state-operational}.
The binary relative entropy appearing below is defined by
\begin{equation}\label{eq:binary-relative-entropy}
 D_{\rm bin}(p\Vert q)\coloneqq
 p\log_2\frac pq+(1-p)\log_2\frac{1-p}{1-q},
 \qquad 0\le p\le1,\quad 0<q<1.
\end{equation}
The terms at $p=0$ or $p=1$ are interpreted by continuity, with
$0\log_2 0=0$. Unlike a one-copy estimate alone, the next argument controls
separable reference states correlated across the repetitions used in
regularization.

\begin{samepage}
\begin{lemma}[Regularized target-test lower bound]\label{lem:regularized-test}
Let $d\ge2$, and let $T$ be a maximally entangled test of target dimension $d$, or a privacy
test of key dimension $d$, and let $\zeta$ be a normal bipartite state with
finite relative entropy of entanglement. Set $p\coloneqq\Tr[T\zeta]$. If $p\ge1/d$, then
\begin{equation}\label{eq:regularized-test}
 E_R^\infty(\zeta)\ge
 D_{\rm bin}(p\Vert1/d)\ge p\log_2d-h_2(p).
\end{equation}
The last inequality, possibly with a nonpositive right side, also holds
for $p<1/d$. A target of dimension one gives the trivial lower bound zero.
\end{lemma}
\end{samepage}
\begin{proof}
Consider the commuting copies $T_1,\ldots,T_s$ and their sum of acceptance
indicators $N_s\coloneqq\sum_{i=1}^sT_i$. For a subset $S$ of copies, the product test
$\prod_{i\in S}T_i$ has acceptance at most $d^{-|S|}$ on an arbitrary
separable state across the grouped laboratories. For a privacy test this
follows because a product of the corresponding private states is a
private state of key dimension $d^{|S|}$; first take the marginal on those
copies. Thus, for $z\ge1$ and every such separable reference $\sigma$,
\begin{equation}\label{eq:test-moment}
 \Tr\!\left[\prod_{i=1}^s(I+(z-1)T_i)\sigma\right]
 \le\sum_{S\subseteq\{1,\ldots,s\}}(z-1)^{|S|}d^{-|S|}
 =(1+(z-1)/d)^s.
\end{equation}
The subset argument allows $\sigma$ to have arbitrary correlations between
copies; no product-reference restriction is used.

For $1/d<a<p$, Markov's inequality and optimization over $z$ imply that
$\sigma$ passes the threshold test $N_s\ge sa$ with probability at most
$2^{-sD_{\rm bin}(a\Vert1/d)}$. The same test on $\zeta^{\otimes s}$ passes
with probability $p_s\to1$ by the law of large numbers. Binary data
processing therefore gives
\begin{equation}\label{eq:test-threshold}
 E_R(\zeta^{\otimes s})\ge
 p_s sD_{\rm bin}(a\Vert1/d)-h_2(p_s).
\end{equation}
Divide by $s$, take $s\to\infty$, and then $a\uparrow p$. Endpoints follow
by continuity (or directly by the all-success test for $p=1$).
If $p=1/d$, the first claim is nonnegativity. The elementary binary formula
gives the last inequality. For $p<1/d$, the function
$p\log_2d-h_2(p)$ is nonpositive: it is convex on $[0,1/d]$ and is
nonpositive at both endpoints. Nonnegativity of REE then suffices.
\end{proof}

The repeated-test construction avoids the incorrect replacement of its
success probability by $p^s$ in a converse. It also avoids any dimension
factor involving the shield. For an operational key code of vanishing
joint error, the locally retained private-state construction makes $p$
tend to one while leaving the original transmitted marginals unchanged.
This is the only property of that construction required for
\cref{thm:multisector}.

\subsection{Numerical families and explicit feasible upper bounds}
\label{app:operational-numerics}
All new examples use the covariance-$I$ vacuum convention and report rates
in bits. The files \texttt{operational\_bounds.py} and
\texttt{generate\_operational\_data.py} generate the covariances, resources,
reference witnesses, and plot data. The source Gaussian states are not
replaced by exponentials of truncated Hamiltonians. The variable $K$ below is
the position block in \cref{eq:paired-reduction}, not the Hamiltonian $H(W)$.

For the phase-invariant examples, $K\succeq I$ in
\cref{eq:paired-reduction} is a proved reduction of the complete PPT and
separable Gaussian optimization. The numerical solver uses real symmetric
$K$, its eigenvalue constraints, and the analytic resolvent gradient.
It evaluates the value separately with the Gaussian Gibbs formula.
After optimization, an explicit scalar multiple of the identity repairs
$K$ to have minimum eigenvalue at least $1+2\times10^{-7}$. This guarantees
a strictly interior quantum covariance matrix and a separable classical-noise
reference in exact arithmetic. Checking doubled resolvent quadrature
provides a consistency test, not an interval proof of a derivative error.
The output is a feasible upper-bound evaluation. The theoretical
geodesic-convexity result does not convert a floating-point stopping flag
into a rigorous global numerical certificate.

For loss-core parameters, the code checks the postprocessing CP matrix and
repairs a candidate by moving it toward the feasible identity core. For
the correlated examples, the noisy-core vector is similarly moved toward
zero until $Q-\diag\boldsymbol r\succeq0$ with slack. Thus outer search
failure can weaken a bound but does not justify returning an infeasible
parameter vector. The resource identity $L_X\Sigma L_X^{\top}=Y$ and the CP
condition on $Y_0$ in \cref{eq:mu-resource} are checked separately from the
entanglement minimization.

The inner core-energy problem is a maximization. A merely feasible
allocation would underestimate that maximum and could invalidate a claimed
upper bound. For an allocation $\boldsymbol x$ with $\sum x_j=E$ and
supporting slopes $d_j$ of the concave functions $\mathcal F_j$ at $x_j$,
we instead use
\begin{equation}\label{eq:allocation-tangent}
 B_{\boldsymbol\tau,\boldsymbol v}(E)
 \le\sum_j[\mathcal F_j(x_j)-d_jx_j]+E\max\left\{0,d_1,\ldots,d_m\right\}.
\end{equation}
For a zero coordinate use its right derivative when finite; a strictly
positive allocation can always be used instead. This is a mathematical
upper certificate on the maximum. Evaluation of its slopes in floating
point still has to be distinguished from interval certification. The
older squashing-allocation comparison uses the same tangent principle.

\subsection{Gaussian and non-Gaussian attainable protocols}
The Gaussian measurement searches use product seeds
\begin{equation}\label{eq:measurement-seed-grid}
 M_X=\bigoplus_j\gamma_j\diag(e^{2r_j},e^{-2r_j}),\qquad
 1\le\gamma_j\le e^7,\quad |r_j|\le3.5.
\end{equation}
Both directions are tested with several initial points. The conditional
Eve covariance is calculated from a Williamson purification of the full
input state. It is not obtained by purifying just the receiver's marginal.
Continuous outcomes can be discretized and truncated before applying the
finite-alphabet coding theorem: refining measurable outcome partitions
converges to the finite Holevo informations, while finite-energy local
cutoffs give finite-dimensional quantum receivers. Therefore every
strictly smaller positive rate in \cref{eq:CQ-key} is attainable within the
finite-round model. The measurement search neither proves optimality over
all multimode seeds nor imposes Gaussian operations on the final coding
protocol.

For the three-mode state, Alice keeps Fock states $\ket0,\ket1$ and applies
the diagonal filter
\begin{equation}\label{eq:state-filters}
 F_z\coloneqq\diag\{\min(1,e^z),\min(1,e^{-z})\}.
\end{equation}
Bob keeps $\ket{00}$ and $\cos\theta\ket{10}+\sin\theta\ket{01}$.
The search uses nine equally spaced $z\in[-2,2]$ and thirteen equally
spaced $\theta\in[-\pi/2,\pi/2]$. Both are contractions, hence valid local
successful Kraus operators. The exact subnormalized $4\times4$ matrix is
computed with \cref{eq:Fock-derivative}; its trace is the success probability.

For the marginal channel protocols, Alice prepares
\begin{equation}\label{eq:hybrid-input}
 \ket{\psi_c}\coloneqq\sqrt{1-c}\ket0_R\ket{M_1}_A
                    +\sqrt c\ket1_R\ket{M_2}_A,
 \qquad e_c\coloneqq(1-c)M_1+cM_2,
\end{equation}
and Bob keeps $\ket{L_1},\ket{L_2}$. We search
$M_1\in\{0,1\}$, $M_1<M_2\le3$, $L_1\in\{0,1,2\}$,
$L_2-L_1=M_2-M_1$, and fifteen $c$ values equally spaced in
$[0.025,0.975]$. This is a finite subset of the Fock-pair constructions in
\cite{MeleLamiGiovannetti2025,BarberPirandola2026}, not a reproduction of
their complete optimizations.
For convenience, the channel matrix elements can be extracted from a
finite-energy squeezed-input output $\omega$: if its squeezed Schmidt
probabilities are $p_j>0$, then
\begin{equation}\label{eq:channel-Fock-extraction}
 \bra\ell\mathcal N(\ket j\bra k)\ket h
 =\frac{\bra{j,\ell}\omega\ket{k,h}}{\sqrt{p_jp_k}}.
\end{equation}
This is an exact coefficient identity. The squeezed state used to extract
coefficients is not the distribution protocol's input. The actual input
energy is $e_c$ in \cref{eq:hybrid-input}, including all failed outputs.

After local phase alignment and Bell twirling, write the normalized Bell
probabilities as $(a,b,c,d)$. One recurrence step, with a chosen Bell-label
pairing, succeeds with probability
\begin{equation}\label{eq:recurrence-probability}
 N\coloneqq(a+b)^2+(c+d)^2,
\end{equation}
and produces, up to a local relabeling, probabilities
\begin{equation}\label{eq:recurrence-update}
 \frac1N(a^2+b^2,\,2cd,\,c^2+d^2,\,2ab).
\end{equation}
Two input pairs are consumed, so the accumulated yield is multiplied by
$N/2$, not by $N$. For a normalized four-component Bell probability vector
$\boldsymbol p$, define its Shannon entropy by
\begin{equation}\label{eq:Bell-Shannon-entropy}
 H_4(\boldsymbol p)\coloneqq-\sum_{j=1}^4p_j\log_2p_j,
 \qquad 0\log_2 0\coloneqq0.
\end{equation}
Final hashing yields $\max\left\{0,1-H_4(\boldsymbol p)\right\}$ ebits per
surviving pair, evaluated at the probability vector after the last
recurrence step. The finite search tests all three pairings at each retained point,
keeps at most 24 candidate probability vectors per depth, and uses at most
nine recurrence rounds. Any selected path is an explicit achievable
protocol. Pruning cannot create an invalid lower bound; it can only miss
a better protocol. Direct hashing of the untwirled conditional state is
also tested. All final yields are multiplied by the initial local-filter
success probability.

For a fixed channel, convex combinations of achieved energy--rate points
are attained by allocating deterministic fractions of channel uses to the
corresponding codes. A small linear program with constraints
$\sum_a p_a=1$ and $\sum_a p_a e_a\le E$ generates the displayed time-sharing
envelopes. The vacuum point $(0,0)$ is included. This procedure is used for
both the marginal Fock-pair protocols and the final Gaussian/non-Gaussian
channel lower-bound envelope. It is not applied between distinct source
states in \cref{fig:operational-three}.

\subsection{Scope of the numerical verification}
The expanded data include 39 two-mode source points, 29 three-mode source
points, two pure-loss channel families, and two correlated-noise channel
families. Seventy-four Gaussian reference optimizations are retained with
their full covariance witnesses. In this run, all terminated successfully;
the largest additional feasibility repair was below $4\times10^{-10}$,
and the largest doubled-quadrature gradient discrepancy was below
$1.2\times10^{-14}$. These figures describe this calculation, not universal
precision guarantees. The complete input parameters, lower-bound protocol
choices, time-sharing weights, CP residuals, and simulator residuals are
stored with the data.

Separate checks compare Gaussian Fock coefficients with thermal and pure
squeezed-state coefficients, verify the exact simulator and core composition
on rectangular and phase-sensitive examples, and compare analytic scalar
derivatives with finite differences. The new plots use exact formulas or
explicit feasible witnesses and achieved protocols. Their decimals remain
floating-point evaluations; the interval proof in \cref{thm:gap} is a
separate, genuinely interval-based computation. In particular, none of the
new plotted curves is labeled an exact unrestricted REE, an exact noisy
channel capacity, or a globally optimal general LOCC lower bound.

The separate regression script \texttt{operational\_checks.py} also checks
48 rectangular, phase-sensitive channel factorizations and two exact-resource
constructions, 45 symmetry identities, 90 scalar derivative and boundary
comparisons, and 548 low-photon matrix elements against elementary thermal
and pure-loss formulas. It re-evaluates all 74 Gaussian reference witnesses
and 60 heralded Fock-pair candidates. These finite checks are recorded with
their parameters and residuals in the data files; they are not an exhaustive
search over channels or a substitute for the theorems.

\end{document}